\documentclass[usenatbib,iop,numberedappendix]{aeb_emulateapj_2010}
\usepackage{amsmath}
\usepackage{amssymb}
\usepackage{xspace}
\usepackage[normalem]{ulem}

\usepackage{natbib}

\usepackage[usenames,dvips]{color}

\def\s{{\rm s}}

\def\Ms{{\rm M}\s} 
\def\yr{{\rm yr}}

\def\m{{\rm m}}

\def\cm{{\rm c}\m} 
\def\km{{\rm k}\m} 
\def\pc{{\rm pc}} 
\def\kpc{{\rm k}\pc} 
\def\Mpc{{\rm M}\pc} 
\def\Gpc{{\rm G}\pc}

\def\Ms{M_\odot} 

\def\eV{{\rm eV}} 
 
\def\MeV{{\rm M}\eV} 
\def\GeV{{\rm G}\eV} 
\def\TeV{{\rm T}\eV}

\def\erg{{\rm erg}} 
 
\def\K{{\rm K}}

\def\G{{\rm G}} 
\def\nG{{\rm n}\G}

\def\rad{{\rm rad}}

\def\nar{New Astron. Rev.}

\def\del#1{{}}

\def\Ms{M_\odot}
\renewcommand{\d}{{\rm d}}
\newcommand{\rmn}{\mathrm}
\newcommand{\e}{{\rm e}}
\newcommand\bmath[1] {\mbox{\boldmath$\rm #1$}}
\def\x{\bmath{x}}
\def\p{\bmath{p}}
\def\E{\bmath{E}}
\def\B{\bmath{B}}
\def\v{\bmath{v}}
\def\j{\bmath{j}}
\def\k{\bmath{k}}
\def\grad{\bmath{\nabla}}

\def\GTS{\Gamma_{\rm TS}}

\def\GIC{\Gamma_{\rm IC}}
\def\GW{\Gamma_{\rm W}}
\def\GO{\Gamma_{\rm M}}
\def\GM{\Gamma_{\rm M,k}}
\def\Tc{T_{\rm crit}}
\def\Dpp{D_{\rm pp}}
\def\O{\mathcal{O}}
\def\nb{n_{b}}
\def\nIGM{n_{\rm IGM}}

\def\Lya{Ly$\alpha$\xspace}

\def\Fermi{{\em Fermi}\xspace}

\def\QLF{\phi_Q}
\def\BLF{\phi_B}

\begin{document}

\title{
The Cosmological Impact of Luminous TeV Blazars I:\\
Implications of Plasma Instabilities for the Intergalactic Magnetic
Field and Extragalactic Gamma-Ray Background}

\author{
Avery E.~Broderick\altaffilmark{1,2,3},
Philip Chang\altaffilmark{1,4},
and
Christoph Pfrommer\altaffilmark{5,1}
}
\altaffiltext{1}{Canadian Institute for Theoretical Astrophysics, 60 St.~George Street, Toronto, ON M5S 3H8, Canada; aeb@cita.utoronto.ca, pchang@cita.utoronto.ca}
\altaffiltext{2}{Perimeter Institute for Theoretical Physics, 31 Caroline Street North, Waterloo, ON, N2L 2Y5, Canada}
\altaffiltext{3}{Department of Physics and Astronomy, University of Waterloo, 200 University Avenue West, Waterloo, ON, N2L 3G1, Canada}
\altaffiltext{4}{Department of Physics, University of Wisconsin-Milwaukee, 1900 E. Kenwood Boulevard, Milwaukee, WI 53211, USA}
\altaffiltext{5}{Heidelberg Institute for Theoretical Studies, Schloss-Wolfsbrunnenweg 35, D-69118 Heidelberg, Germany; christoph.pfrommer@h-its.org}

\shorttitle{The Cosmological Impact of Blazar TeV Emission I}
\shortauthors{Broderick, Chang \& Pfrommer}

\begin{abstract}
Inverse-Compton cascades initiated by energetic gamma rays
($E\gtrsim100\,\GeV$) enhance the GeV emission from bright,
extragalactic TeV sources.  The absence of this emission from bright
TeV blazars has been used to constrain the intergalactic magnetic
field (IGMF), and the stringent limits placed upon the unresolved
extragalactic gamma-ray background (EGRB) by \Fermi has been used to
argue against a large number of such objects at high redshifts.
However, these are predicated upon the assumption that inverse-Compton
scattering is the primary energy-loss mechanism for the
ultra-relativistic pairs produced by the annihilation of the energetic
gamma rays on extragalactic background light photons.  Here we show
that for sufficiently bright TeV sources
(isotropic-equivalent luminosities $\gtrsim10^{42}\,\erg\,\s^{-1}$)
plasma beam instabilities, specifically the 
``oblique'' instability, present a plausible mechanism by which the
energy of these pairs can be dissipated locally, heating the
intergalactic medium.  Since these instabilities typically grow on
timescales short in comparison to the inverse-Compton cooling rate,
they necessarily suppress the inverse-Compton cascades.  As a
consequence, this places a severe constraint upon efforts to limit the
IGMF from the lack of a discernible GeV bump in TeV sources.
Similarly, it considerably weakens the \Fermi limits upon the
evolution of blazar populations.
Specifically, we construct a TeV-blazar luminosity function from
those objects presently observed and find that it is very well
described by the quasar luminosity function at $z\sim0.1$, shifted to
lower luminosities and number densities, suggesting that both classes
of sources are regulated by similar processes.  Extending this
relationship to higher redshifts, we show that the magnitude and
shape of the EGRB above $\sim10\,\GeV$ is naturally reproduced with
this particular example of a rapidly evolving TeV-blazar luminosity
function.
\end{abstract}


\maketitle

\section{Introduction} \label{I}

\begin{deluxetable*}{lccccccccccl}\tabletypesize{\tiny}
\tablecaption{List of TeV Sources with Measured Spectral Properties in Decreasing $100\,\GeV$--$10\,\TeV$ Flux Order\label{tab:TeVsources}}
\tablehead{
\colhead{Name} &
\colhead{$z$} &
\colhead{$D_C$ \tablenotemark{a}} &
\colhead{$f_0$ \tablenotemark{b}} &
\colhead{$E_0$ \tablenotemark{c}} &
\colhead{$\alpha$ \tablenotemark{d}} &
\colhead{$F$ \tablenotemark{e}} &
\colhead{$\log_{10} L$ \tablenotemark{f}} &
\colhead{$\hat{\alpha}$ \tablenotemark{g}} &
\colhead{$\Delta t$ \tablenotemark{h}} &
\colhead{Class \tablenotemark{i}} &
\colhead{\mbox{Reference\hspace{2.8cm}}}
}
\startdata
Mkn 421		                 & 0.030   &  129    & 68    & 1        & 3.32  & $1.7\times10^{3}$ & 45.6 & 3.15 & $2.8\times10^2$ & H & \citet{Chandra+10} \\
1ES 1959+650	                 & 0.047   &  201    & 78    & 1 	& 3.18 	& $1.6\times10^{3}$ & 45.9 & 2.90 & 85              & H & \citet{aharonian+03a} \\
1ES 2344+514	                 & 0.044   &  190    & 120   & 0.5 	& 2.95  & $2.3\times10^{2}$ & 45.0 & 2.82 & $5.0\times10^2$ & H & \citet{albert+07c}\\
Mkn 501 \tablenotemark{j}        & 0.034   &  150    & 8.7   & 1 	& 2.58  & 85                & 44.4 & 2.39 & $1.6\times10^3$ & H & \citet{Huang+09} \\
3C 279		                 & 0.536   & 2000    & 520   & 0.2 	& 4.11 	& 68                & 46.9 & 2.53 & 2.0             & Q & \citet{magic+08} \\
PKS 2155-304	                 & 0.116   &  490    & 1.81  & 1 	& 3.53 	& 64                & 45.4 & 2.75 & $3.3\times10^2$ & H & \citet{hess+10a} \\
PG 1553+113	                 & $>0.09$ &  $>380$ & 46.8  & 0.3      & 4.46  & 41                & $>44.9$ & $<4.29$ & $<5.7\times10^3$ & H & \citet{aharonian+08b}\\
W Comae		                 & 0.102   &  430    & 20    & 0.4 	& 3.68	& 31                & 44.9 & 3.41 & $1.3\times10^3$ & I & \citet{acciari+09a} \\
3C 66A		                 & 0.444   & 1700    & 40    & 0.3 	& 4.1	& 28                & 46.3 & 2.43 & 13              & I & \citet{acciari+09b}\\
1ES 1011+496	                 & 0.212   &  870    & 200   & 0.2 	& 4 	& 26                & 45.5 & 3.66 & $2.5\times10^2$ & H & \citet{albert+07a} \\
1ES 1218+304 \tablenotemark{j}	 & 0.182   &  750    & 11.5  & 0.5 	& 3.07	& 24                & 45.4 & 2.37 & $1.0\times10^2$ & H & \citet{acciari+10b}\\
Mkn 180		                 & 0.045   &  190    & 45    & 0.3 	& 3.25 	& 20                & 44.0 & 3.17 & $8.2\times10^3$ & H & \citet{albert+06a} \\
1H 1426+428	                 & 0.129   &  540    & 2     & 1 	& 2.6 	& 20                & 45.0 & 1.71 & $2.1\times10^2$ & H & \citet{aharonian+02a} \\ 
RGB J0710+591 \tablenotemark{j}	 & 0.125   &  520    & 1.36  &  1 	& 2.69 	& 15                & 44.8 & 1.83 & $3.5\times10^2$ & H & \citet{acciari+10a} \\
1ES 0806+524	                 & 0.138   &  580    & 6.8   & 0.4	& 3.6 	& 10                & 44.7 & 3.21 & $1.4\times10^3$ & H & \citet{acciari+09c} \\
RGB J0152+017 \tablenotemark{j}	 & 0.080   &  340    & 0.57  & 1 	& 2.95 	& 8.5               & 44.1 & 2.45 & $3.0\times10^3$ & H & \citet{aharonian+08a} \\
1ES 1101-232 \tablenotemark{j}	 & 0.186   &  770    & 0.56  & 1 	& 2.94 	& 8.2               & 44.9 & 1.50 & $2.3\times10^2$ & H & \citet{aharonian+07c} \\
1ES 0347-121 \tablenotemark{j}	 & 0.185   &  770    & 0.45  & 1 	& 3.1	& 8.2               & 44.9 & 1.67 & $2.9\times10^2$ & H & \citet{aharonian+07b} \\
IC 310                           & 0.019   &   83    & 1.1   & 1        & 2.0   & 8.1               & 42.8 & 1.90 & $4.8\times10^4$ & H & \citet{alek_etal:10} \\
PKS 2005-489	                 & 0.071   &  300    & 0.1   & 1 	& 4.0 	& 8.0               & 44.0 & 3.56 & $2.3\times10^4$ & H & \citet{aharonian+05a} \\
MAGIC J0223+430	                 & --      &   --    & 17.4  & 0.3 	& 3.1 	& 7.6               & --   & $<3.1$ & --            & R & \citet{aliu+09} \\
1ES 0229+200 \tablenotemark{j}	 & 0.140   &  590    & 0.7   & 1 	& 2.5 	& 6.4               & 44.5 & 1.51 & $4.7\times10^2$ & H & \citet{aharonian+07a} \\
PKS 1424+240	                 & $<0.66$ & $<2400$ & 51    & 0.2	& 3.8 	& 6.3               & $<46.1$ & $>1.42$ & 4.0       & I & \citet{acciari+10c}\\
M87      	                 & 0.0044  &   19    & 0.74  & 1 	& 2.31	& 5.9               & 41.4 & 2.29 & $1.5\times10^6$ & R & \citet{acciari+08a}\\
BL Lacertae	                 & 0.069   &  290    & 0.3   & 1 	& 3.09 	& 5.4               & 43.8 & 2.67 & $8.4\times10^3$ & L & \citet{albert+07b} \\
H 2356-309	                 & 0.165   &  690    & 0.3   & 1 	& 3.09 	& 5.4               & 44.6 & 1.86 & $6.5\times10^2$ & H & \citet{aharonian+06a} \\
PKS 0548-322 \tablenotemark{j}	 & 0.069   &  290    & 0.3   & 1 	& 2.86	& 4.0               & 43.7 & 2.44 & $8.4\times10^3$ & H & \citet{aharonian+10a} \\
Centaurus A	                 & 0.0028  &   12    & 0.245 & 1	& 2.73	& 2.8               & 40.7 & 2.72 & $1.2\times10^7$ & R & \citet{raue+10}\\
\enddata
\tablenotetext{a}{Comoving distance in units of $\Mpc$}
\tablenotetext{b}{Normalization of the observed photon spectrum that we assume to be of the form $dN/dE=f_0 (E/E_0)^{-\alpha}$, in units of $10^{-12}\,\cm^{-2}\s^{-1}\TeV^{-1}$}
\tablenotetext{c}{Energy at which we normalize the spectrum, in units of $\TeV$}
\tablenotetext{d}{Observed spectral index at $E_0$} 
\tablenotetext{e}{Integrated energy flux between $100\,\GeV$ and $10\,\TeV$, in units of $10^{-12}\,\erg\,\cm^{-2}\s^{-1}$}
\tablenotetext{f}{Inferred isotropic integrated luminosity between $100\,\GeV$ and $10\,\TeV$, in units of $\erg\,\s^{-1}$}
\tablenotetext{g}{Inferred intrinsic spectral index at $E_0$}
\tablenotetext{h}{Time delay after which plasma beam instabilities dominate inverse-Compton cooling, in units of $\yr$}    
\tablenotetext{i}{H, I, L, Q, and R correspond to high-energy,
  intermediate-energy, low-energy peaked BL Lacs, flat spectrum radio quasars, and radio galaxies of Faranoff-Riley Type I (FR I), respectively.}
\tablenotetext{j}{Used to place limits upon the IGMF}
\end{deluxetable*}

Imaging atmospheric Cerenkov telescopes, such as H.E.S.S., VERITAS, and MAGIC,\footnote{High Energy
  Stereoscopic System, Major Atmospheric Gamma Imaging Cerenkov Telescope, Very
  Energetic Radiation Imaging Telescope Array System.} have opened
the very-high energy gamma-ray (VHEGR, $E\ge100\,\GeV$) sky, finding a
Universe populated by a variety of energetic, VHEGR sources.
While the majority of observed VHEGR sources are Galactic in origin
(e.g., supernova remnants, etc.), the extragalactic contribution is
dominated by a subset of blazars.
There are presently 46 extragalactic TeV sources
known\footnote{See http://www.mppmu.mpg.de/$\sim$rwagner/sources/ for an up-to-date list.}, of which
28 have well defined spectral energy distributions 
(SEDs), and are collected in Table \ref{tab:TeVsources}.  Of these 28
well-studied objects, 24 are blazars, implying that blazars make up an
overwhelming majority of the bright VHEGR sources.

All of the extragalactic VHEGR emitters are relatively nearby, with
$z\lesssim0.5$ generally, and $z\sim0.1$ typical.
This is a result of the large opacity of the Universe to TeV photons,
which annihilate upon soft photons in the extragalactic background light
(EBL), producing pairs
\citep[see, e.g., ][]{Goul-Schr:67,Sala-Stec:98,Nero-Semi:09}.
Typical mean free paths of VHEGRs range from $30\,\Mpc$ to $1\,\Gpc$
depending upon gamma-ray energy and source redshift, and thus the
absence of high-redshift VHEGR sources is not unexpected.

The pairs produced by VHEGR annihilation are necessarily
ultrarelativistic, with typical Lorentz factors of $10^5$--$10^7$.
The standard assumption is that these pairs lose energy almost
exclusively through inverse-Compton scattering the cosmic microwave
background (CMB) and EBL.  When the up-scattered gamma-ray is itself a
VHEGR the process repeats, creating a second generation of pairs and
up-scattering additional photons.  The result is an inverse-Compton
cascade (ICC) depositing the energy of the original VHEGR in gamma-rays with
energies $\lesssim100\,\GeV$.  This places the ICC gamma rays in the
LAT bands of \Fermi, and thus \Fermi has played a central role in
constraining the VHEGR emission of high-redshift blazars.

Based upon \Fermi observations of $z\simeq0.1$ TeV sources, a number of
authors have now published estimated lower bounds upon the
intergalactic magnetic field
\citep[IGMF; see, e.g.,][]{Nero-Vovk:10,Tave_etal:10a,Tave_etal:10b,Derm_etal:10,Tayl-Vovk-Nero:11,Taka_etal:11,Dola_etal:11}.
Typical numbers range from $10^{-19}\,\G$ to $10^{-15}\,\G$, with
the latter values being of astrophysical interest in the context of
the formation of galactic fields\footnote{After contraction and a
  handful of windings, nG field strengths can be produced from an
  IGMF of $\sim10^{-15}\,\G$.}.
These limits on the IGMF arise from the {\em lack} of the GeV bump
associated with the ICC of the blazar TeV emission, presumably due to
the resulting pairs 
being deflected
significantly under the action of
the IGMF itself.  The wide range in the estimates upon the minimum
IGMF is due primarily to different assumptions about the TeV
blazar duty cycle.

\Fermi has also provided the most precise estimate of the unresolved
extragalactic gamma-ray background (EGRB) for energies between
$200\,\MeV$ and $100\,\GeV$.  Since ICCs reprocess the VHEGR emission of
distant sources into this band, this has been used to constrain the
evolution of the luminosity density of VHEGR sources
\citep[see, e.g., ][]{Naru-Tota:06,Knei-Mann:08,Inou-Tota:09,Vent:10}.
Generally, it has been found that these cannot have exhibited the
dramatic rise in numbers by $z\sim1$--$2$ seen in the quasar
distribution.  That is, the comoving number of blazars must have
remained essentially fixed, at odds with both the physical picture
underlying these systems and with the observed evolution of similarly
accreting systems, i.e., quasars.

Both of these conclusions depend critically upon ICCs dominating the
evolution of the ultra-relativistic pairs.  However, as we will show,
the pairs constitute a cold, highly collimated plasma beam moving
through a dense, stationary background, both of which are susceptible
to collective plasma phenomena.  Such beams are notoriously
unstable; for instance, equal density beams 
typically lose a significant fraction of their energy after
propagating distances measured in plasma skin depths of the background
plasma.  If the VHEGR-generated pairs suffer a similar fate while
propagating through the intergalactic medium (IGM), the cooling of the
pairs would be dominated by plasma instabilities, thereby quenching
the ICCs.  

Here we present a plausible alternative mechanism by which the
energy in the ultra-relativistic pairs can be extracted.  While a
variety of potential plasma beam instabilities exist, we find that the
most relevant for the VHEGR-produced pair beams is the ``oblique''
instability
\citep{Bret-Firp-Deut:04,Bret-Firp-Deut:05,Bret:09,Bret-Grem-Diec:10,Lemo-Pell:10}.
This is a more virulent cousin of the commonly discussed  Weibel and
two stream instabilities.

In Section \ref{sec:PBs} we discuss the formation and  properties of the
ultra-relativistic pair beam, including limits upon its temperature
and density. 
Section \ref{sec:PBIs} presents growth rates for a variety of
plasma instabilities, including the oblique instability.
Particular attention is paid to if and when plasma instabilities
dominate inverse Compton as a means to dissipate the kinetic energy of
the pairs.
The resulting implications for studies of the IGMF are
described in Section \ref{sec:implications IGMF}, including how such
efforts might mitigate the systematic uncertainties arising from plasma
cooling.
The evolution of the luminosity function of VHEGR-emitting blazars is
discussed in Section \ref{sec:BLF}, in which we construct a blazar luminosity
function based upon that of quasars which is consistent with current
TeV source populations and the \Fermi estimates of the EGRB.
Finally, conclusions are contained in Section \ref{sec:C}.

This is the first in a series of three papers that discuss
the potential cosmological impact of the TeV emission from blazars.
Here we provide a plausible mechanism for the local dissipation
of the VHEGR luminosity of bright gamma-ray sources.  In addition to
the particular consequences this has for studies of high-energy
gamma-ray phenomenology, it also provides an novel heating process
within the IGM.  \citet[][ hereafter Paper II]{CBP}
estimates the magnitude of the new heating term, describes the
associated modifications to the thermal history of the IGM, and shows
how this can explain some recent observations of the \Lya
forest.  \citet[][ hereafter Paper III]{PCB} considers the
impact the new heating term has upon the structure and statistics of
galaxy clusters and groups, and upon the ages and properties of dwarf
galaxies throughout the Universe, generally finding that blazar
heating can help explain outstanding questions in both cases.
An additional follow-up paper by \citet{Puchwein+2011} shows that when
combined with most recent estimates of the evolving photoionizing
background and hydrodynamic simulations of cosmological structure
formation, the heating by blazars results in excellent quantitative 
agreement with observations of the mean transmission, one- and
two-point statistics, and line width distribution of high-redshift
\Lya forest spectra.  In particular, these successes depends upon
the peculiar properties of the blazar heating via the dissipation of
plasma instabilities.

For all of the calculations presented below we assume the WMAP7
cosmology with $h_0 = 0.704$, $\Omega_{DM} = 0.227$,
$\Omega_{B} = 0.0456$, and $\Omega_{\Lambda} = 0.728$ \citep{WMAP7_2011}.

\section{The Fate of Very High Energy Gamma Rays and the Properties of the Resulting Pair Beam} \label{sec:PBs}

\subsection{Propagation and Absorption of Very High-Energy Gamma rays}
\begin{figure}
\begin{center}
\includegraphics[width=\columnwidth]{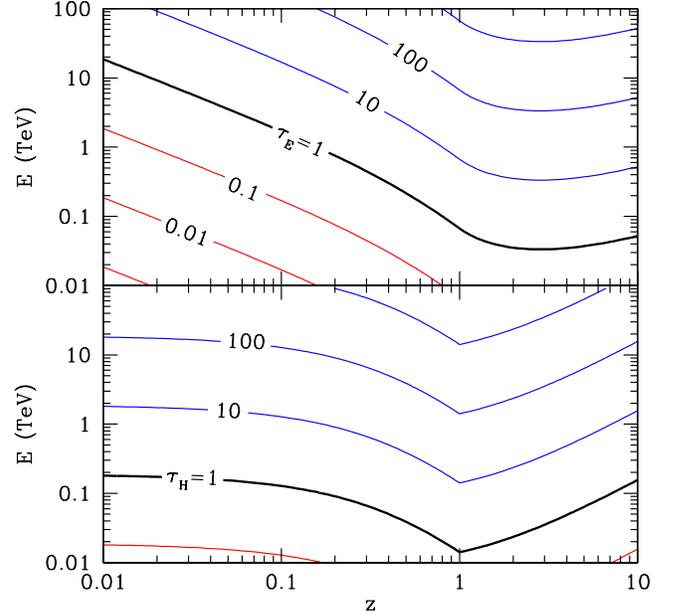}
\end{center}
\caption{Top: Pair-production optical depth of a gamma-ray photon
  emitted with energy $E$ originating at redshift $z$ as it propagates
  to the Earth, $\tau_E$.  Bottom: Pair-production optical depth of a
  gamma ray of energy $E$ propagating across a Hubble
  distance at redshift $z$, $\tau_H$.  In both cases the thick black
  line shows $\tau$ of unity, blue and red lines show optically thick
  and thin regions of the parameter space, respectively.  The sudden
  break in the contours at $z=1$ is due to the assumed EBL evolution
  (see text).}\label{fig:taus}
\end{figure}
Sources of VHEGRs are necessarily attenuated by
the production of pairs upon interacting with the EBL.  Namely, when
the energies of the gamma ray ($E$) and the EBL photon ($E_{\rm EBL}$)
exceed the rest mass energy of the $e^\pm$ pair in the center of mass
frame, i.e., $2 E E_{\rm EBL}(1-\cos\theta) > 4 m_e^2 c^4$, where $\theta$ is
the relative angle of propagation in the lab frame, an $e^\pm$ pair can be
produced with Lorentz factor 
$\gamma\simeq E/2m_e c^2$ \citep{Goul-Schr:67}.
The optical depth the Universe presents to high-energy gamma rays
depends solely upon the energy of the gamma ray and the evolving
spectrum of the EBL\footnote{As a consequence, careful studies of the
  absorption of high-energy gamma rays from extragalactic sources has
  been suggested as a way to probe the EBL
  \citep[see, e.g., ][]{Goul-Schr:67,Stec-deJa-Sala:92,Oh2001}.}.
Detailed estimates for the EBL spectrum and its evolution have
produced an estimated mean free path of TeV photons of
\begin{equation}
\Dpp(E,z)
=
35\left(\frac{E}{1\,\TeV}\right)^{-1}
\left(\frac{1+z}{2}\right)^{-\zeta}\,\Mpc\,,
\label{eq:Dpp}
\end{equation}
where the redshift evolution is due to that of the EBL alone, is dependent
predominately upon the star formation history, and $\zeta=4.5$ for
$z<1$ and $\zeta=0$ for $z\ge1$
\citep{Knei_etal:04,Nero-Semi:09}\footnote{Despite the fact that the
  EBL contribution from starbursts peaks at $z=3$ and declines rapidly
  afterward, galaxies and Type 1 active galactic nuclei compensate for the lost flux
  until $z=1$.  See, for example, Figure 3 from
  \citet{Fran-Rodi-Vacc:08}.}.
The resulting optical depth to VHEGRs emitted with energy $E$ from an
object located at $z$ is
\begin{equation}
\tau_E(E,z) \equiv \int_0^z \frac{c dz'}{\Dpp\left[E(1+z')/(1+z),z'\right] H(z')(1+z')}
\propto E\,,
\end{equation}
(in which $H(z)$ is the Hubble factor), is shown in the top panel of
Figure \ref{fig:taus}.\footnote{Note that the optical depth
  experienced by a photon that is {\em observed} to have energy $E'$
  is given by $\tau_E[E'(1+z),z]$.}  From this it is evident that above
$100\,\GeV$ 
the Universe is optically thick to sources at $z>1$
\citep[cf.][]{Fran-Rodi-Vacc:08}.  A related, and perhaps more
appropriate measure of the optical depth is that across a Hubble length,
$\tau_H\equiv c/(\Dpp H)$, shown in the bottom panel of Figure
\ref{fig:taus}, providing a sense of how opaque the Universe is as a
function of redshift.  In all cases it is clear that at 
$E\gtrsim100\,\GeV$ photons will pair-produce on the EBL
for $z\lesssim10$.  In the future this may not be the case, since for
sufficiently small $z$, $\tau_H$ decreases with decreasing $z$ due to
the dramatic decrease in the EBL photon density associated with both
the Hubble expansion and the slowing of star formation.

\subsection{Temperature of the Ultra-Relativistic Pair Beam}
The pairs produced by the TeV gamma rays are
ultra-relativistic, with typical Lorentz factors of
$E/2m_e c^2\simeq10^6 (E/\TeV)$.  They also necessarily constitute
a cold, highly anisotropic, dilute beam propagating through the IGM.  This
follows immediately from the intergalactic distances traversed by the
gamma rays and the comparatively small EBL photon energies (as seen in
the IGM frame).  Here we estimate the properties of this plasma beam.

The momentum dispersion of the resulting pairs is set by the gamma-ray
spectrum, geometry of the TeV source,  distribution of the EBL
photons, and heating due to pair production.  Of these, only the
last plays a significant role in
setting the transverse momentum dispersion.\footnote{The contribution
  to the transverse momentum dispersion from the finite source size,
  due to the slightly different orientations of photons from opposing
  sides of the TeV emission region of size $\ell$, is
  $p_\perp/p_\parallel\lesssim \ell/2\Dpp$.  With
  $\ell\simeq10^{14}\,\cm$, implied by the X-ray variability
  timescales \citep{Tram_etal:09,Fermi_SED2010}, this gives
  $p_\perp/p_\parallel\lesssim5\times10^{-13} \left[(1+z)/2\right]^\zeta \left(E/\TeV\right)^2$.
  Similarly, since the creation of pairs is dominated by EBL photons
  near the pair-production threshold, the typical energy of the
  relevant EBL photons is roughly $4 m_e^2 c^4/E$ (i.e., twice the
  threshold value for transverse EBL photons), and thus
  $p_\perp/p_\parallel\simeq 4 m_e^2c^4/E^2 \simeq 1\times10^{-12} \left(E/\TeV\right)^{-1}$.}
The center-of-mass frame, i.e., the ``beam frame'', momentum
dispersion resulting from pair production is roughly $m_e c$.
This results in an IGM-frame transverse momentum dispersion of
$p_\perp/p_\parallel\simeq10^{-6}\left(E/\TeV\right)^{-1}$.  With 
$p_\perp,\,m_e c\ll p_\parallel$, the temperature\footnote{The
  temperature of an anisotropic particle distribution is inherently
  ill defined.  Here we identify a temperature with the momentum
  dispersion of the beam using the drifting Maxwell-J\"uttner
  distribution employed by \citet{Bret-Grem-Beni:10} (eq. 1
  therein).} associated with this
transverse momentum dispersion is 
\begin{equation}
\frac{k T_b}{m_e c^2}
\simeq
\frac{p_\parallel}{2 m_e c^2} \left(\frac{p_\perp}{p_\parallel}\right)^2
\simeq
5\times10^{-7}\left(\frac{E}{\TeV}\right)^{-1}\,.
\end{equation}
Since the transverse momentum dispersion of the pairs is much smaller
than that associated with the bulk motion of the beam (i.e., since
$p_\perp/p_\parallel\ll1$), we may safely assume that the beam is
transversely kinematically cold.

\begin{figure}
\begin{center}
\includegraphics[width=\columnwidth]{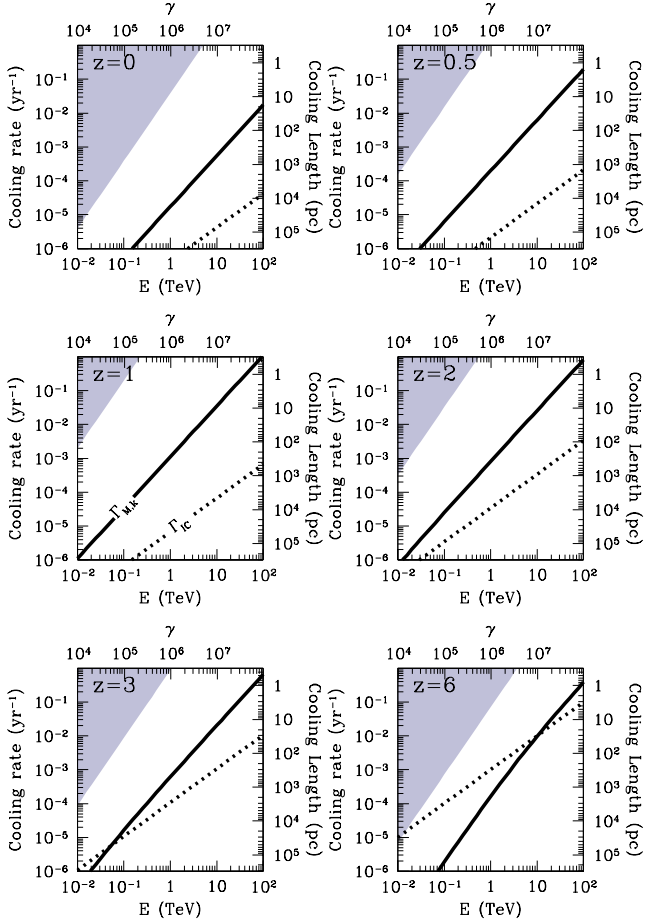}
\end{center}
\caption{Initial pair beam cooling rates due to the kinetic oblique
  instability (thick solid) and inverse Compton scattering (dotted) as a
  function of VHEGR energy ($E$) at a number of redshifts ($z$).  In
  all cases $1+\delta=1$, corresponding to a mean-density region, and
  the isotropic-equivalent luminosity of the source at energy $E$,
  $E L_E$, is $10^{45}\,\erg\,\s^{-1}$, similar to the brightest TeV
  blazars seen from Earth. Finally, we list the initial pair Lorentz
  factor, $\gamma$, and cooling lengthscale along the top and right
  axes, respectively.}\label{fig:Gs}
\end{figure}

\subsection{Density of the Ultra-Relativistic Pair Beam in General}
The density of the pair beam at a given point within the IGM is set by
the rate at which pairs are produced, duration of the TeV emission,
advection of the pairs through the IGM, and the processes by which
they lose their kinetic energy.  That is, the evolution of the density
of pairs per unit Lorentz factor, $n_\gamma$, is governed by the
Boltzmann equation:
\begin{equation}
\frac{\partial n_\gamma}{\partial t}
+
\frac{c}{r^2} \frac{\partial r^2 n_\gamma}{\partial r}
+
\dot{\gamma} \frac{\partial n_\gamma}{\partial \gamma}
=
\dot{n}_\gamma\,,
\end{equation}
where the left-hand side assumes all the pairs are moving away from
the TeV source relativistically ($v^r=c$ and $p^r=\gamma m_e c$), and
the right-hand side corresponds to pair production.  Generally, we may
neglect advection, which alters $n_\gamma$ over timescales of
$\Dpp/c$, much longer than any relevant timescale of interest here
(i.e., $(c/r^2)\partial r^2 n_\gamma/\partial r$ may be neglected).
Furthermore, for most of the potential sources we will consider
(primarily TeV blazars) we will assume that the duration of the TeV
emission is sufficiently long that $n_\gamma$ reaches a steady state
(i.e., $\partial n_\gamma/\partial t=0$).  In this case, we have
$\dot{\gamma} \partial n_\gamma/\partial\gamma\simeq\dot{n}_\gamma$.

Making further progress requires us to define the spectrum of the
pairs, which itself depends upon the spectrum of the gamma rays and
the energy dependence of the cooling processes.  Nevertheless, we may
obtain an estimate of the beam density in the vicinity of a given
Lorentz factor, $\nb\simeq\gamma n_\gamma$, by setting
$\partial n_\gamma/\partial\gamma\simeq-n_\gamma/\gamma=-\nb/\gamma^2$.

The source term is given by twice (since each gamma-ray produces two
leptons) the rate at which high-energy gamma rays with energy $E
\simeq 2\gamma m_e c^2$ annihilate within the region of interest, i.e.,
$\gamma \dot{n}_\gamma = 2 (E dN/dE)/\Dpp = 2 F_E/\Dpp$, where $N$ is
the gamma-ray number flux,
with units of ${\rm photons}~\cm^{-2}\,\s^{-1}$.
Thus, upon defining a cooling rate $\Gamma\equiv-\dot{\gamma}/\gamma$,
we have 
\begin{equation}
\nb \simeq \frac{2 F_E}{\Dpp \Gamma}\,,
\label{eq:nb}
\end{equation}
i.e., the density of pairs of a given energy is determined by
balancing cooling and pair creation.

Generally, $\Gamma$ is a
function of energy and beam density, as well as external factors
(e.g., seed photon density, IGM density, etc.).  Thus this gives a
non-linear algebraic equation to solve for $\nb$, the particulars of
which depend upon the various mechanisms responsible for extracting
the bulk energy of the beam.  In practice, given expressions for
$\Gamma$, associated with the processes discussed in following
section, we solve Equation (\ref{eq:nb}) numerically to obtain
$\nb(E,F_E,z)$.

Which mechanism dominates the cooling of the beam depends upon a
variety of environmental factors and the properties of the pair beam
itself.
Nevertheless, inverse-Compton cooling via the cosmic microwave
background (CMB) provides a convenient lower limit upon $\Gamma$, and
thus an upper limit upon $\nb$.
This is a function of $z$ and $\gamma$ alone, given by
\begin{equation}
\GIC=
\frac{4\sigma_{\rm T} u_{\rm CMB}}{3 m_e c} \gamma
\simeq
1.4\times10^{-20}(1+z)^4\gamma \,\,\s^{-1}\,,
\end{equation}
where $\sigma_{\rm T}$ denotes the Thompson cross section.
The strong redshift dependence arises from the rapid increase in the
CMB energy density with $z$ ($u_{\rm CMB}\propto(1+z)^4$).
Furthermore, since it is $\propto\gamma$, inverse-Compton cooling is
substantially more efficient at higher energies.  The associated
cooling rate is shown as a function of $E$ for a number of redshifts
in Figure \ref{fig:Gs}.

When we set $\Gamma\simeq\GIC$, we obtain the following upper limit
upon the beam density:
\begin{equation}
\begin{aligned}
\nb
&\simeq
\frac{2 F_E}{\Dpp\GIC}
\simeq
\frac{L_E}{2\pi\Dpp^3\GIC}\\
&\simeq
3.7\times10^{-22}
\left(\frac{1+z}{2}\right)^{3\zeta-4}
\left(\frac{E L_E}{10^{45}\erg\,\s^{-1}}\right)
\left(\frac{E}{\TeV}\right)
\,\cm^{-3}\,,
\end{aligned}
\label{eq:nbIC}
\end{equation}
Where we have defined $L_E$ to be the isotropic-equivalent luminosity
(per unit energy) of a source located a distance $\Dpp$ from the
region in question.  Setting $E L_E$ to a typical value
($10^{45}\,\erg\,\s^{-1}$) gives an idea of the typical pair-beam
densities.  Note that despite the large blazar luminosities we
consider, the associated beams are exceedingly dilute, a point
that is of critical importance in the following section.  Since $\GIC$
is independent of $\nb$, this has no implication for inverse-Compton
cooling itself.

\section{Cooling Ultra-Relativistic Pair Beams via Plasma Beam Instabilities}\label{sec:PBIs}

Plasma beams are notoriously unstable, with the instabilities driven
by the anisotropy of the lepton distribution function.  Here we
consider the implications of these instabilities upon the ultimate
fate of the kinetic energy in the TeV-blazar driven pair beams.

\subsection{A Fundamental Limit Upon Plasma Cooling Rates} \label{sec:AFLUPCR}
For collective phenomena to be relevant, it is necessary for many
pairs to be present within each wavelength of the unstably growing
modes.  As we shall see, the relevant scale for the beam plasma
instabilities we describe below is the plasma skin depth of the IGM,
\begin{equation}
\lambda_P 
\equiv
\frac{2\pi c}{\omega_P}
=
\sqrt{\frac{\pi m_e c^2}{e^2 \nIGM}}
=
2.3\times10^{-9}(1+\delta)^{-1/2}(1+z)^{-3/2}\,\pc\,,
\end{equation}
where $\omega_P\equiv\sqrt{4\pi e^2\nIGM/m_e}$ is the IGM plasma
frequency and $\nIGM = 2.2\times10^{-7}(1+\delta)(1+z)^3\,\cm^{-3}$ is
the IGM free-electron number density.
Generally, the growing mode must be uniform on scales considerably
larger than $\lambda_P$, both longitudinally and transversely
(otherwise it is not well-represented as a single Fourier component),
and thus the volume in which many particles must be present is much
larger than that defined by a sphere of diameter $\lambda_P$.
Nevertheless, this gives us a conservative constraint, i.e., we
require
\begin{equation}
\frac{\pi}{6} \lambda_P^3 \,\nb \gg 1\,.
\end{equation}

With Equation (\ref{eq:nb}) this gives a {\em maximum} plasma cooling
rate:
\begin{equation}
\Gamma_{\rm plasma} \ll \frac{\pi}{3} \frac{F_E}{\Dpp}\lambda_P^3\,.
\end{equation}
Plasma processes with cooling rates that exceed this limit necessarily
saturate near this cooling rate.  Such a super-critical process can
potentially operate only until the beam density is driven below the
value at which pairs can support collective phenomena, at which point
they necessarily quench.  However, after plasma cooling ceases, the
plasma beam density rises again (since $\GIC$ is always less than
$\Gamma_{\rm plasma}$ in practice), and thus the super-critical plasma
cooling may resume.  Since the efficiency of the plasma cooling
decreases smoothly to zero at the critical density, this sequence
stabilizes near $\Gamma_{\rm plasma}$.  The associated excluded region
is shown by the grey region in the upper-left corner of Figure
\ref{fig:Gs}.

\begin{figure}
\begin{center}
\includegraphics[width=\columnwidth]{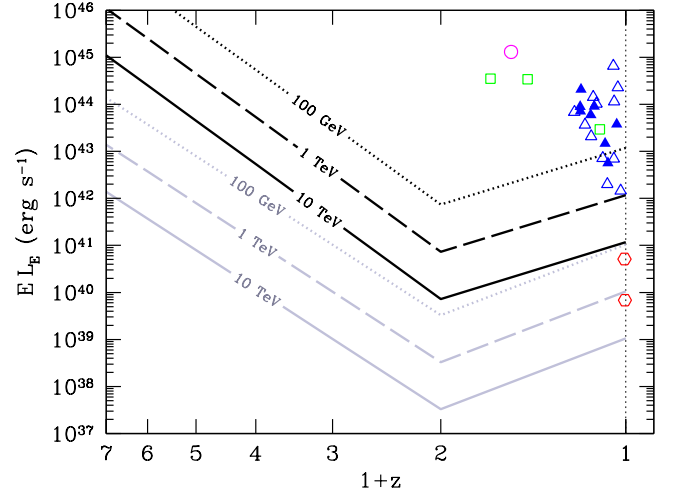}
\end{center}
\caption{Limiting isotropic-equivalent luminosities as a function of
  redshift for various gamma-ray energies for $1+\delta=1$ region of
  the IGM.  Grey lines show the limit defining the applicability
  of the plasma prescription, below which less than a single beam
  lepton is found in a volume $\pi\lambda_P^3/6$.  Below these collective
  phenomena are unimportant.  Black lines show the luminosity at which
  linear plasma cooling rate begins to dominate
  inverse-Compton cooling.  Both 
  limits are sensitive functions of the IGM density, scaling as
  $(1+\delta)^{3/2}$ and $(1+\delta)^{1/2}$, respectively.  Thus in
  low/high density regions these limits become moderately more/less
  permissive.  For reference, the sources listed in Table
  \ref{tab:TeVsources} are also plotted (at $E=1\,\TeV$), with HBL,
  IBL, radio galaxies, and quasars shown by the the blue triangles,
  green squares, red hexagons and magenta circles, respectively.
  Filled points indicate sources that have been used to estimate the
  IGMF (see Section \ref{sec:implications IGMF}).  The only objects
  that fail to meet the plasma-cooling luminosity criterion are the
  radio galaxies M87 and Cen A, both of which are detectable only due
  to their close proximity.
}\label{fig:Nblimits}
\end{figure}

The upper limit upon $\nb$ obtained in Equation (\ref{eq:nbIC})
implies an analogous limit upon the relevant source
isotropic-equivalent luminosities.  Again, this arises from requiring
a sufficient number of beam pairs to be present to support plasma
processes.  Since the beam density is linearly dependent upon the
source luminosity, this gives a constraint upon the latter:
\begin{multline}
E L_E \gg 12\,E \frac{\Dpp^3\GIC}{\lambda_P^{3}}\\
=
3.3\times10^{38}
(1+\delta)^{3/2}
\left(\frac{1+z}{2}\right)^{8.5-3\zeta}
\left(\frac{E}{\TeV}\right)^{-1}\,,
\end{multline}
effectively defining a luminosity cut-off, below which collective
plasma phenomena may be ignored.  This limiting luminosity shown as a
function redshift by the grey lines in Figure \ref{fig:Nblimits} for
various gamma-ray energies.  While the luminosity limit does depend
upon $z$, it is clear that all of the observed TeV blazars
(listed in Table \ref{tab:TeVsources}) are sufficiently luminous for
their resulting pair beams to support collective phenomena at the
redshifts of interest for active galactic nuclei (AGNs).
The only source to fall marginally below the limiting luminosity at
$1\,\TeV$ is the radio galaxy Cen A (the lower-most point in Figure
\ref{fig:Nblimits}), the closest and dimmest object in Table 
\ref{tab:TeVsources}.

\subsection{Cold Plasma Beam Instabilities}
Within
astrophysical contexts there are at least two well known beam
instabilities: two-stream and Weibel, the latter having been suggested
as a mechanism for magnetizing strong shocks \citep{Medv-Loeb:99},
and both implicated in the coupling at collisionless shocks
\citep{Spitkovsky+08}.
  These are, however, simply different limiting
examples of the same underlying filamentary instability, for which the
maximum growth rate exceeds either -- the so-called ``oblique'' mode,
which we discuss below
\citep{Bret-Firp-Deut:04,Bret-Firp-Deut:05,Lemo-Pell:10}.  
Note that since the pair beam is
neutral, it contains its own return current and thus beam
instabilities that arise due to the electron return currents within
the background plasma
\citep[e.g., the Bell and Buneman instabilities, see ][]{Bret:09} are
not relevant.

Here we discuss the nature of these instabilities, and their
growth rates in the cold-plasma limit (i.e., mono-energetic
beams).  While the low-temperature approximation is unlikely to
be applicable in practice for the beams of interest here, it
provides a convenient limit in which to present the relevant
processes within their broader context.  For a similar reason, and
because we have not found analogous derivations elsewhere in the
astrophysical literature, we present the ultra-relativistic pair
beam instability growth rates for the two-stream and Weibel
instabilities in Appendixes \ref{sec:TS} and \ref{sec:W}, only
summarizing the results here.  We defer a discussion of the
more directly relevant warm-plasma oblique instability to the
following section.  Generally, we find that the plasma instabilities
are capable of dominating inverse Compton cooling as a means to
dissipate the bulk kinetic energy of the pair beams from TeV
blazars.

The pair two-stream instability arises due to the interaction of the
anisotropic electron and positron distribution functions with the
comoving background electrostatic wave (i.e., $\bmath{k}\parallel\bmath{p}$,
where $\bmath{k}$ is the electromagnetic mode wavevector and
$\bmath{p}$ is the beam momentum) with wavelength $\sim\lambda_P$ for
$\nb/\nIGM\ll1$, which is generally the case of interest here.
The associated cooling rate\footnote{Note that since we are interested
in the rate at which energy is lost, these are necessarily twice the
instability ``growth'' rates, defined by the e-folding time of the
electromagnetic wave amplitudes.}  in the cold-plasma limit is
\begin{equation}
\GTS
=
\sqrt{\frac{12\pi e^2 \nIGM}{m_e}}\left(\frac{\nb}{2\nIGM}\right)^{1/3}\frac{1}{\gamma}\,,
\end{equation}
which depends only weakly upon the IGM density and the pair
beam density, though decreases rapidly as $\gamma$ becomes large.

The Weibel instability is associated with coupling to a secularly
growing, anharmonic transverse magnetic perturbation (i.e.,
$\bmath{k}\perp\bmath{p}$).  The most rapidly growing wavelength is
again that associated with the background plasma skin depth,
$\sim\lambda_P$, with associated cooling rate in the cold-plasma limit
\begin{equation}
\GW
=
\sqrt{\frac{16\pi e^2 \nb}{m_e \gamma}}\,,
\end{equation}
which depends only upon the pair beam density and the pair Lorentz
factor.  At large $\gamma$ this is suppressed more weakly than 
the two-stream instability,
though is a moderately stronger function of $\nb$.

The computation of the growth rates of the two-stream and Weibel
instabilities are greatly simplified by the particular geometries of
the coupled electromagnetic waves and beam momenta.  However, a
generalized oblique treatment has shown the presence of continuum of
unstable modes
\citep{Bret-Firp-Deut:04,Bret-Firp-Deut:05,Bret:09,Bret-Grem-Diec:10,Lemo-Pell:10},
characterized by the orientation of their wave-vector relative to the bulk
beam velocity.  Of these neither the two-stream nor Weibel are
generally the most unstable.  Rather, generically the most robust
and fastest growing mode occurs at oblique wave-vectors, and thus
referred to as the oblique instability by
\citet{Bret-Grem-Diec:10}.  The cold-plasma cooling rate of this
maximally-growing mode
is
\begin{equation}
\GO
\simeq
\sqrt{\frac{12\pi e^2\nIGM}{m_e}}\left(\frac{\nb}{2\gamma\nIGM}\right)^{1/3}
=
\GTS \gamma^{2/3}
\,.
\end{equation}
This can be much larger than the two-stream and Weibel growth rates
when $\nb/\nIGM\ll1$ and $\gamma\gg1$.

\subsection{Warm Plasma Beam Instabilities}

The cold-plasma oblique instability cooling rates dominate
inverse-Compton cooling by orders of magnitude over the region of
interest.  However, at the very dilute beam densities of relevance
here, the cold-plasma approximation requires exceedingly small beam
temperatures.  Above an IGM-frame temperature of roughly
\begin{equation}
\begin{aligned}
\frac{k\Tc}{m_e c^2}
&\simeq 
\frac{3}{2^{10/3}} \left(\frac{\nb}{\nIGM}\right)^{2/3} \gamma^{1/3}\\
&\simeq
1.0\times10^{-9}
\left(1+\delta\right)^{-2/3}
\left(\frac{1+z}{2}\right)^{2\zeta-14/3}\\
&\qquad\qquad\qquad\times \left(\frac{E L_E}{10^{45}\erg\,\s^{-1}}\right)^{2/3}
\left(\frac{E}{\TeV}\right)\,,
\end{aligned}
\end{equation}
pairs can traverse many wavelengths of the unstable modes over the
cold-instability growth timescale, a situation commonly referred to as
the kinetic regime.  As a consequence, significant phase mixing can
occur, substantially reducing the effective growth rate
\citep{Bret-Grem-Beni:10}.

The way in which finite beam temperatures limit the growth rate
depends sensitively upon the nature of the velocity dispersion and the
modes of interest \citep[see, e.g.,][]{Bret-Firp-Deut:05b}.  For
example, the two-stream instability, associated with wave vectors
parallel to the beam, enters the strongly suppressed ``quasi-linear''
regime when the parallel momentum dispersion is large, though is
insensitive to even large transverse dispersions. 
Conversely, the Weibel instability, associated with wave vectors
orthogonal to the beam, is sensitive to even small
transverse velocity dispersions but unaffected by large parallel
velocity dispersions.  This is simply because a given mode can
tolerate large velocity dispersions within but not across the phase
fronts of the unstable electromagnetic modes.  For the situation of
interest here, dilute beams and cool IGM (i.e.,
$k T_{\rm IGM}/m_e c^2\ll1$), the oblique modes are nearly transverse,
and thus sensitive primarily to large transverse velocity
dispersions.

Nevertheless, even for the small temperatures we have inferred for the
pair beams, we find ourselves in the kinetic regime.  In this case the
oblique instability cooling rate has been numerically measured to be
\begin{equation}
\GM
\simeq
0.4 \frac{m_e c^2}{k T_b} \frac{\nb}{\nIGM} \omega_P
\simeq
0.4 \gamma \sqrt{\frac{4\pi e^2 \nb^2}{m_e \nIGM}}\,,
\end{equation}
where we have set the beam temperature in the beam frame to
$m_e c^2/k$, and thus $kT_b=m_e c^2/\gamma$ \citep{Bret-Grem-Beni:10}.
Both the cold and hot growth rates have been verified explicitly using
particle-in-cell (PIC) simulations, though for somewhat less dilute
beams than those we discuss here \citep{Bret-Grem-Diec:10}.

When the kinetic oblique mode dominates the beam cooling the beam
density is given by
\begin{multline}
\nb \simeq
2.1\times10^{-24}
\left(1+\delta\right)^{1/4}
\left(\frac{1+z}{2}\right)^{(6\zeta+3)/4}\\
\times\left(\frac{E L_E}{10^{45}\,\erg\,\s^{-1}}\right)^{1/2}
\left(\frac{E}{\TeV}\right)^{1/2}
\cm^{-3}\,.
\label{eq:nbGM}
\end{multline}
The associated cooling rate is then
\begin{multline}
\GM
\simeq
3.6\times10^{-11} 
\left(1+\delta\right)^{-1/4}
\left(\frac{1+z}{2}\right)^{(6\zeta-3)/4}\\
\times\left(\frac{E L_E}{10^{45}\,\erg\,\s^{-1}}\right)^{1/2}
\left(\frac{E}{\TeV}\right)^{3/2}
\s^{-1}\,.
\end{multline}
This is a stronger function of gamma-ray energy than inverse-Compton
cooling, implying that it will eventually dominate at sufficiently
high energies, assuming a flat TeV spectrum.  In addition it is a very
weak function of $\delta$, being only marginally faster in
lower-density regions, and thus the cooling of the pairs is largely
independent of the properties of the background IGM.

The rates obtained by numerically solving Equation (\ref{eq:nb}) for $\nb$, with
$\Gamma=\GIC+\GM$, are shown for a number of redshifts in Figure \ref{fig:Gs}.
For the luminosity shown ($EL_E=10^{45}\,\erg\,\s^{-1}$,
typical of the bright TeV blazars) at $z\lesssim4$ plasma cooling dominates
inverse Compton above a TeV.  In the present epoch, $\GM$ is roughly two orders
of magnitude larger than $\GIC$ for bright TeV blazars.

The luminosity dependence of $\GM$ implies a luminosity limit below
which inverse Compton does dominate the linear evolution of the pair beam,
\begin{equation}
E L_E \gtrsim 
7.4\times10^{40}\,
(1+\delta)^{1/2}
\left(\frac{1+z}{2}\right)^{9.5-3\zeta}
\left(\frac{E}{\TeV}\right)^{-1},
\end{equation}
(note that at this luminosity $\Gamma=\GIC+\GM=2\GM$, and thus $\nb$
is half the value shown in Equation (\ref{eq:nbGM})).
This limit is shown as a function of redshift for a number of different
energies by the black lines in Figure \ref{fig:Nblimits}.  Note that
these lie above the corresponding limits associated with the
applicability of the plasma prescription, suggesting that in practice
the beam does support collective phenomena.  The critical luminosity
ranges from $\sim10^{40}\,\erg\,\s^{-1}$ to $10^{45}\,\erg\,\s^{-1}$,
depending upon redshift and energy of interest.  It also depends
upon the over-density as roughly $\propto(1+\delta)^{1/2}$, and thus
at low densities the critical luminosity is moderately smaller.
The only two sources in Table \ref{tab:TeVsources} which fall below
the plasma-cooling luminosity limit are the radio galaxies M87 and Cen
A, both of which are detectable only as a result of their close
proximity.  At $1\,\TeV$, all but a handful of the remaining
sources lie more than an order of magnitude above this limit, and in
the case of the two sources that dominate the TeV flux at Earth,
more than two orders of magnitude above this limit.\\

\subsection{Intuitive Picture of the two-stream, Weibel, and Oblique Instabilities}

Given the technical nature of our discussion above, it is useful to
have a qualitative understanding of these instabilities.  We caution,
however, that intuitive pictures of plasma processes frequently fail
to capture all the relevant physics.
Hence, generalization of these intuitive pictures beyond their
limited range of applicability is potentially misleading.  
This is explicitly illustrated by our examples here:
all of the instabilities discussed in this work belong to the same
family, i.e., instabilities that arise from interpenetrating 
plasmas, but the underlying qualitative pictures for each differ
substantially.

We begin with the Weibel (or filamentation) instability
\citep{Weib:59}, for which a mechanical viewpoint (an initial magnetic
perturbation deflects particles into opposing current streams that
reinforce the perturbed field) can be found in \citet{Medv-Loeb:99},
to which we refer the interested reader.
Here we present picture that although having the virtue of being
simpler is not entirely correct:
because like currents attract, small-scale current perturbations
arising out of the fluctuations within the interpenetrating plasmas
will coalesce preferentially to produce increasingly larger-scale
currents. 
These induce stronger magnetic fields, and thus larger attractive
Lorentz forces between neighboring currents; a positive feedback loop
develops leading to instability. 
This process continues until the associated magnetic field
strengths become sufficiently large to disrupt the currents (in the case
of equal density beams) or until the transverse velocity of the
constituent particles is large enough to efficiently migrate between
current structures on the linear growth timescale, i.e., enter the
kinetic regime.

We should note that the aggregation of currents in the Weibel
instability does not rely upon {\it oscillatory} waves.\footnote{The
  linear analysis of the Weibel instability assumes wave-like
  disturbances in the plasma.  However, there is no associated
  restoring force to these waves and, hence, they do not oscillate.}
Hence, there is no oscillatory component to this instability -- it is
a purely growing mode.  In contrast, the two-stream instability is an
overstable mode, where the oscillatory components are Langmuir (or
plasma) waves with a wavevector parallel to the beam velocity.   
These are longitudinal waves, associated with local charge
oscillations, and are completely described by a propagating perturbation
in the electrical potential.  As particles in the beam traverse
Langmuir waves in the background plasma, they experience
successive periods of acceleration and deceleration, with electrons
(positrons) collecting in minima (maxima) of the electric potential,
where the particle speeds are at their smallest.\footnote{One
  tempting aspect is to think of these particles as almost massless
  and hence imagining that these electric fields are quickly shorted
  out. While this is the case in most astrophysical process, we urge
  the reader to resist this temptation because these instabilities
  occur on short enough timescales that the mass of these charged
  particles plays a crucial role in the physics of the instability.}
This charge-separated bunching of the beam plasma enhances the
background electric perturbation, potentially growing the background
Langmuir wave. 

The bunching within the beam is simply an excitation of Langmuir
waves within the beam plasma itself.  Thus the growth of the
charge perturbations in the background and beam plasmas corresponds to
the resonant coupling between Langmuir waves in the background and
beam.  When the beam density is much less than the background
density, as is the case here, background and beam Langmuir waves only
overlap in frequency, and therefore satisfy the conditions for
resonance, when the wavevector of the latter is parallel to the beam
velocity (see the discussion above Equation (\ref{eq:displin})).
This is always satisfied for the family of comoving beam Langmuir
waves (i.e., waves which in the beam frame move counter to the
background plasma).  
However, of particular importance for the
two-stream instability are the counter-propagating beam Langmuir waves
(i.e., waves which in the beam frame move parallel to the background plasma).
If the beam velocity exceeds the phase velocity of these waves (as
seen in the beam frame), the counter-propagating wave will be dragged
in the direction of the beam (as seen in the background frame), and
therefore has a wavevector which satisfies the resonant condition. 

Nevertheless, the counter-propagating Langmuir wave still carries momentum
in the direction opposite to the beam.  Thus, as the counter-propagating 
wave grows, the momentum, and therefore energy, of the beam-wave
system necessarily decreases.  This implies that the
counter-propagating Langmuir wave is also a {\em negative energy mode}
as seen in the background frame.  
As a consequence, the resonant coupling can transfer energy to the
positive-energy background wave from the negative-energy beam wave,
while growing the amplitudes of both, and thereby leading to instability.
Note that if the phase velocity of the
counter-propagating Langmuir wave is larger than the beam velocity, it
no longer is dragged in the direction of the beam and no longer
satisfies the necessary resonant condition.  For a distribution of
particles, this constraint upon the velocities within the beam
corresponds to the familiar Penrose criterion
\citep{Sturrock94,Boyd-Sand:03}, and is satisfied in the pair beams
resulting from VHEGRs.

The oblique instability encompasses the two-stream and Weibel
instabilities, though the most unstable mode is most similar to the
former in that the intuitive picture focus solely on the
electrostatic forces, ignoring electromagnetic forces.\footnote{The
  picture we present here for the oblique instability is discussed in
  \citet{Nakar+11}.}  In practice, this is a relatively good
approximation and can be used to calculate the growth of these modes in
idealized situations \citep[e.g.][]{Nakar+11,Bret-Firp-Deut:04}.
The qualitative picture proceeds similarly to that for the two-stream
instability described above, with the minor modification that the
perturbing background Langmuir waves now move at an angle $\theta$
relative to the {\it relativistic} beam.  As a consequence, the
resonant Langmuir waves have a phase velocity such that $v_k \approx c
\cos\theta$.  From the simple intuitive picture above, if $\theta$ is
selected such that the projected beam velocity is slightly faster than
the nearly resonant Langmuir wave, an instability develops.

Finally, to understand why the growth rates between the two-stream
instability above and the oblique instability differ, it is useful to
make the approximation that the electric fields generated are in the
direction of the k-vector.  In the two-stream case, the electric field
must slow down (or speed up) particles.  This gets progressively
harder for more relativistic particles.  In the oblique case, the
electric field deflects the particles, changing their {\em projected}
velocity.  While this is also more difficult for more relativistic
particles, this is not nearly as hard as changing the particles
parallel (along the beam) velocity.  Hence, the oblique instability
more easily drives charge density enhancements (and therefore
instabilities) at large $\theta$, i.e., easier deflection, than the
two-stream instability.

\subsection{Non-Linear Saturation} \label{sec:NLS}

We have thus far only treated the linear development of the
relativistic two-stream, Weibel, and oblique instabilities.  However,
the impact pair beams have upon the IGM, gamma-ray cascade emission,
and measures of the IGMF will ultimately depend upon their nonlinear
development.  To address this, however, we are presently forced to
appeal to analytical and numerical calculations of systems in somewhat
different (and less extreme) parameter regimes.

Motivated by the applicability of the Weibel instability in the
context of GRBs, the nonlinear saturation of the relativistic Weibel
instability for equal density plasma beams is well understood 
analytically and numerically. Initially, the Weibel instability
rapidly grows until the energy density of the generated magnetic field
becomes of order the kinetic energy of the two beams
\citep{Silva+03,Frederiksen+04,Chang+08}. Analytically,
\citet{Davidson+72} argued that the Weibel instability would saturate
when the generated magnetic fields become so large that the Larmor
radius of the beam particles is of order the skin depth, i.e., when
the energy of generated magnetic fields is equal to the kinetic energy of two
equal density relativistic beams \citep[see also ][]{Medv-Loeb:99}.
The particles rapidly isotropize with a Maxwellian distribution
\citep{Spitkovsky+08}, i.e., heat, and the magnetic energy then rapidly
decays within an order of a few tens of skin depths \citep{Chang+08}.
Hence for two equal density, relativistic, interpenetrating beams, the
Weibel instability converts anisotropic kinetic energy into heat.
However, as we have already noted, for the pair beams of interest here
the Weibel instability is completely suppressed for tiny transverse
beam temperatures.  Hence, while this instability may initially
operate, it may quickly become suppressed if it results in significant
transverse heating of the beam.

Unlike the Weibel instability, the two-stream and kinetic oblique
instabilities continue to operate, though more slowly than implied by
their cold-plasma limits, in the presence
of substantial beam temperatures.  Due to the geometry of the coupled
modes, the oblique instability is primarily sensitive to transverse
beam-velocity dispersions, though shares the resistance to beam
heating with its two-stream cousin
\citep{Bret-Firp-Deut:05b,Bret:09,Bret-Grem-Diec:10,Lemo-Pell:10}.
Unlike the two-stream instability, the oblique instability in the
parameter range of interest here is also largely insensitive to
longitudinal heating.  Thus, generally, it appears that the oblique
instability is substantially more robust than its more commonly
discussed brethren.

What is less clear a priori is if these instabilities primarily heat
the beam or primarily heat the background plasma.  Here we appeal to
the numerical simulations of \citet{Bret-Grem-Diec:10}, where a mildly
relativistic beam ($\gamma=3$) penetrating into a hot, dense
background plasma (beam-to-background density ratio of $0.1$) was
studied.  In these the oblique instability resulted in a significant
fraction ($\sim 20\%$) of the beam energy heating the background
plasma before the heating of the beam suppressed the oblique
instability in favor of the two-stream instability.  The relative
effectiveness with which the beam heats the background plasma is due
to the efficiency with which the longitudinal electrostatic modes are
dissipated in the background plasma; unlike electromagnetic modes
(e.g., those generated by the Weibel instability), electrostatic modes
are rapidly dissipated via Landau damping.  

We note that \citet{Bret-Grem-Diec:10} found that the heating of the
beam by the oblique instability eventually led to its suppression,
allowing the two-stream instability to grow, continuing the
dissipation of the beam kinetic energy.  
As a result, in their simulations a total of 30\% of the
energy was deposited into the background plasma via a 
{\em combination} of the oblique and two-stream instabilities,
i.e., an additional 10\% of the beam energy was thermalized via the
two-stream instability during and after the suppression of the oblique
instability.

In our case, we expect that much more beam energy (more than the 20\%,
and possibly up to $\sim100\%$) will be deposited into the background
IGM because we are much deeper into the regime in which the
oblique instability dominates, 
i.e., $\gamma\gg1$ and $\nb/\nIGM\ll1$.  In particular, for the case
of interest here, $\gamma\sim10^6$ and
\begin{multline}
\frac{\nb}{\nIGM} \simeq
1.3\times10^{-18}
\left(1+\delta\right)^{-3/4}
\left(\frac{1+z}{2}\right)^{(6\zeta-9)/4}\\
\times\left(\frac{E L_E}{10^{45}\,\erg\,\s^{-1}}\right)^{1/2}
\left(\frac{E}{\TeV}\right)^{1/2}
\cm^{-3}.
\end{multline}
The extremely large Lorentz factor and tiny density ratio make it
computationally prohibitive to assess the beam evolution with
numerical PIC directly.  Nevertheless, because we find ourselves in a
regime in which the oblique instability is much more strongly dominant
than that simulated in \citet{Bret-Grem-Diec:10}, we expect the linear
growth of the kinetic oblique instability to continue for much longer
before the beam changes character and moves out of the
oblique-dominated regime --- potentially once $\sim 100\%$ of the beam
energy has been dissipated.
However, this remains to be studied in future work.

The effect of nonlinear
processes might also effect the evolution of the linear instability.
For instance, \citet{Lesch+87} 
argued that for the relativistic electrostatic two-stream instability,
nonlinear coupling to daughter modes arrest the growth of the linearly
unstable mode at a very low mode energy. Hence, they claim that the
electrostatic two-stream instability can only bleed energy from the
beam at a slow rate.

Utilizing the order of magnitude estimates in \citet{Lesch+87} or the
expression for nonlinear Landau damping in \citet{Melrose80}, we have
found that damping of the pair beam via the relativistic two-stream
instability could be highly suppressed.  The oblique instability may
be similarly suppressed, however, this effect is much more marginal
due to the instabilities much larger growth rate. 
Nevertheless, determining its behavior for the parameters relevant
here is an important unanswered question that is left for future work.
For the purposes of this paper, we will 
rely upon the intuition developed from numerical studies of the
oblique instability and argue that the beam ends up heating the IGM
primarily.

In what follows, we will presume the nonlinear evolution of the ``oblique'' 
or related plasma instabilities lead to the heating of the background IGM, 
i.e., beam cooling.  However, we note that the beam itself may 
be the primary recipient of  this kinetic energy, i.e., beam disruption.  Most 
of our results depend critically on beam cooling and not beam disruption.  This includes 
the effect of blazar heating on the IGM temperature-density relation studied in Paper II, the effects
on structure formation studied in Paper III, the excellent
reproduction of the statistical properties of the high-redshift \Lya
forest found in \citet{Puchwein+2011},  
and the implications on the EGRB and the 
redshift evolution of TeV blazars studied in Section \ref{sec:BLF}.  However, our conclusions 
on the inapplicability of IGMF constraints determined from the non-observation 
of GeV emission from blazars still remains in the presence of beam disruption.
This is because the self-scattering of pairs in the beam would
suppress this GeV emission in similar manner to an IGMF.\\

\subsection{Suppression of the Oblique Instability by an IGMF}\label{SotOIbaI}
The beam instabilities we have discussed have been analyzed primarily
within the context of unmagnetized plasmas.  However, for a variety of
theoretical reasons a weak IGMF is not unexpected.  For example, a
field strength of $10^{-15}\,\G$ is sufficient to explain the observed
$\nG$ galactic fields via compression and winding alone.  Here we
consider the implications that an IGMF has for the instability growth
rates we have described above.

A strong IGMF causes the ultra-relativistic pairs to gyrate, and
therefore to isotropize, suppressing the growth of instabilities that
feed upon the beam anisotropy (e.g., those we have described above).
However, for this to efficiently quench the growth of the plasma beam
instabilities, this isotropization must occur on a timescale
comparable to the instability growth time, i.e., the Larmor frequency
must be comparable to the cooling rate, $\Gamma$.  This condition
gives a lower-limit upon IGMF strengths sufficient to appreciably
suppress the growth of plasma beam instabilities of
\begin{equation}
B 
\gtrsim
10^{-12} 
\left(\frac{\gamma}{10^6}\right)
\left(\frac{\Gamma}{10^{-4}\,\yr}\right)\,\G\,,
\end{equation}
considerably larger than those typical of both, primordial formation
mechanisms \citep{Widrow2002} and implied by galactic magnetic field
estimates, assuming galactic fields are produced by contraction and
winding alone.

Because the VHEGRs emitted by the TeV blazars travel cosmological
distances prior to producing pairs, an IGMF capable of suppressing the
plasma beam instabilities must necessarily have a volume filling
fraction close to unity.  In particular, it must permeate the
low-density regions, where most of the cooling occurs.  However, the
Alfv\'en velocity within those areas is extraordinarily small, roughly
\begin{equation}
v_{\rm A}\simeq 4\times10^{-3}
\left(\frac{B}{10^{-12}\,\G}\right)
\left(1+\delta\right)^{-1/2}\,\km~\s^{-1}.
\end{equation}
The IGM sound speed,
\begin{equation}
c_s \simeq 10
\left(\frac{T}{10^4\,\K}\right)^{1/2}\,\km~\s^{-1}\,,
\end{equation}
is considerably larger, implying convection is much more efficient.
Nevertheless, even after substantial heating via the thermalization of
the TeV blazar emission (Paper II) is taken into account, magnetic
fields will have propagated $\lesssim 100\,\kpc$ over a Hubble time
via convection and much less via diffusion, implying that a pervasive,
sufficiently strong magnetic field can not be produced via ejection
from galactic dynamos.
While galactic winds can produce much faster outflows,
$v_W\sim10^3\,\km~\s^{-1}$, unless they inject a mass comparable to
that contained in the low-density regions over a Hubble time, they are
rapidly slowed via dissipation at shocks in the IGM, again limiting
the spread of galactic fields.  Moreover, we note that since the
magnetic field must be volume filling to suppress the plasma beam
cooling substantially it is insufficient to produce pockets of strong
fields, and thus any galactic origin powered by winds requires a
nearly complete reprocessing of the low-density regions.  However, the
ultimate thermalization of such fast, dense winds would raise the IGM
temperature to $\sim10^8\,\K$, in conflict with the \Lya
forest data and exceeding the entire bolometric output of quasars by
at least a factor of two.  For these reasons we conclude that a
volume-filling strong IGMF would demand a primordial origin.

\section{Implications for IGM Magnetic Field Estimates}\label{sec:implications IGMF}

The existence of plasma processes that can cool the pair
beams associated with TeV blazars has profound consequences for
efforts to constrain the IGMF using the spectra of TeV
blazars.  Here we describe how the reported IGMF limits
have been obtained, the consequences of plasma cooling for these, and
potential strategies for overcoming the constraints it imposes.

The general argument made in efforts to constrain the IGMF based upon
the GeV emission from blazars proceeds as follows: The beamed TeV
blazar emission pair-creates off of the EBL. 
The resulting pairs subsequently up-scatter CMB photons to GeV
energies.  In principle, this should produce an observable GeV excess, or bump, in
the spectra of these objects.  However, in the presence of a
large-scale IGMF, the ultra-relativistic pairs can be deflected
significantly, directing the beamed GeV emission away from
Earth.
Thus, it is argued, the lack of a discernible GeV bump in a
number of TeV blazars implies a lower limit upon the IGM field
strength (as a function of TeV jet opening angle and variability
timescale).  The crucial components of the argument are
\begin{enumerate}
\item The TeV emission is beamed with the typical opening angles
  inferred from radio observations of AGN jets.
\item The variability timescale within the TeV source is long in
  comparison to the geometric time delays between the original TeV and
  inverse-Compton produced GeV gamma rays due to the orbit of the
  pairs through some angle $\vartheta$, $\Delta t\sim 10^6\,
  (\Dpp/80\,\Mpc) (\vartheta/0.1\,\rad)^2\,\yr$ \citep[see][]{Derm_etal:10}.
\item The pairs produced by TeV absorption on the EBL cool primarily
  via inverse-Compton scattering the CMB (and therefore evolve
  only due to inverse-Compton cooling and orbiting within the
  large-scale IGMF).
\end{enumerate}

The first of these is supported indirectly by the lack of TeV emission
from non-blazars.  The implied TeV source stability required by the
second is at odds with the variability observed in blazars at longer
wavelengths.  However, since the TeV-GeV delay is a strong function of  
deflection angle, given an empirical limit upon the TeV blazar
variability timescale \citep[presently $4\,\yr$, ][]{Derm_etal:10} it
is possible to produce a substantially weaker constraint upon the IGMF
($\sim10^{-19}\,\G$)
\citep{Derm_etal:10,Tayl-Vovk-Nero:11,Taka_etal:11}.

\begin{figure}
\begin{center}
\includegraphics[width=0.95\columnwidth]{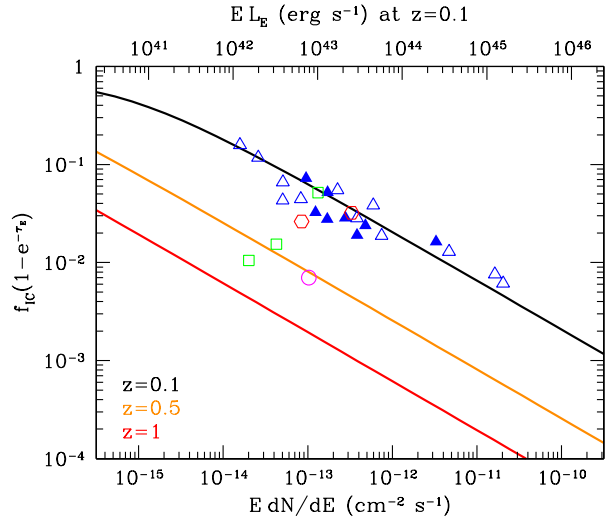}
\end{center}
\caption{Fraction of the energy in the pair beam lost via
  inverse-Comptonization of the CMB at a number of different
  redshifts.  In all cases the injection photon energy (as would have
  been observed at $z=0$) is
  $E=1.85\,\TeV$, resulting in $3\,\GeV$ up-scattered CMB photons, the
  energy at which the \Fermi/LAT instrument is most sensitive.
  For reference, the sources listed in Table
  \ref{tab:TeVsources} are also plotted (at $E=1\,\TeV$), with HBL,
  IBL, radio galaxies, and quasars shown by the the blue triangles,
  green squares, red hexagons and magenta circles, respectively.
  Filled points indicate sources that have been used to estimate the
  IGMF, corresponding to (in increasing flux)
  PKS 0548-322, 1ES 0347-121, 1ES 1101-232, RGB J0152+017, 1ES
  0229+200, 1ES 1218+304, RGB J0710+591, and Mkn 501.
  For reference, the inferred isotropic luminosity is shown in the top
  axis for sources located at $z=0.1$.
 }\label{fig:fIC}
\end{figure}

Plasma cooling provides a fundamental limitation for these methods,
however, by violating the third condition explicitly.
Figures \ref{fig:Nblimits} and \ref{fig:fIC} imply
that for all but the dimmest and highest-redshift ($z\gtrsim4$)
gamma-ray blazars only a small fraction of the pair energy is lost to
inverse-Compton on the CMB
($f_{\rm IC}\equiv\GIC/\left(\GIC+\GM\right)$).\footnote{Here we 
  implicitly assume that the nonlinear saturation of the plasma
  instabilities extract energy from the pair beam at the linear
  growth rate.}
In particular, the black lines in Figure \ref{fig:Nblimits} shows
where $f_{\rm IC}=0.5$, i.e., roughly 50\% of the TeV-photon power is
ultimately converted into heat via plasma instabilities.
As a result, the putative GeV component is typically much less
luminous than otherwise expected, reducing the significance of
non-detections substantially.  

This may be seen explicitly in Figure \ref{fig:fIC}, which shows
$f_{\rm IC}$ as a function of gamma-ray flux for $E=1.85\,\TeV$
(corresponding to a Comptonized-CMB photon energy of approximately
$3\,\GeV$), at number of source redshifts.  Typical values for the TeV
blazars collected in Table \ref{tab:TeVsources}
\citep[including those employed by ][]{Nero-Vovk:10,Tave_etal:10a,Tave_etal:10b,Derm_etal:10,Tayl-Vovk-Nero:11,Taka_etal:11,Dola_etal:11}
lie in the range $10^{-3}$ to $3\times10^{-2}$, implying correspondingly
small GeV Comptonization signals.

More importantly, when the beam evolution is dominated by plasma
instabilities, over the inverse-Compton cooling timescale the pair 
distribution necessarily becomes isotropized.  As a consequence, the
angular distribution of the resulting GeV gamma rays (i.e., the
orientation of the GeV ``beam'') is no longer indicative of the beam
propagation through a large-scale magnetic field.  That is, one
expects large-angle deviations regardless of the IGMF
strength.

Unfortunately, subject to the caveats of the preceding section, plasma
instabilities appear to be the dominant cooling mechanism 
for the pair beams associated with the TeV blazars that have been used
to constrain the IGMF thus far.  The isotropic-equivalent luminosities
of these sources range from 
$2.5\times10^{43}\,\erg\,\s^{-1}$ to $10^{45}\,\erg\,\s^{-1}$, placing
them well within the plasma-instability dominated regime.  As a
result, the reported IGMF limits inferred from TeV
blazars are presently unreliable.

Nevertheless, it may be possible to avoid the limitations imposed by
plasma instabilities.  At sufficiently low pair densities the cooling
rates associated with the beam instabilities described in Section
\ref{sec:PBIs} fall below that due to inverse Compton.  Moreover, at
some point the plasma prescription breaks down altogether, suggesting
that any plasma instabilities that operate on the skin-depth of the
IGM are strongly suppressed.  There are two distinct ways in which low
pair densities can arise: low intrinsic luminosities and very short
timescale events.

Below isotropic-equivalent luminosities of roughly
$\sim10^{42}\,\erg\,\s^{-1}$ inverse-Compton cooling dominates the
beam instabilities we describe in Section \ref{sec:PBIs} at
$\sim1\,\TeV$.  Below $10^{39}\,\erg\,\s^{-1}$ the plasma prescription
itself breaks down altogether.  While no TeV blazars with
isotropic-luminosities below $10^{39}\,\erg\,\s^{-1}$ are known, there
are a handful of very nearby sources below $10^{42}\,\erg\,\s^{-1}$.
The two dimmest sources, the radio galaxies M87 and Cen A, however, are observable only
due to their proximity (i.e., they have proper distances $\ll\Dpp$),
intrinsically preventing them from providing a significant constraint
on the IMGF.  The highest source in Figure \ref{fig:fIC}, and thus
presumably the source with the largest fractional inverse-Compton
signal, is PKS 2005-489, a dim, soft TeV blazar at $z=0.071$.  Even in
this case, however, the inverse-Compton cooling timescale is roughly 4 times longer than the
plasma cooling timescale.  Hence, probing the IGMF with TeV blazars
will require observing considerably dimmer objects than those used
thus far.  Since doing so will likely require substantial increases in
detector sensitivities\footnote{As described in Section
  \ref{sec:AELB}, the incompleteness of the TeV observations are
  well-described by a single flux limit, corresponding to an
  isotropic-equivalent luminosity limit at $10^2\,\Mpc$ of roughly
  $10^{43}\,\erg\,\s^{-1}$.  The flux limit of \Fermi is estimated to
  be roughly $4\times10^{43}\,\erg\,\s^{-1}$ at $z\simeq0.1$
  \citep[see Figure 23 of ][]{Fermi_AGNCatalogue2010}.} 
we will not consider this possibility any further here.

Low pair densities may also be produced by limiting the duration of
the VHEGR emission.  Equation (\ref{eq:nb}) was derived assuming that
the pairs had reached a steady state between their formation via
VHEGRs annihilating upon the EBL and their cooling via inverse-Compton
and plasma processes.  However, it takes roughly a cooling timescale
for the pair beam densities to saturate at this level, which can be as
long as $10^5\,\yr$ at some $E$ and $z$.  Therefore, generally there
is a lag between the onset of the VHEGR emission and the time at which
plasma processes begin to dominate the cooling of the beam.

We can estimate this by setting $\nb\simeq 2 F_E \Delta t/\Dpp$ where
$\Delta t$ is the duration of the emission.  This is the case when
the pair density is initially zero (i.e., it has been many cooling
timescales since the last period of VHEGR emission) and
$\Gamma \Delta t\ll 1$.  Inserting this into the various cooling rates
and setting $\GM=\GIC$ gives an estimate for the maximum duration for
which the pair density remains sufficiently low that inverse-Compton
cooling dominates the cooling:
\begin{multline}
\Delta t \lesssim 5.3\,
\left(1+\delta\right)^{1/2}
\left(\frac{1+z}{2}\right)^{(11-6\zeta)/2}\\
\times
\left(\frac{E L_E}{10^{45}\,\erg\,\s^{-1}}\right)^{-1}
\left(\frac{E}{\TeV}\right)^{-2}
\,\yr\,,
\end{multline}
which may be found for the observed TeV sources in Table \ref{tab:TeVsources}.
Typically $\Delta t\simeq300\,\yr$ for the blazars that have been used to
constrain the IGM, though $\Delta t$ ranges from $10^2\,\yr$ to
$10^4\,\yr$ for TeV blazars generally.
Note that it is not sufficient for emission to vary on $\Delta t$; such
variations will be temporally smoothed on the much larger cooling timescale.
Rather, it is necessary for the source to have been quiescent
(i.e. isotropic-equivalent luminosity considerably less than
$10^{42}\,\erg\,\s^{-1}$) for periods long in comparison to
$\Gamma^{-1}$ and have turned on less than a time
$\Delta t\ll\Gamma^{-1}$ ago.  As a consequence, only IGMF estimates
using new TeV blazars can potentially avoid plasma instabilities in
this way.

A natural example of a class of transient gamma-ray sources is
gamma-ray bursts (GRBs).  For GRBs an isotropic-equivalent energy of
roughly $10^{54}\,\erg$ is required for the plasma instabilities to grow
efficiently, comparable to the total  energetic output of the
brightest events.  Using GRBs to probe the IGMF has been
suggested previously, and limits based upon the lack of a delayed GeV
component in some GRBs already exist,  
finding field strengths above $\sim10^{-22}\,\G$ 
\citep[see, e.g., ][]{Plag:95,Dai-Lu:02,Guet-Gran:03,Taka_etal:08}.
Future efforts should be able to probe field strengths of
$10^{-15}\,\G$ for $z\lesssim0.2$
\citep[see Figure 4 of ][]{Taka_etal:08}.
This is predicated, however, upon the existence of a significant VHEGR
component in the prompt or afterglow emission of GRBs.

\citet{Ando+10} have attempted to directly detect the
GeV signal in the vicinity of AGNs that do not exhibit TeV emission.
That is, since an IGMF can in principle induce large-angle deviations
in the propagation direction of the up-scattered GeV photon relative
to the original VHEGR, one may search for a GeV excess at large
distances from non-blazars.  This is made much more difficult by the
extraordinarily low surface gamma-ray brightness due to the large
$\Dpp$, random source orientations, and potential dilution associated
with the beam diffusion.  Nevertheless, \citet{Ando+10}
claim a marginal detection of this component in stacked \Fermi
images, though the \Fermi point-spread function appears to be capable
of completely explaining their result \citep{Nero-Semi-Tiny-Tkac:11}.
If such a GeV-halo were found around TeV-bright sources with the
anticipated luminosity, it would argue strongly against plasma
instabilities dominating the beam evolution.  However, we note that
for many TeV-dim sources inverse-Compton cooling still dominates the
beam evolution, and thus in stacked images of these sources GeV-halos
may still be present.

Note that, as discussed in Section \ref{SotOIbaI}, a strong IGMF
($\sim10^{-12}\,\G$) could, in principle, suppress the growth of the
plasma beam instabilities significantly.  As we argue there, a
sufficiently pervasive and strong IGMF would necessarily be primordial
in origin.  However, few constraints upon such a primordial field
exist, especially since we have argued that the lack of an inverse
Compton GeV bump cannot be used as a constraint on the IGMF.  Hence, we  
are left with constraints from Big Bang Nucleosynthesis 
\citep[$B\lesssim10^{-7}\,\G$,][and references therein]{Kandus+11},
from the CMB 
\citep[$B\lesssim 10^{-9}\,\G$,][]{Barrow+97}, and limits on the
Faraday rotation of distance radio sources
\citep[$B\lesssim 10^{-9}\,\G$,][and references therein]{Vallee11}. 
A stronger limit can be obtained from limits upon the deflection angle
($<2^\circ$) of ultra-high energy cosmic rays ($E\gtrsim10^{20}\,\eV$)
from distant sources at a distance $d > \lambda_B$, which upon assuming a magnetic correlation
length, $\lambda_B$, gives
$B\lesssim 10^{-11}\left(E/10^{20}\,\eV\right)\left(\lambda_B/100\,\Mpc\right)^{-1}\left(d/\lambda_B\right)^{-1/2}\,\G$
\citep[see for instance][]{Waxman+96,de_Angelis+08}, though this depends strongly upon the assumed
structure of the IGMF and small-scale fields may be significantly
stronger than this \citep{Kotera+08}.
Proposed mechanisms by which a primordial IGMF can be theoretically
generated are essentially unconstrained, yielding B-field strengths
between $B\sim10^{-9}$--$10^{-30}\,\G$ \citep{Kandus+11}.

However, in the remainder of this work and in Paper II and Paper III,
we show that the effect of the "oblique" (or similar) instability may
potentially explain many different observed phenomena in cosmology and
high energy astrophysics.  Taking the beam instabilities we present
here at face value, that so many observational puzzles can be
simultaneously explained by the local dissipation of TeV blazar
emission would imply the strongest upper limit upon primordial
magnetism to date.  Namely, the apparent impact of TeV blazars upon
the large-scale cosmological environment places a constraint on the
IGMF of $B\lesssim 10^{-12}\,\G$.

\section{Implications for the Blazar Luminosity Function} \label{sec:BLF}

The EGRB corresponds to the diffuse gamma-ray background
($E>10\,\MeV$) not associated with Galactic sources.  This has been
measured at high energies by EGRET \citep{EGRET_EGRB1998,Stro-Mosk-Reim:04} and more
recently strongly constrained between $200\,\MeV$ and $100\,\GeV$ by
\Fermi \citep{Fermi_EGRBApJ2010,Fermi_EGRB2010}.  The most recent \Fermi estimate of the
EGRB is a featureless power-law, somewhat softer than that found by
EGRET, and does not exhibit the localized high-energy excess claimed
by a reanalysis of the EGRET data \citep{Stro-Mosk-Reim:04}.

While bright, nearby blazars are resolved, and therefore excluded from
the EGRB, the remainder is thought to arise nearly exclusively from
distant gamma-ray emitting blazars, upon which the \Fermi EGRB places
severe constraints \citep{Fermi_EGRBApJ2010}.  Beyond $z\simeq0.25$
\Fermi cannot detect 
blazars with isotropic-equivalent luminosities comparable to the
objects listed in Table \ref{tab:TeVsources}, and thus the vast
majority of such objects contribute to the \Fermi EGRB.

A significant EGRB above $100\,\GeV$ is not expected due to
annihilation upon the EBL.  However, if they operate efficiently, ICCs
can reprocesses the VHEGR emission of distant sources into the \Fermi
EGRB energy range (i.e., $\lesssim100\,\GeV$).  Thus, a number of
efforts to constrain the VHEGR emission from extragalactic sources
based upon the EGRB can be found in the literature
\citep{Naru-Tota:06,Knei-Mann:08,Inou-Tota:09}.
These have typically found that the comoving number density of VHEGR-emitting
blazars could not have been much higher at high-$z$ than it is today.
Even with moderate evolutions \citep[e.g., that in ][]{Naru-Tota:06,Inou-Tota:09}
consistent with the EGRET EGRB, substantially over-produce the \Fermi
EGRB, and are therefore believed to be excluded \citep{Vent:10}.

However, here we show that this conclusion is predicated upon the
high-efficiency of the ICC.  We have already shown that for bright
VHEGR sources plasma beam instabilities extract the kinetic energy of
the first generation of pairs much more rapidly than inverse-Compton
scattering.  As a consequence, the ICC is typically quenched,
substantially limiting the contributions of these sources to the
EGRB.  Thus, even with a dramatically evolving blazar population,
e.g., similar to that of quasars, it is possible for these objects to
be consistent with the \Fermi EGRB.  
While our discussion of the EGRB and the evolution of blazars is
predicated on the extraction of the pair-beam kinetic energy by plasma
beam instabilities, our conclusions are not.  Rather they will
continue to hold as long as the pair beams are locally dissipated --
via plasma beam instabilities or some other equally powerful
mechanism.

\subsection{An Empirical Estimate of the TeV Blazar Luminosity Function}

Blazars dominate the extragalactic gamma-ray sky, and thus represent
the best studied VHEGR source class.  Since we are interested
exclusively in those objects responsible for the bulk of the VHEGR
emission, we necessarily concentrate upon the subset of blazars that
are luminous VHEGR emitters.  Of the 28 TeV sources listed in Table
\ref{tab:TeVsources}, 23 are peaked at very-high energies (the HBL and
hard IBL sources, for a full definition see below), and comprise what we
will call collectively TeV blazars. 
This necessarily is limited to $z\simeq0.1$ as a consequence of the
annihilation of VHEGRs upon the EBL.  Nevertheless, we will find that
this is remarkably similar to the local quasar luminosity function ($\QLF$),
and thus attempt to construct a blazar luminosity function by analogy:
\begin{equation}
\BLF(z,L)\equiv\frac{d^2 \Phi_B}{dz\, d\log_{10}L}\,,
\end{equation}
in which $\Phi_B$ is the comoving number density of blazars at
redshifts less than $z$ with isotropic-equivalent luminosities below $L$.
This is different from previous efforts to
empirically constrain $\BLF$ from the EGRB \citep[see, e.g., ][]{Naru-Tota:06,Inou-Tota:09} in at
least two ways.  First, we begin with an empirically determined local
$\BLF$ and attempt to extend this by placing the TeV blazars within the
broader context of accreting supermassive black holes, rather than
beginning with the EGRB and working backwards to infer an acceptable
$\BLF$.  Second, we are primarily concerned with $\BLF$ of TeV blazars,
and specifically do not consider the contributions from other kinds of
objects.  While this does not represent a significant oversight in
terms of the high-energy contributions to the EGRB, which almost
certainly arises from the VHEGR-emitting blazars, it does mean that
our conclusions regarding the luminosity function of the TeV blazars
do not necessarily apply to all \Fermi AGN (e.g., the FSRQs, see below).

\subsubsection{Placing TeV blazars in Context}
The TeV blazars presumably fit within the broader context of the
blazars observed by \Fermi specifically, and AGNs generally.  For this
reason, here we briefly review the physical classification scheme 
based on the widely accepted AGN standard paradigm that provides a
unified picture of the emission emission properties of these objects
\citep[e.g., ][]{Urry+1995}.
Specifically, we summarize the classes of objects believed to be capable
of producing significant TeV luminosities and the potential physical
processes responsible for the observed emission.  Based upon these we
then assess the implications for the number, variability, and redshift
evolution of the TeV blazars.

In general there exist two main classes of AGNs that differ in their
accretion mode and in the physical processes that dominate the
emission.
\begin{enumerate}
\item {\em Thermal/disk-dominated AGNs:} Infalling matter assembles in a thin
  disk and radiates thermal emission with a range of temperatures. The
  distributed black-body emission is then Comptonized by a hot corona above
  the disk that produces power-law X-ray emission. Hence the emission is a
  measure of the accretion power of the central object.  This class of objects
  are called QSOs or Seyfert galaxies and make up about 90\% of AGNs. They
  preferentially emit in the optical or X-rays and do not show significant
  nuclear radio emission. None of these sources have so far been
  unambiguously detected by 
  \Fermi or imaging atmospheric Cherenkov telescopes because the Comptonizing
  electron population is not highly relativistic and emits
  isotropically, i.e. there is no beaming effect that boosts the emission.
\item {\em Non-thermal/jet-dominated AGNs:} The non-thermal emission from the
  radio to X-ray is synchrotron emission in a magnetic field by highly energetic
  electrons that have been accelerated in a jet of material ejected from the
  nucleus at relativistic speed. The same population of electrons can also
  Compton up-scatter any seed photon population either provided by the
  synchrotron emission itself or from some other external radiation field such
  as UV radiation from the accretion disk. Hence the SED of these
  objects shows two distinct peaks. The luminosity 
  of these non-thermal emission components probes the jet power of these
  objects. Observationally, this leads to the class of radio-loud AGNs which can
  furthermore be subdivided into blazars and non-aligned non-thermal dominated
  AGNs depending on the orientation of their jets with respect to the line of
  sight.
\end{enumerate}
There are no known sources above $600\,\GeV$ that correspond to AGNs with jets
pointed at large angles ($\approx 15\degr - 40\degr$, see \citealt{Urry+1995})
with respect to the line of sight \citep[for an example of a non-aligned AGN,
NGC1275, that shows a very steep high-energy spectrum, emitting a negligible
number of VHEGRs, see][]{Mariotti+2010}.  Hence we turn our attention to
blazars, which can be powerful TeV sources.

Blazars can further be subdivided into two main subclasses depending
upon their optical spectral properties: flat spectrum radio quasars
(FSRQ) and BL Lacs.  FSRQs, defined by broad optical emission lines,
have SEDs that peak at energies below $1\,\eV$, implying a maximum
particle energy within the jet and limiting the inverse-Compton
scattered photons mostly to the soft gamma-ray band.  
It is presumably for this reason that no continuous TeV component has
been detected in an FSRQ
(note, however, that TeV flares from FSRQs have been detected in two cases
\citep{MAGIC+2008,Mariotti_FSRQ2010}).

In contrast, BL Lacs or Blazars of the BL Lac type
\citep{Massaro+2009} can be copious TeV emitters.
These are
very compact radio sources and have a broadband SED similar to that of
strong lined blazars, though lack the broad emission lines that define
those.  Depending upon the peak energy in the synchrotron spectrum,
which approximately reflects the maximum particle energy within the
jet, they are classified as \mbox{low-,} intermediate-, or high-energy peaked BL Lacs,
respectively called LBL, IBL, and HBL 
\citep{Padovani+1995, Fermi_SED2010}.\footnote{The source classes of
  HSP/ISP/LSP used in recent \Fermi publications are very similar to
  the commonly used HBL/IBL/LBL classes. Hence we identify these with
  each other, respectively, though minor differences may be found in
  the literature.  Nevertheless, where we refer to the number counts
  observed by \Fermi specifically, we will refer to the classes
  HSP/ISP/LSP in keeping with their notation.}
While LBLs peak in the far-IR or IR band, they exhibit a flat or
inverted X-ray spectrum due to the dominance of the 
inverse-Compton component \citep[see Fig 15 of][ for a visualization of the SED
of BL Lacs]{Fermi_SED2010}.  The synchrotron component of IBLs peaks in the
  optical which moves their inverse-Compton peak into the gamma-ray band of
  \Fermi.  HBLs are much more powerful particle accelerators, with the
synchrotron peak reaching into the UV or, in some cases, the soft X-ray
bands. The inverse-Compton peak can then reach TeV energies
\citep{Ghisellini+2008,Tavecchio+2008,Fermi_SED2010}.\footnote{For the highest
  energies scattering occurs in the Klein-Nishina regime, which results in
  steeper inverse-Compton spectra.}

In the gamma-ray band, the subclass of IBLs that emit VHEGRs are
almost indistinguishable from the HBLs, suggesting that the
location of the synchrotron peak does not uniquely characterize the
VHEGR emission from these sources (e.g., due to variations among
individual blazars in the magnetic field strength within the
synchrotron emitting region and the origins and properties of the
seed photons that are ultimately Comptonized). 
Hence we identify HBLs and VHEGR-emitting IBLs with the single source
class of TeV blazars.
We note that
there is presently no evidence for the hypothetical class of
ultra-HBLs that were proposed to have a very energetic synchrotron
component extending to $\gamma$-rays \citep{Ghisellini+1999}.  If such
a population of bright and numerous sources exists, \Fermi should have
seen it \citep{Fermi_AGNCatalogue2010}.  The ultra-HBLs may have escaped
detection from \Fermi thus far by being either intrinsically dim 
$\gamma$-ray sources or very rare objects \citep{Costamante+2007}.

TeV blazars have a redshift distribution that is peaked at low redshifts
extending only up to $z=0.7$. This is most likely entirely a flux
selection effect; TeV blazars are intrinsically less luminous than
LBLs and FSRQs, 
with an observed isotropic-equivalent luminosity range of
$10^{44} - 2\times10^{46}~\rmn{erg~s}^{-1}$, with the highest redshift
TeV blazars also being among the most luminous objects
\citep[see Figures 23 and 24 in][]{Fermi_AGNCatalogue2010}.
That TeV blazars should be intrinsically less luminous than FSRQs is not
entirely unexpected, however.  \citet{Ghisellini+2009} have argued that
the physical distinction between FSRQs and TeV blazars has its origin in the
the different accretion regimes of the two classes of objects.  Using
the gamma-ray luminosity as a proxy for the bolometric luminosity, the
boundary between the two subclasses of blazars can be associated with
the accretion rate threshold (nearly 1\% of the Eddington rate)
separating optically thick accretion disks with nearly Eddington
accretion rates from radiatively inefficient accretion flows.
The spectral separation in hard (BL Lacs) and soft (FSRQs) objects
then results from the different radiative cooling suffered by the
relativistic electrons in jets propagating into different surrounding
media \citep{Ghisellini+2009}.  Hence in this model, TeV blazars cannot reach
higher luminosities than approximately
$2\times10^{46}\,\erg\,\s^{-1}$ since they are limited by the nature
of inefficient accretion flows that power these jets and by the
maximum black hole mass, $\sim10^{10}\,\Ms$.

\subsubsection{An Empirical Local $\BLF$(z,L)} \label{sec:AELB}

\begin{figure}
\begin{center}
\includegraphics[width=\columnwidth]{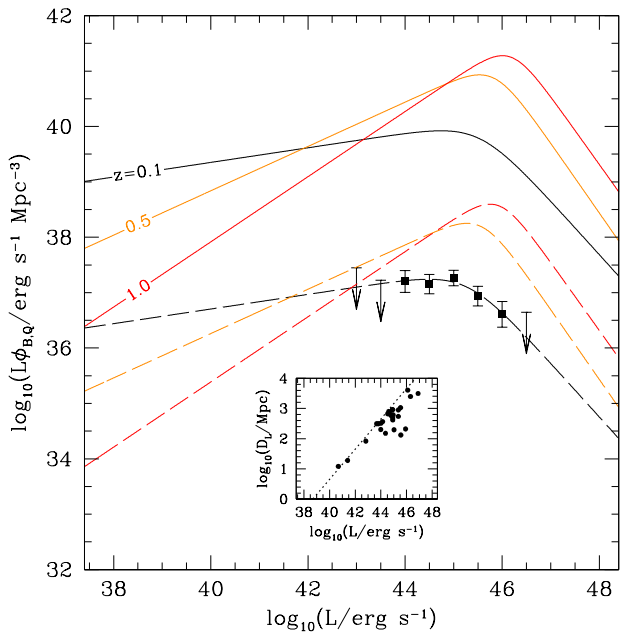}
\end{center}
\caption{Comparison between the luminosity-weighted quasar and
  TeV-blazar luminosity functions ($L\QLF(z,L)$ and $L\BLF(z,L)$, respectively).
  The solid lines show the absolute $L\QLF$ (in comoving Mpc), while
  the dashed lines show $L\QLF$ rescaled in magnitude by
  $2.1\times10^{-3}$ and shifted to lower luminosities by a factor of
  $0.55$.  In both cases the black, orange, and red lines correspond
  to $\QLF(0.1,L)$, $\QLF(0.5,L)$, and $\QLF(1,L)$, respectively.  The
  points and upper-limits show $\BLF$ of the HBL and IBL sources
  listed in Table \ref{tab:TeVsources}, assuming the relevant limits
  for PKS 1553+113 and PKS 1424+240.
  Presented in the inset is the TeV source luminosity distance as a
  function of source luminosity for all of the sources in Table
  \ref{tab:TeVsources} with redshift estimates (including limits).
  The dotted line shows the distance-dependence of the flux
  limit we employ in the completeness correction.}\label{fig:BLF}
\end{figure}

While we have attempted to place the observed TeV blazars, which we
have identified with the HBLs and VHEGR-emitting IBLs, into the
broader context of AGNs using the unified model, due to the
substantial distinctions in accretion rate, emission properties,
object morphology and geometry, it is not obvious that any of the
properties of TeV blazars should be similar to those 
of AGNs more generally.  Nevertheless, evidence for a simple
connection between the two populations can be found in the
similarity between their the luminosity functions (a fact we will exploit later
in estimating the redshift evolution of the TeV blazars).  Here we
define the luminosity for the purposes of defining $\BLF$ to be the
isotropic-equivalent value associated with emission between
$100\,\GeV$ and $10\,\TeV$.  While this may be considered to be a
VHEGR luminosity, because most TeV blazars are peaked within this
band, this corresponds to the majority of the emission from these
sources.  

The objects listed in Table \ref{tab:TeVsources} were chosen because
they have well defined SEDs, based upon a combination of VERITAS,
H.E.S.S., and MAGIC observations.
These 28 sources have VHEGR spectra that are well fit by the form,
\begin{equation}\label{eq:spectra}
\frac{dN}{dE} = f_0 \left(\frac {E}{E_0}\right)^{-\alpha},
\end{equation}
where $f_0$ is the normalization in units of
$\cm^{-2}\,\s^{-1}\,\TeV^{-1}$.  The gamma-ray energy flux is 
trivially related to $dN/dE$ by $F_E=E dN/dE \propto E^{1-\alpha}$,
from which we obtain a VHEGR flux,
\begin{equation}
F =  E_0 f_0 \int_{100\,\GeV}^{10\,\TeV} dE\,\left(\frac{E}{E_0}\right)^{1-\alpha}\,,
\end{equation}
and for sources with a measured redshift a corresponding
isotropic-equivalent luminosity, $L=4\pi D_L^2 F$, where $D_L$ is the
luminosity distance.\footnote{Since the VHEGR spectra of TeV blazars
  typically are peaked above $100\,\GeV$, this overestimates the
  luminosity by a factor of order unity.}
The resulting $f_0$, $E_0$, $\alpha$, $F$, and $L$ are
collected in Table \ref{tab:TeVsources}.  In addition we list the
redshift, inferred comoving distance,
and absorption-corrected intrinsic spectral index, defined at $E_0$,
obtained via
\begin{equation}
\hat{\alpha}
=
-\left.\frac{d\ln E^{-\alpha} e^{\tau_E[E(1+z),z]}}{d\ln E}\right|_{E_0}
\simeq
\alpha - \tau_{E}\left[E_0(1+z),z\right]\,.
\end{equation}
For high-redshift sources $\hat{\alpha}$ can be
substantially less than $2$, implying that an intrinsic spectral
upper-cutoff must exist.

To produce $\phi_B$, we must account for a variety of selection effects inherent
in the sample listed in Table \ref{tab:TeVsources}.  The objects in Table
\ref{tab:TeVsources} were originally selected for study for a variety
of source-specific reasons, e.g., existing well known sources, extremely
high X-ray to radio flux ratio in the Sedentary High energy peaked BL
Lac catalog, hard spectrum sources in the \Fermi point source catalog,
and flagged as promising by the \Fermi-LAT collaboration.  In
addition, the source selection suffered from the usual problems
associated with surveys (e.g., scheduling conflicts with other
targets, moon, bad weather, etc.).  As a consequence, this sample is
somewhat inhomogeneous.  Nevertheless, in lieu of a less-biased
sample, we will treat it as homogeneous and correct for the selection
effects were possible, focusing upon those due to the sky coverage and
duty cycle of TeV observations, and those due to sensitivity limits of
current imaging atmospheric Cerenkov telescopes.

To estimate the sky
completeness and duty cycle of this set of objects we rely upon the all-sky
$\GeV$ gamma-ray observations of HBL and IBL sources by \Fermi
\citep{Fermi_AGNCatalogue2010}.  Outside of the Galactic plane, \Fermi observes
118 high-synchrotron peaked (HSP) blazars and a total of 46 intermediate-synchrotron
peaked (ISP) blazars.  Roughly half of the latter are likely to emit
VHEGRs as indicated by their flat spectral index between 0.1 and 100 GeV
($\alpha\lesssim2$; see the spectral index distribution of Figure 14
in \citet{Fermi_AGNCatalogue2009}).  Of these potential 141 TeV
blazars, only 22 have also been coincidentally identified as TeV
sources, whereas there are a total of 33 known TeV blazars (29 HBL, 4
IBL).  If these 141 sources are all 
$\TeV$ emitters, but have not been detected due to incomplete sky coverage of
current TeV instruments, then the selection factor is
$\eta_{\rm sel}=141/33=4.3$. 
In addition, the duty cycle of coincident $\GeV$ and $\TeV$
emission is $\eta_{\rm duty}=33/22=1.5$.\footnote{Note that we implicitly assume
  that the luminosity distribution of observed TeV sources reflects the true
  distribution after accounting for flux incompleteness, and thus constant
  correction factors (independent of luminosity) can be used to estimate the
  true distribution. We show that this assumption is fulfilled in Figure
  \ref{fig:BLF}.}  Finally, by excluding the Galactic plane for galactic
latitudes $b<10\degr$, this is an underestimate by roughly
$\eta_{\rm sky}=1.17$.

We make a rough attempt to correct for the sensitivity
limit of the TeV instruments, inferring a flux limit by fitting the
upper-envelope of the TeV-blazar flux--luminosity distance
distribution.  Somewhat surprisingly, despite the heterogeneous nature
of the TeV observations, are remarkably well described by a single
flux limit: $4.19\times10^{-12}\,\erg\,\cm^{-2}\,\s^{-1}$.  This
process and the associated limit are shown explicitly in the inset of
Figure \ref{fig:BLF}.  From this we obtain a maximum redshift, and
therefore comoving volume, associated with each luminosity.
We then construct $\BLF$ by counting the number of TeV blazars
per unit $\log_{10}L$ in each logarithmic luminosity bin and dividing
by the comoving volume.
The resulting $\BLF$, weighted by luminosity (and therefore showing
the luminosity density in comoving units), is shown in Figure
\ref{fig:BLF}.  It peaks at $\sim4\times10^{44}\,\erg\,\s^{-1}$,
implying that, as expected, these objects are systematically dimmer
than most other AGN, and exhibits the broken-power law shape typical
of AGN luminosity functions.  More importantly, despite a handful of
sources with redshifts $\sim0.5$, the objects in Table \ref{tab:TeVsources}
are all nearby, and therefore our $\BLF$ corresponds to that in the local
Universe, i.e., $z\sim0.1$.
Based upon this, the inferred present-day TeV-blazar luminosity
density is roughly $1\times10^{38}\,\erg\,\s^{-1}\,\Mpc^{-3}$.

Also shown in Figure \ref{fig:BLF} is the $L\QLF$ obtained by
\citet{Hopkins+07}.  After rescaling $L\QLF$ to lower luminosities
($0.55$) and lower luminosity densities ($2.1\times10^{-3}$), it
provides a remarkably good fit to our $L\BLF$ (reduced-$\chi^2$ of
0.13 with 3 degrees of freedom), i.e., we find
\begin{equation}
\BLF(0.1,L) \simeq 3.8\times10^{-3} \QLF(0.1,1.8 L)\,,
\label{eq:BLFQLF}
\end{equation}
where an explicit expression for $\QLF$ is given in Appendix
\ref{app:QLF}.  While there is considerable uncertainty in $L\BLF$,
especially at low luminosities, it clearly does not fit $L\QLF$ at
higher $z$.  This suggests two immediate conclusions:
\begin{enumerate}
\item The bolometric output of TeV blazars and quasars are regulated by
similar mechanisms, presumably accretion, despite the large
difference in luminosity and the details of the emission processes
between the two populations.
\item TeV blazars and quasars are contemporaneous elements in a single
AGN distribution; specifically, TeV-blazar activity does not lag that
of quasars.
\end{enumerate}

\subsubsection{Extending $\BLF(z,L)$ to high $z$}

Based upon the strong similarities between $\BLF$ and $\QLF$, and the
associated implications, we make the conservative {\em theoretical}
assumption that the redshift evolution of the TeV blazars follows that
of quasars.
That is, we suppose that Equation (\ref{eq:BLFQLF}) holds at all $z$.
This implies that the integrated comoving TeV-blazar
isotropic-equivalent luminosity density is given by
\begin{equation}
\Lambda_B(z) \equiv \int_{-\infty}^\infty L \BLF(z,L)\,d\log_{10}L 
=
\eta_B \Lambda_Q(z)\,,
\end{equation}
where the constant of proportionality, $\eta_B\simeq2.1\times10^{-3}$,
is then set by the comparison between $\BLF$ and $\QLF$ at $z=0.1$.
As a consequence, in comoving units, the TeV-blazar luminosity density
would be roughly an order of magnitude larger at $z\simeq2$,
$\sim1\times10^{39}\,\erg\,\s^{-1}\,\Mpc^{-3}$.

\subsection{\Fermi TeV Blazar Counts and the Local Evolution of the  Blazar Luminosity Function}

\begin{figure}
\begin{center}
\includegraphics[width=0.95\columnwidth]{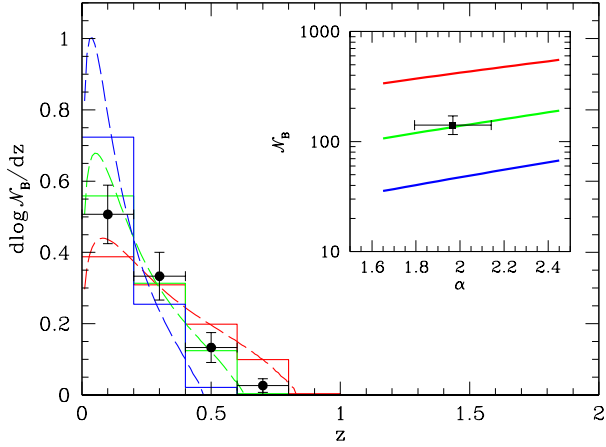}
\end{center}
\caption{The normalized number of TeV blazars (presumably the hard
  gamma-ray blazars comprised of hard ISPs and all HSPs) that
  are expected to have been observed by \Fermi as a function of
  redshift.  The dashed lines and solid histograms show the ideal and
  binned distributions for $\alpha=1.95$, though neither varies
  significantly with $\alpha$.  The total number of objects is sensitive
  to the spectral index, and shown as a function of $\alpha$ in the
  inset.  Different colors correspond to different normalizations
  between the $100\,\GeV$--$10\,\TeV$ and $100\,\MeV$--$100\,\GeV$
  luminosities: $\eta_{\rm min}=0.78$ (red), 1.6 (green), and 3.1
  (blue), i.e., the former is the latter luminosity multiplied by
  $\eta_{\rm min}$).  For reference, the normalized number of hard
  gamma-ray blazars (HSPs and half of the ISPs) observed by \Fermi are
  shown by the black circles, and the black square in the inset shows
  the average spectral index and total number observed, with horizontal
  error bars giving the $1$-$\sigma$ range.
}\label{fig:FHBLc}
\end{figure}

The redshift distribution of the \Fermi BL Lac sample
\citep[i.e., the First LAT Catalog, 1LAC; ][]{Fermi_AGNCatalogue2010}
is peaked at $z<0.2$ and falls rapidly thereafter.  Inasmuch as the
\Fermi HSP and ISP counts directly probes the low-$z$ $\BLF$, it may appear that
they are inconsistent with the rapidly evolving $\BLF$ described in the
previous section.  Indeed, an analysis of the first three months of \Fermi
observations, suggested that the BL Lac population does not grow
substantially with redshift \citep{Fermi_AGNCatalogue2009}.
However, as seen in Figure \ref{fig:FHBLc}, this is not necessarily
the case.

Assuming, as we have, that the number of TeV blazars is equal to the
the number of hard \Fermi gamma-ray blazars, corresponding to ISPs
with $\alpha\lesssim2$ (comprising roughly half of that class) and
HSPs, the expected number observed inside of a given redshift is:
\begin{multline}
{\mathcal N}_B(z) = 
\int_0^z dz' \frac{dD}{dz'} 4\pi D_A^2 \\
\times\int_{\log_{10}L_{\rm min}(z')}^{\log_{10}L_{\rm max}} (1+z')^3 \BLF(z',L) \,\d\log_{10}L\,,
\end{multline}
where $D\equiv\int c dt'=\int c dz'/\left[H(z')(1+z')\right]$
and $D_A=D_L/(1+z)^2$ are the proper and angular diameter distances,
respectively, $L_{\rm max}\simeq2\times10^{46}\,\erg\,\s^{-1}$ is the
intrinsic upper-cutoff of the TeV blazar isotropic-equivalent luminosity
function, and
\begin{equation}
\begin{aligned}
L_{\rm min}(z)
&=
F_{\rm min} 4\pi D_L^2(z) (1+z)^{\alpha-2}\\
&\simeq
4\times10^{46}\,\eta_{\rm min}\left(\frac{1+z}{2}\right)^{\alpha-2} \left[\frac{D_L(z)}{D_L(1)}\right]^{2}
\,\erg\,\s^{-1}
\end{aligned}
\end{equation}
is lower-limit set by the \Fermi flux limit
\citep[see Figure 23 of ][and surrounding discussion]{Fermi_AGNCatalogue2010}.
The factor $\eta_{\rm min}$ is a correction relating the $100\,\GeV$--$10\,\TeV$
isotropic-equivalent luminosities we employ to define $\BLF$ to the
$100\,\MeV$--$100\,\GeV$ luminosities used by \Fermi to define the
flux limit.   
The dependence upon $\alpha$ arises from the limited spectral coverage
of both definitions, though here we fix $\alpha=3$ for all objects
based upon the sources that dominate the TeV flux at Earth.
Note that since the $\QLF$ from \citet{Hopkins+07} diverges at small
$L$, the lower-luminosity cutoff is critical to getting both the total
number of TeV blazars and the shape of their redshift distribution correct.
The resulting ${\mathcal N}_B$ is shown in Figure \ref{fig:FHBLc} for three
different choices of $\eta_{\rm min}$: $0.78$, $1.6$, and $3.1$, of
which $\eta_{\rm min}=1.6$ is most similar to the 1LAC hard gamma-ray blazars.

Generally, our $\BLF$ does an excellent job of reproducing the
overall number of 1LAC hard gamma-ray blazars and the dominance of
nearby objects in their redshift distribution.  This is despite the
strong redshift evolution of $\BLF$ implied by its relationship with
$\QLF$.  The reason for this is the flat distribution at low $L$, the
steep drop-off at high $L$ (a result of which is that the shape is only
marginally sensitive to the cutoff at $L_{\rm max}$) and the rapidly
growing $L_{\rm min}$ due to the fixed flux limit.

However, we note that the comparison between $\BLF$ and the 1LAC hard
gamma-ray blazar statistics assumes that those with measured
redshifts, comprising roughly half of the 1LAC 
hard gamma-ray blazar sample, are representative of the hard gamma-ray
blazar population as a whole.  This appears not to 
be the case; as we discuss in detail in Appendix \ref{app:FHSPzs}, it
is clear that the 1LAC HSPs with and without measured redshifts are
not drawn from the same underlying $\alpha$-distribution.  Based upon
a similar analysis for 1LAC BL Lacs generally,
\citet{Fermi_AGNCatalogue2010} argued that the objects without
redshifts are more consistent with $z>0.5$ 
population.  However, this conclusion does not extend to HSPs, for
which there are only three sources in the clean 1LAC HSP sample with
$z>0.5$.  Rather, in Appendix \ref{app:FHSPzs}, we show that the 1LAC
HSPs without redshifts are distributed both in spectral index and flux
much more similarly with nearby HSPs ($z<0.25$) than more distant HSPs
($z\ge0.25$).  If this is because these sources are intrinsically
under-luminous, nearby objects, it could marginally improve the
already remarkable comparison between the number of objects implied by
our $\BLF$ and those observed.
However, because the number of predicted
1LAC hard gamma-ray blazars is strongly dependent upon the flux limit,
even a marginal increase in either $\BLF$ at low luminosities or the
effective flux limit results in a $d{\mathcal N}_B/dz$ that is more
strongly weighted at low $z$, potentially allowing even more dramatically
evolving luminosity functions.  For this reason we conclude that our
$\BLF$ is broadly consistent with the 1LAC hard gamma-ray blazar
distribution, and that \Fermi does not, at present, exclude a rapidly
evolving $\BLF$ in the recent past.

\subsection{The TeV Blazar Contribution to the \Fermi EGRB}

\begin{figure}
\begin{center}
\includegraphics[width=0.95\columnwidth]{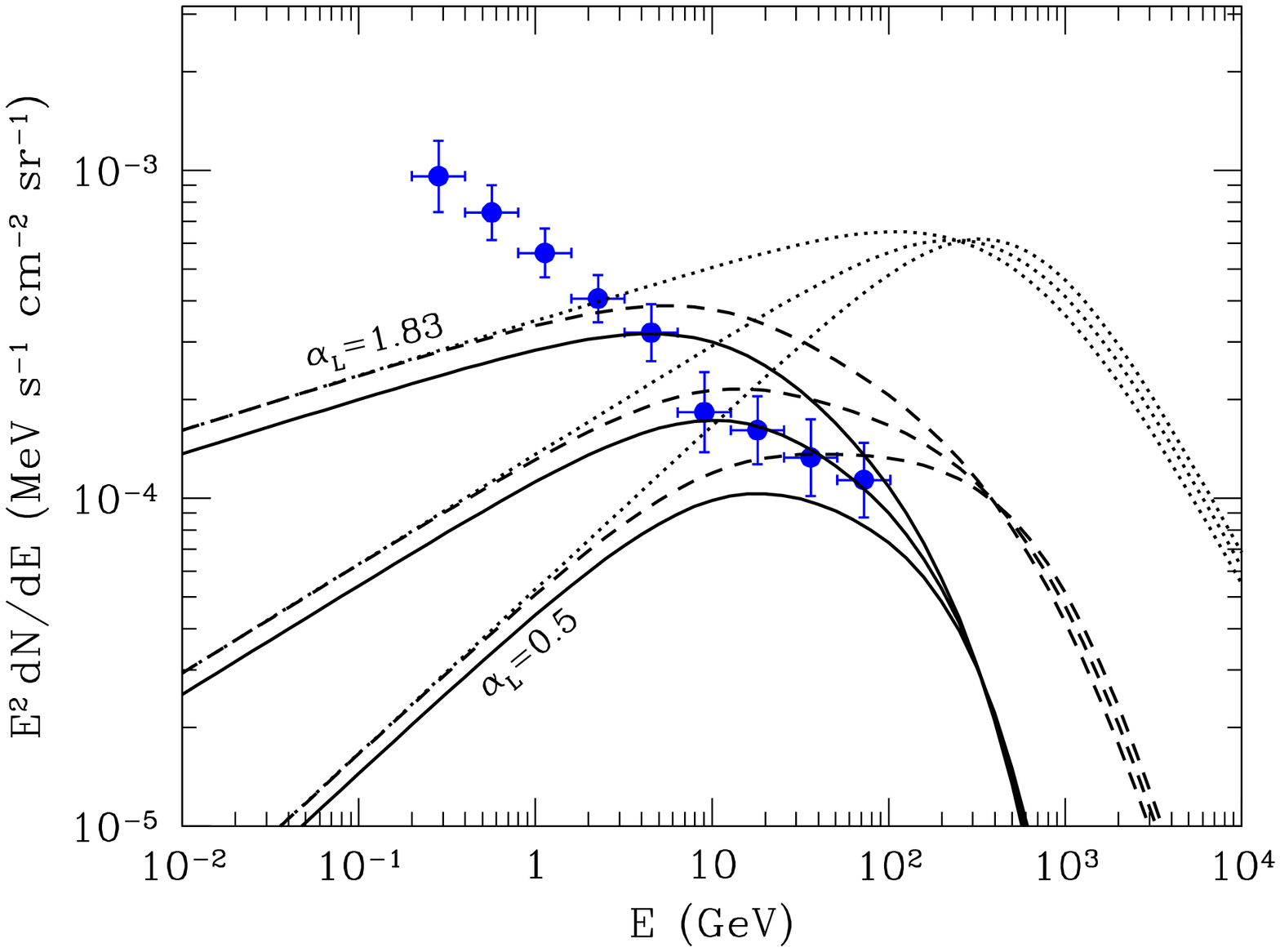}\\
\includegraphics[width=0.95\columnwidth]{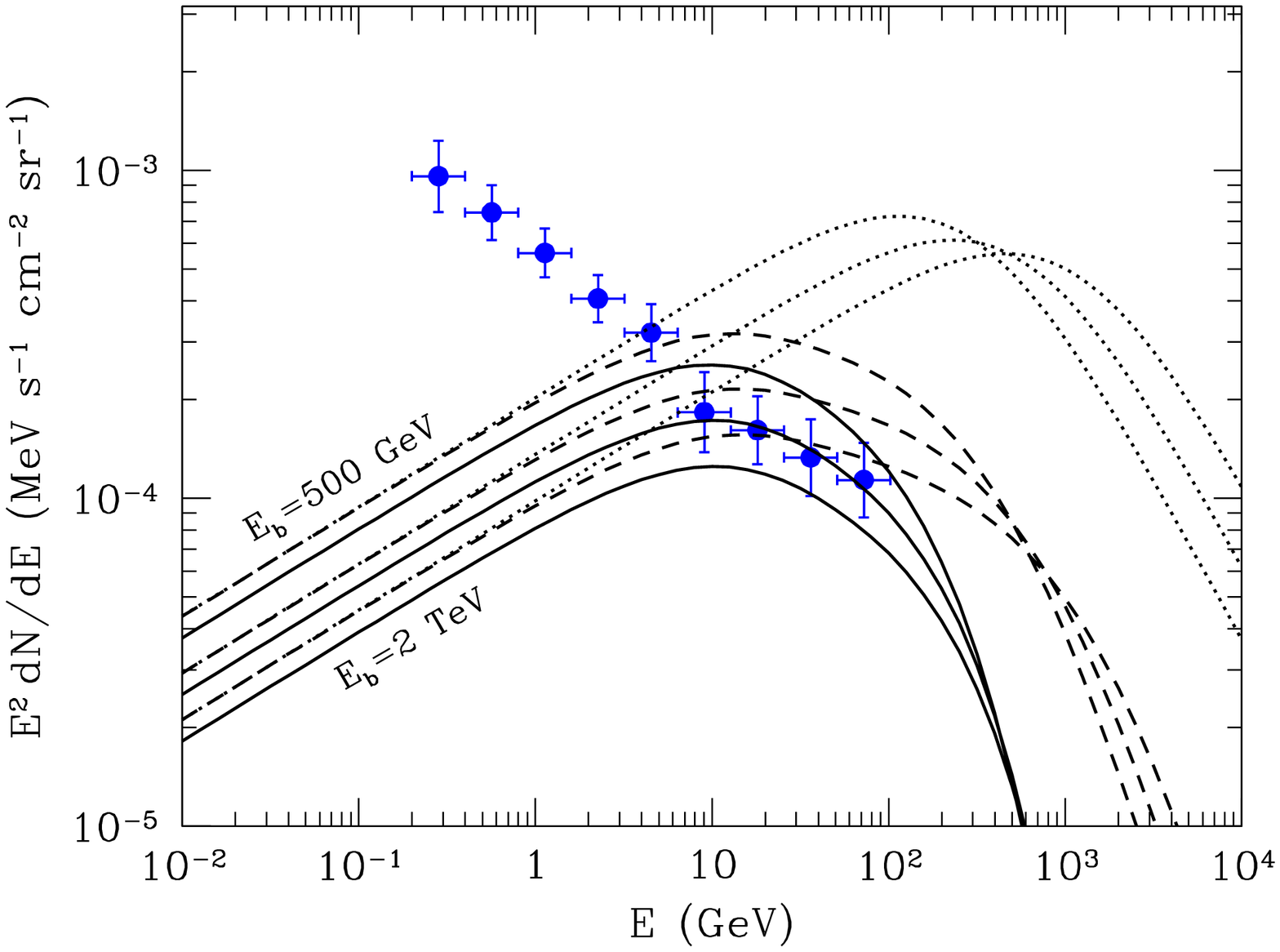}\\
\includegraphics[width=0.95\columnwidth]{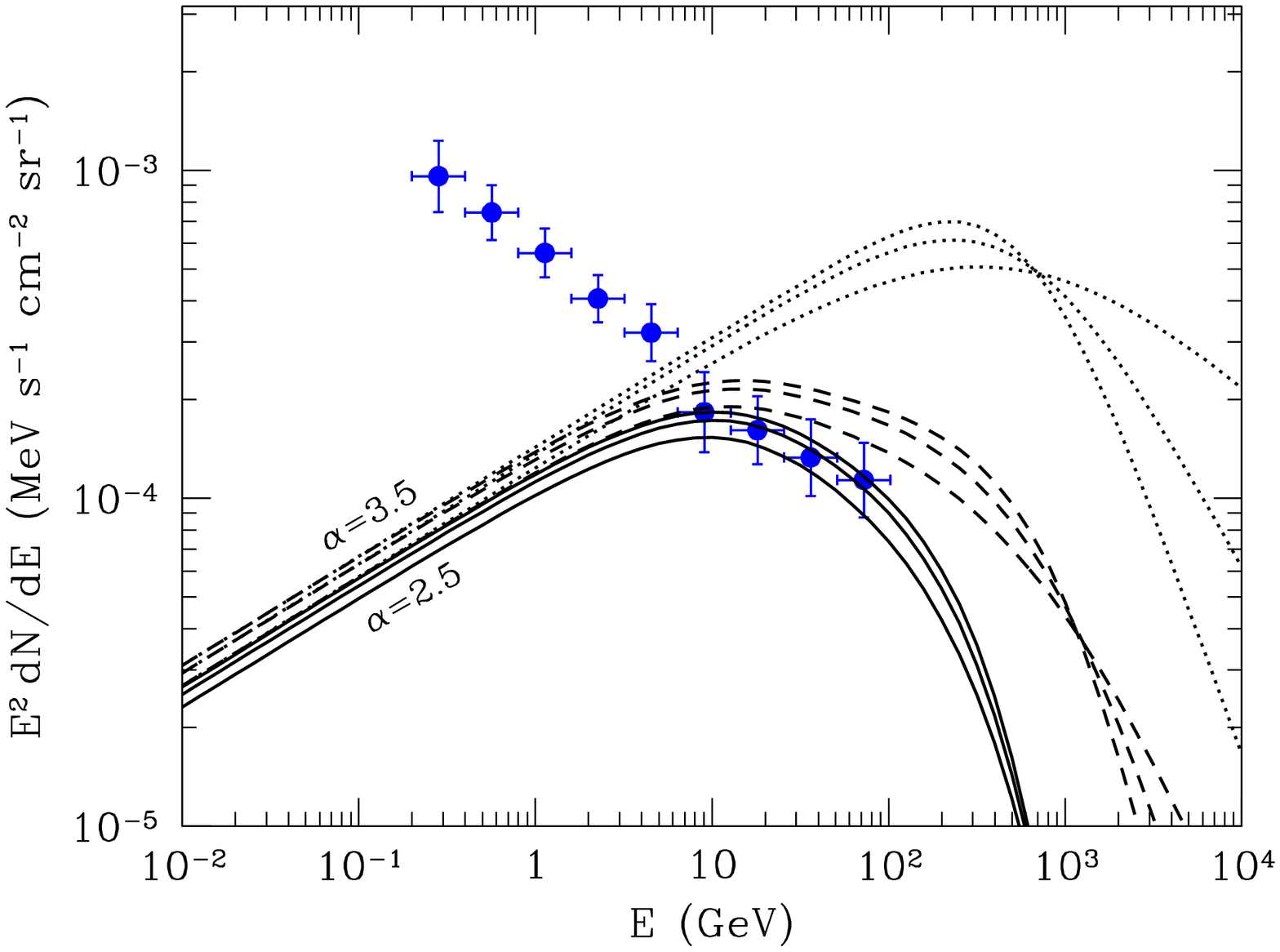}
\end{center}
\caption{Contribution of TeV blazars to the EGRB.  The solid lines
  show the contribution from TeV blazars beyond $z=0.25$, roughly the set for
  which \Fermi can no-longer resolve them.  The dashed lines show the
  contribution from all TeV blazars, including those which \Fermi should have
  been able to detect individually and to remove.  The dotted lines show
  the contribution from all TeV blazars when the annihilation upon the EBL is
  neglected.  Different curves correspond to different parameters of
  the intrinsic spectrum assumed for the blazar populations.  Top:
  Variations in the low-energy spectral index (top to bottom,
  $\alpha_L=1.83$, $1.67$, $1.5$).  Middle: Variations in the break energy
  (top to bottom, $E_b=500\,\GeV$, $1\,\TeV$, $2\,\TeV$).  Bottom:
  Variations in the high-energy spectral index (top to bottom,
  $\alpha=3.5$, $3$, $3.5$).  Unless otherwise specified, the
  remaining parameters are $\alpha_L=1.63$, $E_b=1\,\TeV$, and
  $\alpha=3$.  Finally, the \Fermi measurement of the EGRB reported in
  \citet{Fermi_EGRB2010} is shown by the blue points.
  Note that in our scenario, the integral of the difference between the
  unabsorbed (dotted) and absorbed (dashed) energy fluxes, which dominates the
  total energy budget of these sources, has been dissipated into heat
  within the IGM.
}\label{fig:EGRB}
\end{figure}

While the statistics of the 1LAC hard gamma-ray blazar sample probes
the evolution of $\BLF$ at $z\lesssim0.5$, the \Fermi EGRB is
sensitive to TeV Blazars at high $z$.  The contribution to the EGRB
from TeV blazars given our $\BLF$ is shown in Figure \ref{fig:EGRB}.
Because ICCs are suppressed, computing the EGRB flux requires only
summing the individual intrinsic $\GeV$ spectra of the TeV blazars.
For simplicity, to estimate the TeV blazar contribution to the EGRB we
assume all TeV blazars have identical intrinsic spectra, given by a
broken power law,
\begin{equation}
F_E = L\hat{F}_E \propto\frac{1}{\left(E/E_b\right)^{\alpha_L-1} + \left(E/E_b\right)^{\alpha-1}}\,,
\end{equation}
where the normalization is set such that the 
$\int_{100\,\GeV}^{10\,\TeV} \hat{F}_E dE = 1$ (i.e.,
$\int_{100\,\GeV}^{10\,\TeV} L\hat{F}_E dE = L$, corresponding to the
luminosity used to define $\BLF$,  and thus determine $\eta_B$), $E_b$
is the energy of the spectral break, and $\alpha_L<\alpha$ are the low and
high-energy spectral indexes, respectively.\footnote{Note that
  $\BLF(z,L)$ was determined assuming an unbroken spectrum, and thus
  in this case would {\em over estimate} the total flux from the TeV
  blazars.  Nevertheless, here this is not accounted for, i.e., we
  renormalize the broken power-law spectrum to our previous power-law
  estimate of the TeV blazar luminosity density.  As a consequence,
  our estimates of the EGRB are also {\em over estimates}.}
With this, after performing the integral over $L$, the EGRB flux is then
\begin{equation}
\begin{aligned}
E^2 \frac{dN}{dE}(E,z)
&=
\frac{\eta_B}{4\pi} \int_z^\infty\! dz' \,\frac{dD}{dz'} \,4\pi D_A^2 \frac{\tilde{\Lambda}_Q(z') \hat{F}_{E'}}{4\pi D_L^2}\e^{-\tau_E[E',z']}\\
&=
\frac{\eta_B}{4\pi} \int_z^\infty\! dz' \frac{c\tilde{\Lambda}_Q(z')}{H(z')(1+z')^5}  \,\hat{F}_{E'}\,\e^{-\tau_E[E',z']}\,,
\end{aligned}
\end{equation}
where $E'=E(1+z')$, and
$\tilde{\Lambda}_Q = \Lambda_Q (1+z)^3$ is the physical quasar luminosity density.
We choose fiducial values for
the spectral parameters of $\alpha_L=1.33$, $E_b=1\,\TeV$, and
$\alpha=3$, typical of the TeV blazars
\citep[see Table \ref{tab:TeVsources} and ][]{Ghis:11}.  Finally,
since \Fermi resolves individual gamma-ray blazars with
isotropic-equivalent luminosities $\sim10^{45}\,\erg\,\s^{-1}$,
roughly the location of the peak in $\BLF$, for $z\lesssim0.25$, we
construct the EGRB from sources at larger redshift.

If $E_b$ is sufficiently high ($\gtrsim2\times10^2\,\GeV$) the EGRB is
remarkably insensitive to the particular values of $E_b$ and
$\alpha$ (see the bottom two panels of Figure \ref{fig:EGRB}).  This
is a direct result of the annihilation of the VHEGRs upon the EBL,
effectively removing the relevant portion of the blazar SED from the
EGRB.  More important is the low-energy spectral index.  However, for
values of $\alpha_L$ that are consistent with the Comptonization
models for the VHEGR emission, the anticipated EGRB is
consistent with the \Fermi result.   Thus, we find that
despite our dramatically evolving $\BLF$, we are are able to satisfy the
limits imposed by the \Fermi EGRB.  Moreover, for reasonable
spectral-parameter values it is possible to accurately reproduce both
the magnitude and shape of the high-energy EGRB (i.e., at energies
$>10\,\GeV$).\footnote{Note that 
  since we are only presenting the EGRB from the TeV blazars, we
  necessarily neglect the contributions from the FSRQs and LBLs which
  dominate the EGRB at low energies.}
At first this may appear in conflict with other studies
that have performed more sophisticated analyses of the gamma-ray
emission from blazars \citep[e.g.,
][]{Naru-Tota:06,Knei-Mann:08,Inou-Tota:09,Vent:10} that have
typically found that such a strongly evolving $\BLF$ over-produces the
\Fermi EGRB.  However, this is not the case for three reasons.

First, the AGN populations typically considered in computations of the
EGRB include the FSRQs and LBLs, which dominate at low energies.  The
TeV blazars of interest here are only the 
high-energy tail of the blazar distribution, and it is these alone
that we fix to the quasar luminosity function (indeed, it is for only
these objects that the arguments surrounding Figure \ref{fig:BLF} have
been made).  Moreover, the TeV blazars themselves are generally
dimmer, both bolometrically and within the \Fermi LAT energy band
($200\,\MeV$--$100\,\GeV$), than the LBLs and FSRQs, which have
typical isotropic-equivalent luminosities of
$3\times10^{47}\,\erg\,\s^{-1}$ \citep{Naru-Tota:06}.  

Second, and perhaps most importantly, the suppression of the ICCs
means that the VHEGR emission is lost to heat in the IGM, not
reprocessed to below $100\,\GeV$.  Since the TeV blazar flux is presumed to
be dominated by emission above $100\,\GeV$, the non-radiative
dissipation of this emission substantially reduces the impact of the
TeV blazars upon the EGRB.  For all of the spectral parameters we considered
the un-absorbed fluxes exceed the $30\,\GeV$ EGRB by a significant
margin, implying that the suppression of the ICCs is crucial to
bringing the TeV blazar contributions in line with the \Fermi EGRB.  Thus,
the lack of the ICCs generally appears necessary to reconcile the
blazar and quasar luminosity functions.

Third, the VHEGR spectra from the most luminous TeV blazars
necessarily peaks near or above TeV energies.  For a number of the
sources in Table \ref{tab:TeVsources}, the intrinsic VHEGR spectrum
(adjusted for annihilation on the EBL) is inverted, indicating that
this break is {\em above} a TeV.  The brightest TeV blazar in Table
\ref{tab:TeVsources}, Mkn 421, has an inverted spectrum below
$100\,\GeV$ \citep{Fermi_AGNCatalogue2010}, implying
$E_b\sim500\,\GeV$.  The second brightest, 1ES 1959+650, is consistent
with being flat below $100\,\GeV$, and appears to peak near $1\,\TeV$
when in the high-state \citep{Dani_etal:05}.  Similarly, 1ES 2344+514
peaks near $\sim300\,\GeV$ \citep{albert+07c} and Mkn 501 is inverted
below $100\,\GeV$, implying a peak above that
value \citep{Fermi_AGNCatalogue2010}.  Thus the sources likely to
dominate the VHEGR background, and by extension the EGRB, all turn
over near $100\,\GeV$--$1\,\TeV$.  This is expected if the the VHEGR
emission arises from inverse-Compton scattering the synchrotron bump
\citep[see, e.g.,][ and references therein]{Ghis:11}.  The consequence
of the spectral break is the suppression of the contribution from the
TeV blazars to the EGRB below $E_b$.
Were the TeV blazar contribution to the EGRB dominated by sources with
$E_b\lesssim2\times10^2\,\GeV$ it would exceed the \Fermi EGRB at
$E\lesssim10\,\GeV$.

A more complete analysis of the EGRB should include a variety of spectra,
including a distribution of break energies, smoothly connecting the
HBL, IBL and LBL populations.  However, this is beyond the scope of this
paper, which is primarily concerned with the fate of the VHEGR
emission, absent in the LBLs and FSRQs.  Nevertheless a recent
comprehensive model, which includes the contributions of the FSRQs and
BL Lacs observed by \Fermi, as well as starburst galaxies, has had
considerable success fitting the \Fermi EGRB with an even more extremely
evolving $\BLF$; fixing it to the luminosity function of radio galaxies 
\citep{Cavadini+2011,Stec-Vent:11}.  Of particular relevance here is that
\citet{Cavadini+2011} and \citet{Stec-Vent:11} explicitly ignored the
potential contributions of ICCs.  Unlike their analysis, however, we
find that the TeV blazars are capable of reaching the highest \Fermi
bands despite the annihilation upon the EBL.

\section{Conclusions}\label{sec:C}
The cold, highly anisotropic beams of ultra-relativistic $e^\pm$ pairs
produced by the annihilation of VHEGRs upon the EBL are unstable to
plasma beam instabilities.  More importantly, for a wide range of
parameters relevant for the observed TeV blazars these instabilities
may be capable of isotropizing, and potentially extracting the
kinetic energy 
of, the pairs at a rate orders of magnitude faster than inverse-Compton
scattering.

This has far reaching consequences for efforts to constrain the IGMF 
using empirical limits upon the GeV emission from known TeV sources.
Typically, $\sim300\,\yr$ after the onset of TeV emission the pair
beam density has grown sufficiently for plasma beam instabilities to
dominate its evolution, randomize the beam, and potentially suppress the
inverse-Compton signal upon which the IGMF limits are based.  Note
that due to the beam disruption by the instabilities, this occurs even
if the plasma instabilities do not ultimately cool the pairs.
As a consequence, the present
constraints upon the IGMF, obtained by the non-observation of an
inverse-Compton GeV bump in the spectra of bright TeV blazars are
inherently unreliable.  Nevertheless, the sudden appearance of a
TeV-bright blazar or intrinsically transient sources (e.g., GRBs)
provide a means to temporarily avoid the consequences of plasma beam
instabilities during the growth of the pair beam.  Alternatively,
observing particularly dim sources, $L\lesssim10^{42}\,\erg\,\s^{-1}$,
limits the beam density directly, again avoiding the complications
imposed by plasma processes.  Finally, the presence of these 
plasma instabilities in the pair beams of TeV blazars, which manifest 
themselves through their impact on the IGM (see Paper II, Paper
III, and \citet{Puchwein+2011}), implies the most stringent upper
limit to date on the IGMF: $B\lesssim 10^{-12}\,\G$. 

If the plasma instabilities can efficiently convert the pair beam
kinetic energy into heat in the IGM, as we anticipate based upon
existing numerical simulations and the arguments in Section
\ref{sec:NLS}, they would necessarily suppress the development of
ICCs, and thus prevent the reprocessing of the VHEGR emission from
bright sources to GeV energies.
The lack of ICCs, independent of the mechanism that facilitates the
local dissipation of the pair kinetic energy, would greatly weaken the
constraints upon the evolution of the blazar population derived from
the unresolved EGRB measured by \Fermi.  By introducing a spectral
break near $1\,\TeV$ and eliminating the reprocessed VHEGR emission,
we find that the \Fermi EGRB is consistent with a TeV blazar (and
therefore presumably hard gamma-ray blazar) luminosity function fixed
to that of quasars, normalized by comparing objects in the local Universe 
($z\simeq0.1$), and motivated by the remarkable similarity between
them in the local Universe.  This conclusion is relatively insensitive
to the particular parameters governing the VHEGR spectra of the TeV
blazars, requiring only that the VHEGR emission is produced via
inverse-Compton scattering.  For a wide range of spectral parameters,
we are able to match the magnitude and shape of the \Fermi EGRB at high
energies, with the low-energy component presumably arising from the
FSRQs and LBLs.  This $\BLF$, and perhaps an even more rapidly
evolving luminosity function, is also consistent with the observed
redshift distribution of the 1LAC HSP and hard-ISP sample.

Matching the high-energy EGRB (above $10\,\GeV$) requires more TeV
blazars than are currently observed, though a comparable number to
those inferred once the sky-coverage and GeV duty-cycle completeness
corrections are included.  Based upon these factors, and our rough
estimate of the EGRB, we predict that upcoming surveys performed with
next-generation Cerenkov telescope arrays 
\citep[see, e.g.,][]{CTA:10} should find roughly
$2\times10^{2}$ sources above $4.2\times10^{-12}\,\erg\,\s^{-1}\,\cm^{-2}$ (our
estimate of the effective flux limit of current imaging Cerenkov telescopes), and a handful
of additional sources comparable to the brightest TeV blazars
observed.  
In addition, based upon the current number
of known TeV blazars and our estimate of $\BLF(z,L)$, the improved
anticipated sensitivities of these instruments,
$5$--$10$ time larger than current arrays,
should result in the detection of $1.5\times10^{3}$--$3\times10^{3}$ 
additional TeV blazars, with median luminosities
$\sim3\times10^{45}\,\erg\,\s^{-1}$.  These should allow more precise
estimates for their gamma-ray SEDs and a better characterization of
$\BLF(z,L)$, especially for low-luminosity objects.

Unlike inverse-Compton cooling, the plasma beam instabilities deposit the
energy locally, heating the IGM.  Moreover, the homogeneity of the EBL, and the
weak dependence of the plasma cooling rates upon the IGM density, result in a
uniform {\em volumetric} heating, in clear contrast to either photoionization
heating or mechanical feedback from AGN.  While we shall defer a detailed
discussion of the consequences of this heating to Papers II, III, and
\citet{Puchwein+2011} here we note that this unusual heating
prescription naturally explains a number of 
heretofore outstanding questions, including the inverted equation-of-state
(temperature-density relation) for low density regions in the IGM (Paper II),
the suppression of dwarf galaxies and their histories,
the segregation of galaxy clusters and groups into cool core and
non-cool core populations (Paper III), and the quantitative
properties of the high-redshift \Lya forest \citep{Puchwein+2011}.
As a consequence, despite the fact that our estimates of the plasma cooling rates
are limited to the linear regime (though with some numerical support), there
are a variety of observational reasons to believe that plasma cooling,
or an analogous mechanism, does in fact dominate the evolution of the
ultra-relativistic pair beams.

\acknowledgements We thank Tom Abel, Marco Ajello, Marcelo Alvarez,
Arif Babul, Roger Blandford, James Bolton, Mike Boylan-Kolchin, Luigi
Costamante, Andrei Gruzinov, Peter Goldreich, Martin Haehnelt, Andrey Kravtsov, Ue-li
Pen, Ewald Puchwein, Volker Springel, Chris Thompson, Matteo Viel,
Marc Voit, and Risa Wechsler for useful discussions.  We are indebted
to Peng Oh for his encouragement and useful suggestions. We thank
Steve Furlanetto for kindly providing technical expertise. These
computations were performed on the Sunnyvale cluster at CITA.
A.E.B. and P.C. are supported by CITA. A.E.B. gratefully acknowledges
the support of the Beatrice D. Tremaine Fellowship.  C.P. gratefully
acknowledges financial support of the Klaus Tschira Foundation and
would furthermore like to thank KITP for their hospitality during the
galaxy cluster workshop.  This research was supported in part by the
National Science Foundation under Grant No. NSF PHY05-51164.

\begin{appendix}

\section{Instability Growth Rates}\label{sec:instabilities}
Here we compute the growth rates for the various plasma instabilities
discussed in the text within the context of relativistically moving
pair plasmas.  In all cases we make use of the kinetic theory
description of the underlying plasmas.  This necessarily assumes that
the plasma, and in particular the beam plasma, is sufficiently dense
that it is well described by a distribution function on the relevant
scales.  This is equivalent to requiring that many particles within a
characteristic energy range be present on the plasma scale of the IGM,
i.e., the conditions laid out in Section \ref{sec:AFLUPCR}.
In our analysis the Vlasov equation will play a central role, which
owing to the relativistic nature
of the calculation, we express in terms of the canonical pair,
$\x$ and $\p$, making use of the Lorentz invariance of the
electron/positron distribution functions $f^\mp(\x,\p)$ and the
phase-space volume element $d^3\!x\,d^3\!p$.  In what follows we set
$c=1$ unless otherwise specified.

\subsection{Relativistic Pair Two-Stream Instability} \label{sec:TS}
The two-stream instability arises due to the excitation of
negative-energy electrostatic waves in the beam and target plasmas.
These waves carry away both the energy and momentum of the beam.
Specifically, we compute the growth rate of the electrostatic wave
moving in the direction of the beam in the absence of a background
magnetic field\footnote{Where the beam energy density is large in comparison
to that associated with the background magnetic field this is well
justified.}.  In this case, the Vlasov equations for the electrons and
positrons are:
\begin{equation}
\begin{aligned}
\frac{D f^-}{Dt} - e\E\cdot\frac{\partial f^-}{\partial\p} &=  0\\
\frac{D f^+}{Dt} + e\E\cdot\frac{\partial f^+}{\partial\p} &=  0\,,
\end{aligned}
\end{equation}
where $D/Dt\equiv\partial/\partial t + \v\cdot\grad$ is the Stokes
derivative and $\E$ is the net electric field.  Linearizing these and
Fourier transforming in $t$ and $\x$ gives
\begin{equation}
\begin{aligned}
-i(\omega-\k\cdot\v) f^-_1 - e \E_1\cdot\frac{\partial f^-_0}{\partial\p}&=0\\
-i(\omega-\k\cdot\v) f^+_1 + e \E_1\cdot\frac{\partial f^+_0}{\partial\p}&=0\,,
\end{aligned}
\end{equation}
where $f^\mp_1$ are the perturbations to the electron and positron
distribution functions, $\E_1$ is the electrostatic wave field and we
have assumed that the background distributions, $f^\mp_0$, are
isotropic.  As a result we have
\begin{equation}
f^\mp_1 = \frac{\pm ie}{\omega-\k\cdot\v} \E_1\cdot\frac{\partial f^\pm_0}{\partial\p}\,.
\end{equation}

At this point we may compute the dielectric tensor associated with the
plasma response, however it will suffice to consider Gauss's law.
Thus we now compute the perturbed charge density:
\begin{equation}
\begin{aligned}
\rho_1
&=
\int \left(e f^+_1 - e f^-_1\right) \d^3\!p\\
&=
-ie^2\E_1\cdot \int \frac{1}{\omega-\k\cdot\v} \left(
  \frac{\partial f^-_0}{\partial\p}
  +
  \frac{\partial f^+_0}{\partial\p}
\right)\d^3\!p\\
&=
ie^2\E_1\cdot \int 
\left( f^-_0  + f^+_0 \right)
\left(\frac{\partial}{\partial\p}
\frac{1}{\omega-\k\cdot\v}
\right)\d^3\!p\\
&=
\frac{ie^2}{m_e}\E_1\cdot \int 
\left( f^-_0  + f^+_0 \right)
\frac{(\k - \k\cdot\v\v)}{\gamma(\omega-\k\cdot\v)^2}\,\d^3\!p\,.
\end{aligned}
\end{equation}

It is now necessary to specify the $f^\mp_0$.  We idealize the target
ionic plasma as cold and the pair beam plasma as mono-energetic,
yielding
\begin{equation}
f^-_0 = n_t \delta^3(\p) + \frac{\nb}{2}\delta^3(\p-\p_b)
\quad{\rm and}\quad
f^+_0 = \frac{\nb}{2} \delta^3(\p-\p_b)\,,
\label{eq:tsfpm0}
\end{equation}
where $n_t$ and $\nb$ are the lepton number densities in the target
and beam, respectively.  After performing the trivial integrals we
then obtain
\begin{equation}
\begin{aligned}
\rho_1
&=
\frac{i e^2}{m_e} \E_1 \cdot \left(
  n_t \frac{\k}{\omega^2}
  +
  \nb \frac{\k-\k\cdot\v_b\v_b}{\gamma_b(\omega-\k\cdot\v_b)^2}
\right)\\
&=
\frac{i e^2}{m_e} \E_1\cdot\k \left(
  \frac{n_t}{\omega^2}
  +
  \frac{\nb}{\gamma_b^3(\omega-k v_b)^2}
\right)\,,
\end{aligned}
\end{equation}
were in the final equality we used the fact that $\k\parallel\v_b$.
From Gauss's law we have $i \k\cdot\E_1 = 4\pi\rho_1$, and therefore
\begin{equation}
1 - \frac{\omega_{P,t}^2}{\omega^2} - \frac{\omega_{P,b}^2}{\gamma_b^3(\omega-kv_b)^2} = 0\,,
\end{equation}
where $\omega_{P,t}^2\equiv 4\pi e^2 n_t/m_e$ and 
$\omega_{P,b}^2\equiv 4\pi e^2 \nb/m_e$ are the plasma frequencies
associated with the target and beam plasmas.

This explicitly provides the dispersion relation, quadratic in
$\omega$ (one electrostatic wave traveling in each direction for each
plasma).  When $\nb=0$, $\omega=\omega_{P,t}$.  When $\nb\ll n_t$, as
is the case of interest here, we may solve the dispersion relation
perturbatively.  We do this by setting $\omega=\omega_{P,t}(1+\eta)$,
with $\eta\ll1$, which gives:
\begin{equation}
2 \eta - \frac{\omega_{P,b}^2}{\gamma_b^3(\omega_{P,t}-kv_b+\eta\omega_{P,t})^2}
=0\,,
\end{equation}
which is no longer independent of $k$.  When
$|\omega_{P,t}-kv_b|\gg|\eta\omega_{P,t}|$, $\eta$ is real and thus
there is no instability.  On the other hand, where
$k\simeq\omega_{P,t}/v_b$, we have
\begin{equation}
2\eta - \frac{\omega_{P,b}^2}{\gamma_b^3\omega_{P,t}^2\eta^2} = 0\,,
\label{eq:displin}
\end{equation}
and therefore $\eta^3 = \nb/2n_t\gamma_b^3$.  This has three
solutions:
\begin{equation}
\eta = \frac{1}{\gamma_b} \left(\frac{\nb}{2n_t}\right)^{1/3}
\left\{
1\,,\quad
-\frac{1}{2}-\frac{\sqrt{3}}{2}i\,,\quad
-\frac{1}{2}+\frac{\sqrt{3}}{2}i
\right\}
\end{equation}
the first of which is oscillatory, the second is decaying and the
third is growing with timescale
\begin{equation}
\Im\left(\omega\right)
=
\Im\left(\omega_{P,t}\eta\right)
=
\frac{\sqrt{3}}{2}\left(\frac{\nb}{2n_t}\right)^{1/3}\frac{\omega_{P,t}}{\gamma_b}\,.
\label{eq:TS}
\end{equation}
This differs from that associated with non-relativistic, ionic beams
only by the factor of $1/\gamma_b$, arising due to time-dilation
within the beam.

Note that since the energy within the electrostatic wave is
proportional to $E_1^2$, the rate at which energy is removed from the
beam is $\GTS\equiv2\Im\left(\omega\right)$.

\subsection{Relativistic Pair Weibel Instability} \label{sec:W}
The relativistic version of the Weibel instability has been discussed
in some detail in the literature for the case of anisotropic, though
symmetric beams
\citep{Weib:59,Frie:59,Yoon-Davi:87,Medv-Loeb:99}.  Here we consider a
the case of a dilute pair beam incident upon a much denser target
plasma.  This is essentially identical to the two-stream instability
discussed previously, though coupling instead to an electromagnetic
mode with $\k\perp\v_b$.

Again we begin with the linearized Vlasov equations for the electrons
and positrons:
\begin{equation}
\begin{aligned}
\frac{Df^-_1}{Dt} - e \left(\E_1+\v\times\B_1\right)\cdot\frac{\partial f^-_0}{\partial\p} &= 0\\
\frac{Df^+_1}{Dt} + e \left(\E_1+\v\times\B_1\right)\cdot\frac{\partial f^+_0}{\partial\p} &= 0\,,
\end{aligned}
\end{equation}
where we have $\B_1 = \k\times\E_1/\omega$ from Faraday's law.  Fourier
transforming in $t$ and $\x$ and solving for $f^\mp_1$ gives 
\begin{equation}
\begin{aligned}
f_1^\mp
&=
\pm
\frac{ie\left[\E_1+\v\times\left(\k\times\E_1\right)/\omega\right]}{\omega-\k\cdot\v}
\cdot\frac{\partial f^\mp_0}{\partial\p}\\
&=
\pm
\frac{ie}{\omega}
\left(\E_1 + \frac{\v\cdot\E_1}{\omega-\k\cdot\v}\k\right)
\cdot\frac{\partial f^\mp_0}{\partial\p}\,.
\end{aligned}
\end{equation}
The associated perturbation to the current is
\begin{equation}
\begin{aligned}
\j_1
&=
\int \left( e f^+_1 - e f^-_1 \right)\v d^3\!p\\
&=
-\frac{ie^2}{\omega}\int 
\frac{\partial (f^+_0+f^-_0)}{\partial\p}
\cdot
\left(\E_1 + \frac{\v\cdot\E_1}{\omega-\k\cdot\v}\k\right)
\v d^3\!p\\
&=
\frac{ie^2}{\omega}\int
\frac{f_0}{m_e \gamma}
\left[
\hat{\bmath{1}}
+
\frac{\k\v+\v\k}{\omega-\k\cdot\v}
-
\frac{\omega^2-k^2}{(\omega-\k\cdot\v)^2}
\v\v
\right]
\cdot
\E_1
d^3\!p\,,
\end{aligned}
\label{eq:Wj1}
\end{equation}
in which we've defined $f_0\equiv f_0^+ + f_0^-$.

Choosing the $f_0^\mp$ given by Equation (\ref{eq:tsfpm0}) and
assuming $\k\parallel\v_b\parallel\E_1$ we recover the standard
two-stream instability.  For our purposes here, however, the
computations may be substantially simplified by boosting into a frame
in which the target plasma is not at rest.  In this frame we have
$f_0$ given by,
\begin{equation}
f_0 = n_t' \delta^3(\p'-\p_t') + \nb' \delta^3(\p'-\p_b')\,,
\end{equation}
with $\p_t',\,\p_b'\parallel\v_b'$, where we will take care at the end
to properly relate all of these quantities to their target-frame
analogs.  In particular, we choose this frame such that
$\tilde{n}_t v_t'=\tilde{n}_b v_b'$, where $\tilde{n}_t$ and
$\tilde{n}_b$ are the proper densities of the target and beam plasmas,
respectively.  If $v_b$ is the beam plasma velocity in the target (or
lab) frame, this gives
\begin{equation}
v_t'
=
\xi v_b'
=
\xi \frac{v_b-v_t'}{1-v_b v_t'}\,,
\label{eq:vtp}
\end{equation}
where $\xi\equiv\tilde{n}_b/\tilde{n}_t$ and is in our case much less
than unity.  Note that this is only the center of momentum frame if
$\xi=1$ (for which $v_t'=v_b'$, the case most commonly discussed).
If we choose $\k'\perp\v_b'$ this choice of frame causes the terms
linear in $\v_b'$ in Equation (\ref{eq:Wj1}) to vanish identically,
yielding
\begin{equation}
\j_1'
=
\frac{ie^2}{m_e\omega'}
\left[\left(
\frac{n_t'}{\gamma_t'}+\frac{\nb'}{\gamma_b'}
\right)\, \hat{\bmath{1}}
-
\frac{\omega'^2-k'^2}{\omega'^2}
\left(\frac{n_t'}{\gamma_t'}\v_t'\v_t' + \frac{\nb'}{\gamma_b'}\v_b'\v_b'\right)
\right]\cdot\E_1'\,.
\end{equation}
Using the inhomogeneous wave equation, obtained from Faraday's and Ampere's law,
\begin{equation}
(k'^2 - \omega'^2) \E_1' = 4 \pi i \omega' \j_1',
\end{equation}
and choosing $\E_1'\parallel\v_b'$  produces the dispersion relation
\begin{equation}
\omega'^2-k'^2
-
\frac{\omega_{P,t}'^2}{\gamma_t'^3}
-
\frac{\omega_{P,b}'^2}{\gamma_b'^3}
-
\frac{k'^2}{\omega'^2}
\left(\frac{\omega_{P,t}'^2 v_t'^2}{\gamma_t'} + \frac{\omega_{P,b}'^2 v_b'^2}{\gamma_b'}\right)
=
0\,.
\end{equation}

At this point we make use of the fact that in our case $\xi\ll1$,
which allows a perturbative solution of Equation (\ref{eq:vtp}):
$v_t'=\xi v_b + \O(\xi^2)$ and therefore $v_b'=v_b(1-\xi)+\O(\xi^2)$.
Furthermore, $\omega' = \omega + \O(\xi^2)$ and $k'=k$ for the
geometry under consideration.  As consequence,
$\omega_{P,t}'v_t'^2/\gamma_t'\simeq\omega_{P,t}\xi^2v_b^2=\xi\omega_{P,b}v_b^2/\gamma_b\ll\omega_{P,b}v_b^2/\gamma_b$
and
$\omega_{P,t}'^2/\gamma_t'^3 \simeq \omega_{P,t}^2 = \xi^{-1} \omega_{P,b}^2 \gg \omega_{P,b}'^2/\gamma_b'^3$,
which implies that
\begin{equation}
\omega^2 - k^2
- \omega_{P,t}^2
- \frac{k^2}{\omega^2} \frac{\omega_{P,b}^2 v_b^2}{\gamma_b}(1-2\xi)
=
0\,.
\end{equation}
This has a purely imaginary solution:
\begin{equation}
\omega = i \sqrt{\frac{k^2+\omega_{P,t}^2}{2}} \sqrt{
\left[
1
+
\frac{4 k^2 \omega_{P,b}^2 v_b^2(1-2\xi)}{(k^2+\omega_{P,t}^2)^2\gamma_b}
\right]^{1/2}
-
1
}\,.
\end{equation}
For $k\ll\omega_{P,t}$ this rises linearly with $k$, saturating for
$k>\omega_{P,t}\gg\omega_{P,b}v_b^2/\sqrt{\gamma_b}$, giving the
growth rate
\begin{equation}
\Im\left(\omega\right) \simeq \frac{\omega_{P,b} v_b}{\sqrt{\gamma_b}}(1-\xi) + \O(\xi^2)\,.
\end{equation}
Given that for our application $\xi\ll1$, we will drop terms
first order in $\xi$ as well.  Note that the plasma frequency that
enters into the growth rate is that of the {\em beam}, which is much
lower than that associated with the target plasma.  Nevertheless, the
scale of the rapidly growing perturbations is limited to that
associated with the {\em target} plasma, which are general much
smaller than that associated with the perturbations in the beam
alone.  Again, the rate at which energy is sapped from the beam is
then $\GW=2\Im\left(\omega\right)$.

\section{An Explicit Expression for the Quasar Luminosity Function} \label{app:QLF}
In the interests of completeness, here we reproduce the $\QLF$ from
\citet{Hopkins+07}, corresponding to the ``Full'' case in that paper,
that we employ.  See \citet{Hopkins+07} for how this $\QLF$ was obtained,
and caveats regarding its application.

\begin{deluxetable*}{cccccccc}\tabletypesize{\tiny}
\tablecaption{Parameters of the Quasar Luminosity Function from \citet{Hopkins+07}\label{tab:QLF}}
\tablehead{
\multicolumn{2}{c}{Normalization}
&
\multicolumn{2}{c}{$\log_{10} L_*$}
&
\multicolumn{2}{c}{$\gamma_1$}
&
\multicolumn{2}{c}{$\gamma_2$}
}
\startdata
$\log_{10}\phi_*$ \tablenotemark{a} & $-4.825\pm0.060$
&
$\left(\log_{10} L_*\right)_0$ \tablenotemark{b} & $13.036\pm0.043$
&
$\gamma_{1,0}$ & $0.417\pm0.055$
&
$\gamma_{2,0}$ & $2.174\pm0.055$\\
&&
$k_{L,1}$ & $0.632\pm0.077$
&
$k_{\gamma_1}$ & $-0.623\pm0.132$
&
$k_{\gamma_2,1}$ & $1.460\pm0.096$\\
&&
$k_{L,2}$ & $-11.76\pm0.38$
&
&&
$k_{\gamma_2,2}$ & $-0.793\pm0.057$\\
&&
$k_{L,3}$ & $-14.25\pm0.80$
&&
&&\\
\enddata
\tablenotetext{a}{In units of comoving $\Mpc^{-3}$}
\tablenotetext{b}{In units of $L_\odot\equiv3.9\times10^{33}\,\erg\,\s^{-1}$}
\end{deluxetable*}

The form of $\QLF$ is assumed to be a broken power law:
\begin{equation}
\QLF(z,L) = \frac{\phi_*}{[L/L_*(z)]^{\gamma_1(z)}+[L/L_*(z)]^{\gamma_2(z)}}\,,
\end{equation}
where the location of the break ($L_*(z)$) and the power laws
($\gamma_1(z)$ and $\gamma_2(z)$) are functions of redshift.  These
are given by,
\begin{equation}
\begin{aligned}
\log_{10} L_*(z) &= \left(\log_{10} L_*\right)_0 + k_{L,1}\xi + k_{L,2}\xi^2 + k_{L,3}\xi^3\\
\gamma_1(z) &= \gamma_{1,0} 10^{k_{\gamma_1}\xi}\\
\gamma_2(z) &= 2 \gamma_{2,0} \left( 10^{k_{\gamma_2,1}\xi} + 10^{k_{\gamma_2,2}\xi} \right)^{-1}\\
\end{aligned}
\end{equation}
where
\begin{equation}
\xi \equiv \log_{10}\left(\frac{1+z}{3}\right)\,.
\end{equation}
The values of the relevant parameters are given in Table
\ref{tab:QLF}, and where any estimate of the uncertainty is made we
assume these are independently, normally distributed.

\section{Inferring the Redshift Distribution of the Fermi HSPs without Measured Redshifts}\label{app:FHSPzs}

\begin{figure}
\begin{center}
\includegraphics[width=0.95\columnwidth]{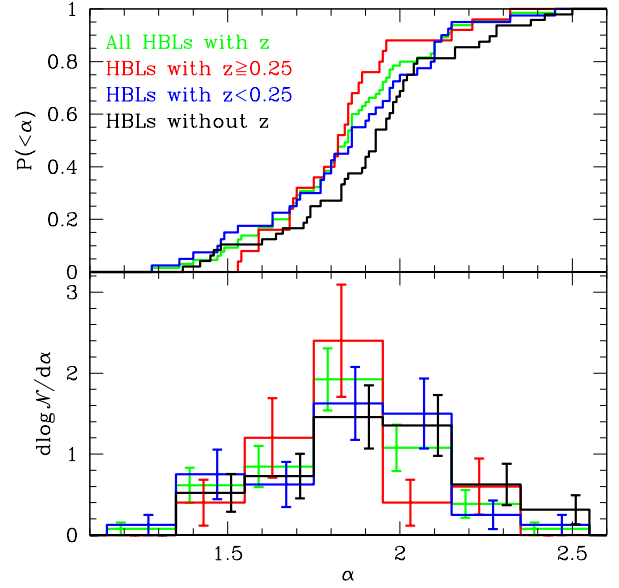}
\end{center}
\caption{Cumulative (top) and differential (bottom) number of clean Fermi
  HSPs as a function of measured spectral index.  Different line colors
  correspond to different subpopulations of the Fermi HSP sample:
  all HSPs with measured redshifts (green), HSPs with measured
  redshifts $\ge0.25$ (red), HSPs with measured redshifts $<0.25$
  (blue), and HSPs without redshift measurements (black).  In the
  bottom panel, error bars denote the Poisson uncertainty only.
}\label{fig:HSPalpha}
\end{figure}

\begin{figure}
\begin{center}
\includegraphics[width=0.95\columnwidth]{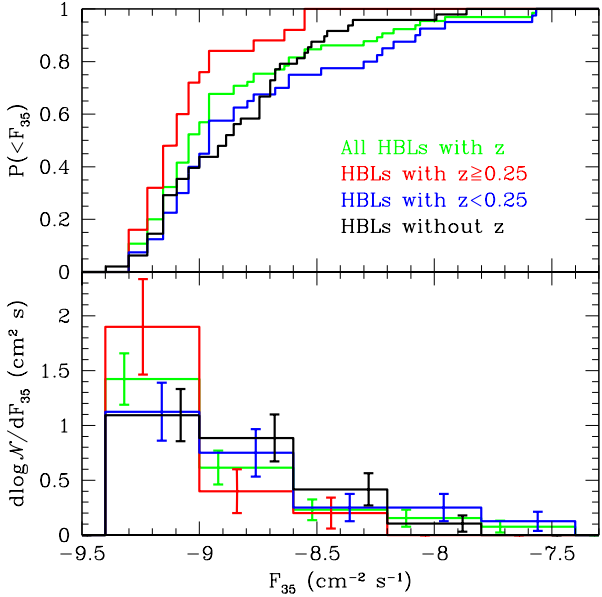}
\end{center}
\caption{Cumulative (top) and differential (bottom) number of clean Fermi
  HSPs as a function of integrated flux between $1\,\GeV$ and
  $100\,\GeV$.  Different line colors correspond to different subpopulations
  of the Fermi HSP sample: all HSPs with measured redshifts (green),
  HSPs with measured redshifts $\ge0.25$ (red), HSPs with measured
  redshifts $<0.25$ (blue), and HSPs without redshift measurements
  (black).  In the bottom panel, error bars denote the Poisson
  uncertainty only.
}\label{fig:HSPflux}
\end{figure}

\begin{figure}
\begin{center}
\includegraphics[width=0.95\columnwidth]{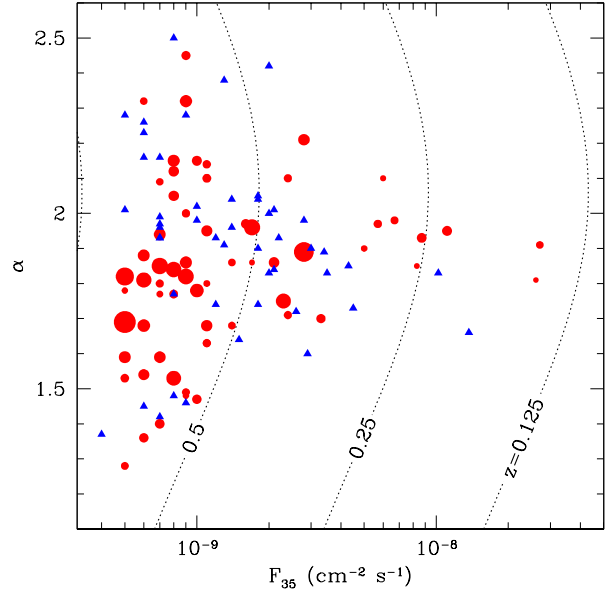}
\end{center}
\caption{Redshift and spectral index--flux distribution of the clean
  Fermi HSP sample.  HSPs with measured redshifts are shown by the red
  circles, with the circle size linearly related to $z$ (ranging
  between $z=0$ and $0.7$, i.e., large $z$ is denoted by large points
  and small $z$ by small points).  HSPs without redshifts are shown by
  the blue triangles.  For reference, the dotted black lines show
  lines of constant redshift for a source with luminosity
  $2\times10^{46}\,\erg\,\s^{-1}$ between $100\,\MeV$ and $100\,\GeV$
  \citep[the definition of $L_\gamma$ in ][note the difference with
    the definition of $F_{35}$]{Fermi_AGNCatalogue2010}.
}\label{fig:HSPfaz}
\end{figure}

\begin{deluxetable}{lcccc}\tabletypesize{\tiny}
\tablecaption{Comparison Between HSPs With and Without Measured Redshifts\label{tab:KSHSP}}
\tablehead{
\colhead{Redshift} &
\multicolumn{2}{c}{Spectral Index} &
\multicolumn{2}{c}{Flux between $1\,\GeV$--$100\,\GeV$}\\
\colhead{Population} &
\colhead{$P_{\rm KS}$} &
\colhead{$\langle\alpha\rangle$} &
\colhead{$P_{\rm KS}$} &
\colhead{$\langle\log_{10}\left(F_{35}/\cm^{-2}\s^{-1}\right)\rangle$}
}
\startdata
All HSPs with $z$'s  & $0.04$  & $2.84\pm0.03$ & $0.07$  & $-8.89\pm0.01$\\
HSPs with $z\ge0.25$ & $0.01$  & $2.83\pm0.05$ & $0.007$ & $-9.06\pm0.03$\\
HSPs with $z<0.25$   & $0.45$   & $2.84\pm0.03$ & $0.42$   & $-8.78\pm0.02$\\
HSPs without $z$'s   & ---     & $2.91\pm0.03$ & ---     & $-8.86\pm0.02$\\
\enddata
\end{deluxetable}

Of the 118 HSPs detected by Fermi and reported in the First LAT AGN
Catalog
\citep[1LAC, ][, there called HSPs]{Fermi_AGNCatalogue2010}, 
113 are members of the clean sample (meaning that there are no
ambiguities surrounding their detection), and of these only 65 have
measured redshifts.  This is a somewhat larger fraction that for 1LAC
BL Lacs generally, though this leaves nearly half of the HSP
population with their redshifts undetermined.  Based upon comparisons
between the spectral index and flux distributions between the HSPs
without redshifts and those at various redshift ranges, here we
argue that these are likely to be located nearby.

Already, based upon comparisons between the spectral index
distributions (SIDs) of the 1LAC BL Lacs at large, it is clear that the
objects with and without measured redshifts are not drawn from the
same underlying population \citep{Fermi_AGNCatalogue2010}.
Similarities between the SIDs of the unknown-$z$ objects and the
$z>0.5$ subset of those with measured redshifts,
\citet{Fermi_AGNCatalogue2010} has suggested that the BL Lacs without
redshifts may be biased towards higher redshifts.  However, we note
that there are only three HSPs with $z>0.5$, and thus this conclusion
is relevant for ISPs and LSPs only.

The SIDs of the unknown-$z$ HSPs is shown in Figure
\ref{fig:HSPalpha}, compared with the SIDs of HSPs with redshifts
\citep[cf. Figure 22 of ][]{Fermi_AGNCatalogue2010}.  The
Kolmogorov-Smirnov (KS) probability that these are drawn from the same
parent population is $0.04$, indicating that this is unlikely at the
2-$\sigma$ level.  In addition, we show the SIDs of HSPs with $z<0.25$
(blue line) and $z\ge0.25$ (red line), with corresponding KS
probabilities of $0.45$ and $0.01$, respectively (these are collected
in Table \ref{tab:KSHSP}).  Thus, in contrast to the 1LAC BL Lac
sample at large, the HSPs without redshifts appear to have SIDs that
are strongly inconsistent with the population observed to have high
redshifts, and indistinguishable from those at low
redshifts.  Nevertheless, the HSPs with unknown-$z$s still tend to be
softer than those with measured redshifts in either range.

A similar analysis may be performed upon the reported flux measure,
$F_{35}$, corresponding to the flux between $1\,\GeV$ and
$100\,\GeV$.  Figure \ref{fig:HSPflux} shows the flux distributions
(FDs) of the unknown-$z$, all measured $z$, $z\ge0.25$, and $z<0.25$
HSP populations, and is analogous to Figure \ref{fig:HSPalpha}.  As
before, the HSPs without redshifts have FDs that are marginally
inconsistent with being drawn from the same parent population as the
complete set of HSPs with measured redshifts, having a KS probability
of $0.07$.  The inconsistency with the high-$z$ HSP FD is even more
striking than for the SIDs, with a KS probability less than $0.007$,
i.e., that the high-$z$ FD of the unknown-$z$ and $z\ge0.25$ HSPs are
from the same distribution is excluded.  However, again, we find that
the FDs of the low-$z$ and unknown-$z$ populations are 
indistinguishable, having a KS probability of $0.42$.

That the SIDs and FDs both favor a relationship between the unknown-$z$
and low-$z$ HSP populations provides some confidence that this may, in
fact, be the case.  This is supported by the redshift distribution in
the $\alpha$-$F_{35}$ plane, depicted in Figure \ref{fig:HSPfaz}.
HSPs with high redshifts are clustered at low fluxes and hard
$\alpha$.  This is not unexpected given an upper limit upon the
luminosity of HSPs.  The dotted lines in Figure \ref{fig:HSPfaz} show
constant redshift curves in the $F_{35}$--$\alpha$ plane for a
$100\,\MeV$--$100\,\GeV$ luminosity of $2\times10^{46}\,\erg\,\s^{-1}$
\citep[the definition of $L_\gamma$ in ][note the distinction with the definition of $F_{35}$]{Fermi_AGNCatalogue2010}.
If all HSPs have intrinsic luminosities less than
$2\times10^{46}\,\erg\,\s^{-1}$, no sources at a given $z$ should be
found to the right of the associated line.  Since the volume of the
visible Universe is dominated by $z\sim1$, the majority of high-$z$
objects should then be found up against the instrumental flux limit,
i.e., at low fluxes and nearly flat spectra.  Sources with harder
spectra are likely to have higher bolometric gamma-ray luminosities
since they are likely to be more below the inverse-Compton peak,
biasing high-$z$ sources towards harder spectra.

However, the HSPs without redshifts are noticeably absent within this
high-$z$ dominated region.  This is responsible for the fact that they
more closely share their SID and FD with the low-$z$ HSPs.
The presence of HSPs without redshifts at a variety of $\alpha$s and
$F_{35}$s suggests that this is not a result of a selection effect
upon either.
It is nonetheless possible that there is some instrumental effect
which prevents the measurement of high redshifts in
{\em intrinsically} bright, soft objects, in which case a population of
high-luminosity HSPs with high (and therefore unmeasured) $z$s would
necessarily be located in regions with predominantly low measured
$z$s.  However, this is belied by the broad distribution of
luminosities and spectral indexes among the HSPs with measured redshifts.
Thus, we adopt the simpler explanation: the HSPs with unknown
redshifts have and SID and FD similar to low-$z$ objects because they
are intrinsically dim, nearby objects.

\end{appendix}

\bibliographystyle{apj}

\begin{thebibliography}{109}
\expandafter\ifx\csname natexlab\endcsname\relax\def\natexlab#1{#1}\fi

\bibitem[{{Abdo} {et~al.}(2009){Abdo}, {Ackermann}, {Ajello}, {Atwood},
  {Axelsson}, {Baldini}, {Ballet}, {Barbiellini}, {Bastieri}, {Baughman},
  {Bechtol}, {Bellazzini}, {Blandford}, {Bloom}, {Bonamente}, {Borgland},
  {Bouvier}, {Bregeon}, {Brez}, {Brigida}, {Bruel}, {Burnett}, {Caliandro},
  {Cameron}, {Caraveo}, {Casandjian}, {Cavazzuti}, {Cecchi}, {Charles},
  {Chekhtman}, {Chen}, {Cheung}, {Chiang}, {Ciprini}, {Claus}, {Cohen-Tanugi},
  {Colafrancesco}, {Collmar}, {Cominsky}, {Conrad}, {Costamante}, {Cutini},
  {Dermer}, {de Angelis}, {de Palma}, {Digel}, {do Couto e Silva}, {Drell},
  {Dubois}, {Dumora}, {Farnier}, {Favuzzi}, {Fegan}, {Ferrara}, {Finke},
  {Focke}, {Foschini}, {Frailis}, {Fuhrmann}, {Fukazawa}, {Funk}, {Fusco},
  {Gargano}, {Gasparrini}, {Gehrels}, {Germani}, {Giebels}, {Giglietto},
  {Giommi}, {Giordano}, {Giroletti}, {Glanzman}, {Godfrey}, {Grenier},
  {Grondin}, {Grove}, {Guillemot}, {Guiriec}, {Hanabata}, {Harding}, {Hartman},
  {Hayashida}, {Hays}, {Healey}, {Horan}, {Hughes}, {J{\'o}hannesson},
  {Johnson}, {Johnson}, {Johnson}, {Johnson}, {Kadler}, {Kamae}, {Katagiri},
  {Kataoka}, {Kerr}, {Kn{\"o}dlseder}, {Kocian}, {Kuehn}, {Kuss}, {Lande},
  {Latronico}, {Lemoine-Goumard}, {Longo}, {Loparco}, {Lott}, {Lovellette},
  {Lubrano}, {Madejski}, {Makeev}, {Massaro}, {Mazziotta}, {McConville},
  {McEnery}, {McGlynn}, {Meurer}, {Michelson}, {Mitthumsiri}, {Mizuno},
  {Moiseev}, {Monte}, {Monzani}, {Moretti}, {Morselli}, {Moskalenko}, {Murgia},
  {Nolan}, {Norris}, {Nuss}, {Ohsugi}, {Omodei}, {Orlando}, {Ormes}, {Ozaki},
  {Paneque}, {Panetta}, {Parent}, {Pelassa}, {Pepe}, {Pesce-Rollins}, {Piron},
  {Porter}, {Rain{\`o}}, {Rando}, {Razzano}, {Razzaque}, {Reimer}, {Reimer},
  {Reposeur}, {Reyes}, {Ritz}, {Rochester}, {Rodriguez}, {Romani}, {Ryde},
  {Sadrozinski}, {Sanchez}, {Sander}, {Saz Parkinson}, {Scargle}, {Schalk},
  {Sellerholm}, {Sgr{\`o}}, {Shaw}, {Smith}, {Smith}, {Spandre}, {Spinelli},
  {Starck}, {Strickman}, {Suson}, {Tajima}, {Takahashi}, {Takahashi}, {Tanaka},
  {Taylor}, {Thayer}, {Thayer}, {Thompson}, {Tibaldo}, {Torres}, {Tosti},
  {Tramacere}, {Uchiyama}, {Usher}, {Vilchez}, {Villata}, {Vitale}, {Waite},
  {Winer}, {Wood}, {Ylinen}, \& {Ziegler}}]{Fermi_AGNCatalogue2009}
{Abdo}, A.~A., {et~al.} 2009, \apj, 700, 597

\bibitem[{{Abdo} {et~al.}(2010{\natexlab{a}}){Abdo}, {Ackermann}, {Ajello},
  {Atwood}, {Baldini}, {Ballet}, {Barbiellini}, {Bastieri}, {Baughman},
  {Bechtol}, {Bellazzini}, {Berenji}, {Blandford}, {Bloom}, {Bonamente},
  {Borgland}, {Bregeon}, {Brez}, {Brigida}, {Bruel}, {Burnett}, {Buson},
  {Caliandro}, {Cameron}, {Caraveo}, {Casandjian}, {Cavazzuti}, {Cecchi}, {{\c
  C}elik}, {Charles}, {Chekhtman}, {Cheung}, {Chiang}, {Ciprini}, {Claus},
  {Cohen-Tanugi}, {Cominsky}, {Conrad}, {Cutini}, {Dermer}, {de Angelis}, {de
  Palma}, {Digel}, {di Bernardo}, {E Silva}, {Drell}, {Drlica-Wagner},
  {Dubois}, {Dumora}, {Farnier}, {Favuzzi}, {Fegan}, {Focke}, {Fortin},
  {Frailis}, {Fukazawa}, {Funk}, {Fusco}, {Gaggero}, {Gargano}, {Gasparrini},
  {Gehrels}, {Germani}, {Giebels}, {Giglietto}, {Giommi}, {Giordano},
  {Glanzman}, {Godfrey}, {Grenier}, {Grondin}, {Grove}, {Guillemot}, {Guiriec},
  {Gustafsson}, {Hanabata}, {Harding}, {Hayashida}, {Hughes}, {Itoh},
  {Jackson}, {J{\'o}hannesson}, {Johnson}, {Johnson}, {Johnson}, {Johnson},
  {Kamae}, {Katagiri}, {Kataoka}, {Kawai}, {Kerr}, {Kn{\"o}dlseder}, {Kocian},
  {Kuehn}, {Kuss}, {Lande}, {Latronico}, {Lemoine-Goumard}, {Longo}, {Loparco},
  {Lott}, {Lovellette}, {Lubrano}, {Madejski}, {Makeev}, {Mazziotta},
  {McConville}, {McEnery}, {Meurer}, {Michelson}, {Mitthumsiri}, {Mizuno},
  {Moiseev}, {Monte}, {Monzani}, {Morselli}, {Moskalenko}, {Murgia}, {Nolan},
  {Norris}, {Nuss}, {Ohsugi}, {Omodei}, {Orlando}, {Ormes}, {Paneque},
  {Panetta}, {Parent}, {Pelassa}, {Pepe}, {Pesce-Rollins}, {Piron}, {Porter},
  {Rain{\`o}}, {Rando}, {Razzano}, {Reimer}, {Reimer}, {Reposeur}, {Ritz},
  {Rochester}, {Rodriguez}, {Roth}, {Ryde}, {Sadrozinski}, {Sanchez}, {Sander},
  {Parkinson}, {Scargle}, {Sellerholm}, {Sgr{\`o}}, {Shaw}, {Siskind}, {Smith},
  {Smith}, {Spandre}, {Spinelli}, {Starck}, {Strickman}, {Strong}, {Suson},
  {Tajima}, {Takahashi}, {Takahashi}, {Tanaka}, {Thayer}, {Thayer}, {Thompson},
  {Tibaldo}, {Torres}, {Tosti}, {Tramacere}, {Uchiyama}, {Usher}, {Vasileiou},
  {Vilchez}, {Vitale}, {Waite}, {Wang}, {Winer}, {Wood}, {Ylinen}, {Ziegler},
  \& {Fermi-LAT Collaboration}}]{Fermi_EGRB2010}
---. 2010{\natexlab{a}}, \prl, 104, 101101

\bibitem[{{Abdo} {et~al.}(2010{\natexlab{b}}){Abdo}, {Ackermann}, {Ajello},
  {Antolini}, {Baldini}, {Ballet}, {Barbiellini}, {Bastieri}, {Baughman},
  {Bechtol}, {Bellazzini}, {Berenji}, {Blandford}, {Bloom}, {Bonamente},
  {Borgland}, {Bouvier}, {Bregeon}, {Brez}, {Brigida}, {Bruel}, {Burnett},
  {Buson}, {Caliandro}, {Cameron}, {Caraveo}, {Carrigan}, {Casandjian},
  {Cavazzuti}, {Cecchi}, {{\c C}elik}, {Charles}, {Chekhtman}, {Cheung},
  {Chiang}, {Ciprini}, {Claus}, {Cohen-Tanugi}, {Conrad}, {Costamante},
  {Cutini}, {Dermer}, {de Angelis}, {de Palma}, {Silva}, {Drell}, {Dubois},
  {Dumora}, {Farnier}, {Favuzzi}, {Fegan}, {Focke}, {Fukazawa}, {Funk},
  {Fusco}, {Gargano}, {Gasparrini}, {Gehrels}, {Germani}, {Giglietto},
  {Giommi}, {Giordano}, {Glanzman}, {Godfrey}, {Grenier}, {Grove}, {Guiriec},
  {Hadasch}, {Hayashida}, {Hays}, {Healey}, {Horan}, {Hughes}, {Itoh},
  {J{\'o}hannesson}, {Johnson}, {Johnson}, {Johnson}, {Kamae}, {Katagiri},
  {Kataoka}, {Kawai}, {Kn{\"o}dlseder}, {Kuss}, {Lande}, {Latronico}, {Lee},
  {Lemoine-Goumard}, {Llena Garde}, {Longo}, {Loparco}, {Lott}, {Lovellette},
  {Lubrano}, {Madejski}, {Makeev}, {Mazziotta}, {McConville}, {McEnery},
  {Meurer}, {Michelson}, {Mitthumsiri}, {Mizuno}, {Monte}, {Monzani},
  {Morselli}, {Moskalenko}, {Murgia}, {Nolan}, {Norris}, {Nuss}, {Ohsugi},
  {Omodei}, {Orlando}, {Ormes}, {Ozaki}, {Paneque}, {Panetta}, {Parent},
  {Pelassa}, {Pepe}, {Pesce-Rollins}, {Piron}, {Porter}, {Rain{\`o}}, {Rando},
  {Razzano}, {Reimer}, {Reimer}, {Ritz}, {Rochester}, {Rodriguez}, {Romani},
  {Roth}, {Sadrozinski}, {Sander}, {Saz Parkinson}, {Scargle}, {Sgr{\`o}},
  {Shaw}, {Smith}, {Spandre}, {Spinelli}, {Starck}, {Strickman}, {Strong},
  {Suson}, {Tajima}, {Takahashi}, {Takahashi}, {Tanaka}, {Thayer}, {Thayer},
  {Thompson}, {Tibaldo}, {Torres}, {Tosti}, {Tramacere}, {Uchiyama}, {Usher},
  {Vasileiou}, {Vilchez}, {Vitale}, {Waite}, {Wang}, {Winer}, {Wood}, {Yang},
  {Ylinen}, {Ziegler}, \& {Fermi-LAT Collaboration}}]{Fermi_EGRBApJ2010}
---. 2010{\natexlab{b}}, \apj, 720, 435

\bibitem[{{Abdo} {et~al.}(2010{\natexlab{c}}){Abdo}, {Ackermann}, {Ajello},
  {Allafort}, {Antolini}, {Atwood}, {Axelsson}, {Baldini}, {Ballet},
  {Barbiellini}, {Bastieri}, {Baughman}, {Bechtol}, {Bellazzini}, {Berenji},
  {Blandford}, {Bloom}, {Bogart}, {Bonamente}, {Borgland}, {Bouvier},
  {Bregeon}, {Brez}, {Brigida}, {Bruel}, {Buehler}, {Burnett}, {Buson},
  {Caliandro}, {Cameron}, {Cannon}, {Caraveo}, {Carrigan}, {Casandjian},
  {Cavazzuti}, {Cecchi}, {{\c C}elik}, {Celotti}, {Charles}, {Chekhtman},
  {Chen}, {Cheung}, {Chiang}, {Ciprini}, {Claus}, {Cohen-Tanugi}, {Conrad},
  {Costamante}, {Cotter}, {Cutini}, {D'Elia}, {Dermer}, {de Angelis}, {de
  Palma}, {De Rosa}, {Digel}, {Silva}, {Drell}, {Dubois}, {Dumora}, {Escande},
  {Farnier}, {Favuzzi}, {Fegan}, {Ferrara}, {Focke}, {Fortin}, {Frailis},
  {Fukazawa}, {Funk}, {Fusco}, {Gargano}, {Gasparrini}, {Gehrels}, {Germani},
  {Giebels}, {Giglietto}, {Giommi}, {Giordano}, {Giroletti}, {Glanzman},
  {Godfrey}, {Grandi}, {Grenier}, {Grondin}, {Grove}, {Guiriec}, {Hadasch},
  {Harding}, {Hayashida}, {Hays}, {Healey}, {Hill}, {Horan}, {Hughes},
  {Iafrate}, {Itoh}, {J{\'o}hannesson}, {Johnson}, {Johnson}, {Johnson},
  {Johnson}, {Kamae}, {Katagiri}, {Kataoka}, {Kawai}, {Kerr}, {Kn{\"o}dlseder},
  {Kuss}, {Lande}, {Latronico}, {Lavalley}, {Lemoine-Goumard}, {Llena Garde},
  {Longo}, {Loparco}, {Lott}, {Lovellette}, {Lubrano}, {Madejski}, {Makeev},
  {Malaguti}, {Massaro}, {Mazziotta}, {McConville}, {McEnery}, {McGlynn},
  {Michelson}, {Mitthumsiri}, {Mizuno}, {Moiseev}, {Monte}, {Monzani},
  {Morselli}, {Moskalenko}, {Murgia}, {Nolan}, {Norris}, {Nuss}, {Ohno},
  {Ohsugi}, {Omodei}, {Orlando}, {Ormes}, {Ozaki}, {Paneque}, {Panetta},
  {Parent}, {Pelassa}, {Pepe}, {Pesce-Rollins}, {Piranomonte}, {Piron},
  {Porter}, {Rain{\`o}}, {Rando}, {Razzano}, {Reimer}, {Reimer}, {Reposeur},
  {Ripken}, {Ritz}, {Rodriguez}, {Romani}, {Roth}, {Ryde}, {Sadrozinski},
  {Sanchez}, {Sander}, {Saz Parkinson}, {Scargle}, {Sgr{\`o}}, {Shaw},
  {Siskind}, {Smith}, {Spandre}, {Spinelli}, {Starck}, {Stawarz}, {Strickman},
  {Suson}, {Tajima}, {Takahashi}, {Takahashi}, {Tanaka}, {Taylor}, {Thayer},
  {Thayer}, {Thompson}, {Tibaldo}, {Torres}, {Tosti}, {Tramacere}, {Ubertini},
  {Uchiyama}, {Usher}, {Vasileiou}, {Vilchez}, {Villata}, {Vitale}, {Waite},
  {Wallace}, {Wang}, {Winer}, {Wood}, {Yang}, {Ylinen}, \&
  {Ziegler}}]{Fermi_AGNCatalogue2010}
---. 2010{\natexlab{c}}, \apj, 715, 429

\bibitem[{{Abdo} {et~al.}(2010{\natexlab{d}}){Abdo}, {Ackermann}, {Agudo},
  {Ajello}, {Aller}, {Aller}, {Angelakis}, {Arkharov}, {Axelsson}, {Bach},
  {et~al.}}]{Fermi_SED2010}
---. 2010{\natexlab{d}}, \apj, 716, 30

\bibitem[{{Acciari} {et~al.}(2009{\natexlab{a}}){Acciari}, {Aliu}, {Arlen},
  {Bautista}, {Beilicke}, {Benbow}, {B{\"o}ttcher}, {Bradbury}, {Buckley},
  {Bugaev}, {Butt}, {Byrum}, {Cannon}, {Celik}, {Cesarini}, {Chow}, {Ciupik},
  {Cogan}, {Colin}, {Cui}, {Dickherber}, {Duke}, {Ergin}, {Falcone}, {Fegan},
  {Finley}, {Finnegan}, {Fortin}, {Fortson}, {Furniss}, {Gall}, {Gibbs},
  {Gillanders}, {Grube}, {Guenette}, {Gyuk}, {Hanna}, {Hays}, {Holder},
  {Horan}, {Hui}, {Humensky}, {Imran}, {Kaaret}, {Karlsson}, {Kertzman},
  {Kieda}, {Kildea}, {Konopelko}, {Krawczynski}, {Krennrich}, {Lang},
  {LeBohec}, {Maier}, {McCann}, {McCutcheon}, {Millis}, {Moriarty},
  {Mukherjee}, {Nagai}, {Ong}, {Otte}, {Pandel}, {Perkins}, {Petry}, {Pohl},
  {Quinn}, {Ragan}, {Reyes}, {Reynolds}, {Roache}, {Rose}, {Schroedter},
  {Sembroski}, {Smith}, {Steele}, {Swordy}, {Theiling}, {Toner}, {Valcarcel},
  {Varlotta}, {Vassiliev}, {Wagner}, {Wakely}, {Ward}, {Weekes}, {Weinstein},
  {White}, {Williams}, {Wissel}, {Wood}, \& {Zitzer}}]{acciari+09c}
{Acciari}, V., {et~al.} 2009{\natexlab{a}}, \apjl, 690, L126

\bibitem[{{Acciari} {et~al.}(2008){Acciari}, {Beilicke}, {Blaylock},
  {Bradbury}, {Buckley}, {Bugaev}, {Butt}, {Celik}, {Cesarini}, {Ciupik},
  {Cogan}, {Colin}, {Cui}, {Daniel}, {Duke}, {Ergin}, {Falcone}, {Fegan},
  {Finley}, {Finnegan}, {Fortin}, {Fortson}, {Gibbs}, {Gillanders}, {Grube},
  {Guenette}, {Gyuk}, {Hanna}, {Hays}, {Holder}, {Horan}, {Hughes}, {Hui},
  {Humensky}, {Imran}, {Kaaret}, {Kertzman}, {Kieda}, {Kildea}, {Konopelko},
  {Krawczynski}, {Krennrich}, {Lang}, {LeBohec}, {Lee}, {Maier}, {McCann},
  {McCutcheon}, {Millis}, {Moriarty}, {Mukherjee}, {Nagai}, {Ong}, {Pandel},
  {Perkins}, {Pohl}, {Quinn}, {Ragan}, {Reynolds}, {Rose}, {Schroedter},
  {Sembroski}, {Smith}, {Steele}, {Swordy}, {Syson}, {Toner}, {Valcarcel},
  {Vassiliev}, {Wakely}, {Ward}, {Weekes}, {Weinstein}, {White}, {Williams},
  {Wissel}, {Wood}, \& {Zitzer}}]{acciari+08a}
{Acciari}, V.~A., {et~al.} 2008, \apj, 679, 397

\bibitem[{{Acciari} {et~al.}(2009{\natexlab{b}}){Acciari}, {Aliu}, {Aune},
  {Beilicke}, {Benbow}, {B{\"o}ttcher}, {Boltuch}, {Buckley}, {Bradbury},
  {Bugaev}, {Byrum}, {Cannon}, {Cesarini}, {Ciupik}, {Cogan}, {Cui},
  {Dickherber}, {Duke}, {Falcone}, {Finley}, {Fortin}, {Fortson}, {Furniss},
  {Galante}, {Gall}, {Gibbs}, {Gillanders}, {Grube}, {Guenette}, {Gyuk},
  {Hanna}, {Holder}, {Hui}, {Humensky}, {Kaaret}, {Karlsson}, {Kertzman},
  {Kieda}, {Konopelko}, {Krawczynski}, {Krennrich}, {Lang}, {Le Bohec},
  {Maier}, {McArthur}, {McCann}, {McCutcheon}, {Millis}, {Moriarty}, {Ong},
  {Otte}, {Pandel}, {Perkins}, {Pichel}, {Pohl}, {Quinn}, {Ragan}, {Reyes},
  {Reynolds}, {Roache}, {Rose}, {Sembroski}, {Smith}, {Steele}, {Theiling},
  {Thibadeau}, {Varlotta}, {Vassiliev}, {Vincent}, {Wakely}, {Ward}, {Weekes},
  {Weinstein}, {Weisgarber}, {Williams}, {Wissel}, {Wood}, {Pian},
  {Vercellone}, {Donnarumma}, {D'Ammando}, {Bulgarelli}, {Chen}, {Giuliani},
  {Longo}, {Pacciani}, {Pucella}, {Vittorini}, {Tavani}, {Argan},
  {Barbiellini}, {Caraveo}, {Cattaneo}, {Cocco}, {Costa}, {Del Monte}, {De
  Paris}, {Di Cocco}, {Evangelista}, {Feroci}, {Fiorini}, {Froysland},
  {Frutti}, {Fuschino}, {Galli}, {Gianotti}, {Labanti}, {Lapshov},
  {Lazzarotto}, {Lipari}, {Marisaldi}, {Mastropietro}, {Mereghetti}, {Morelli},
  {Morselli}, {Pellizzoni}, {Perotti}, {Piano}, {Picozza}, {Pilia},
  {Porrovecchio}, {Prest}, {Rapisarda}, {Rappoldi}, {Rubini}, {Sabatini},
  {Soffitta}, {Trifoglio}, {Trois}, {Vallazza}, {Zambra}, {Zanello}, {Pittori},
  {Santolamazza}, {Verrecchia}, {Giommi}, {Colafrancesco}, {Salotti},
  {Villata}, {Raiteri}, {Aller}, {Aller}, {Arkharov}, {Efimova}, {Larionov},
  {Leto}, {Ligustri}, {Lindfors}, {Pasanen}, {Kurtanidze}, {Tetradze},
  {Lahteenmaki}, {Kotiranta}, {Cucchiara}, {Romano}, {Nesci}, {Pursimo},
  {Heidt}, {Benitez}, {Hiriart}, {Nilsson}, {Berdyugin}, {Mujica}, {Dultzin},
  {Lopez}, {Mommert}, {Sorcia}, \& {de la Calle Perez}}]{acciari+09a}
---. 2009{\natexlab{b}}, \apj, 707, 612

\bibitem[{{Acciari} {et~al.}(2009{\natexlab{c}}){Acciari}, {Aliu}, {Arlen},
  {Beilicke}, {Benbow}, {B{\"o}ttcher}, {Bradbury}, {Buckley}, {Bugaev},
  {Butt}, {Byrum}, {Cannon}, {Celik}, {Cesarini}, {Chow}, {Ciupik}, {Cogan},
  {Cui}, {Daniel}, {Dickherber}, {Ergin}, {Falcone}, {Fegan}, {Finley},
  {Fortin}, {Fortson}, {Furniss}, {Gall}, {Gibbs}, {Gillanders}, {Godambe},
  {Grube}, {Guenette}, {Gyuk}, {Hanna}, {Hays}, {Holder}, {Horan}, {Hui},
  {Humensky}, {Imran}, {Kaaret}, {Karlsson}, {Kertzman}, {Kieda}, {Kildea},
  {Konopelko}, {Krawczynski}, {Krennrich}, {Lang}, {LeBohec}, {Maier},
  {McCann}, {McCutcheon}, {Millis}, {Moriarty}, {Mukherjee}, {Nagai}, {Ong},
  {Otte}, {Pandel}, {Perkins}, {Petry}, {Pizlo}, {Pohl}, {Quinn}, {Ragan},
  {Reyes}, {Reynolds}, {Roache}, {Rose}, {Schroedter}, {Sembroski}, {Smith},
  {Steele}, {Swordy}, {Theiling}, {Toner}, {Varlotta}, {Vassiliev}, {Wagner},
  {Wakely}, {Ward}, {Weekes}, {Weinstein}, {Williams}, {Wissel}, {Wood}, \&
  {Zitzer}}]{acciari+09b}
---. 2009{\natexlab{c}}, \apjl, 693, L104

\bibitem[{{Acciari} {et~al.}(2010{\natexlab{a}}){Acciari}, {Aliu}, {Beilicke},
  {Benbow}, {Boltuch}, {B{\"o}ttcher}, {Bradbury}, {Bugaev}, {Byrum},
  {Cesarini}, {Ciupik}, {Cogan}, {Cui}, {Dickherber}, {Duke}, {Falcone},
  {Finley}, {Finnegan}, {Fortson}, {Furniss}, {Galante}, {Gall}, {Gibbs},
  {Guenette}, {Gillanders}, {Godambe}, {Grube}, {Hanna}, {Hui}, {Humensky},
  {Imran}, {Kaaret}, {Karlsson}, {Kertzman}, {Kieda}, {Krawczynski},
  {Krennrich}, {Lang}, {LeBohec}, {Maier}, {McArthur}, {McCann}, {Moriarty},
  {Nagai}, {Ong}, {Otte}, {Pandel}, {Perkins}, {Pichel}, {Pohl}, {Quinn},
  {Ragan}, {Reyes}, {Reynolds}, {Roache}, {Rose}, {Schroedter}, {Sembroski},
  {Smith}, {Steele}, {Swordy}, {Theiling}, {Thibadeau}, {Vassiliev}, {Vincent},
  {Wakely}, {Weekes}, {Weinstein}, {Weisgarber}, {Williams}, \& {VERITAS
  Collaboration}}]{acciari+10b}
---. 2010{\natexlab{a}}, \apjl, 709, L163

\bibitem[{{Acciari} {et~al.}(2010{\natexlab{b}}){Acciari}, {Aliu}, {Arlen},
  {Aune}, {Bautista}, {Beilicke}, {Benbow}, {B{\"o}ttcher}, {Boltuch},
  {Bradbury}, \& et~al.}]{acciari+10c}
---. 2010{\natexlab{b}}, \apjl, 708, L100

\bibitem[{{Acciari} {et~al.}(2010{\natexlab{c}}){Acciari}, {Aliu}, {Arlen},
  {Aune}, {Bautista}, {Beilicke}, {Benbow}, {B{\"o}ttcher}, {Boltuch},
  {Bradbury}, \& et~al.}]{acciari+10a}
---. 2010{\natexlab{c}}, \apjl, 715, L49

\bibitem[{{Aharonian} {et~al.}(2002){Aharonian}, {Akhperjanian}, {Barrio},
  {Beilicke}, {Bernl{\"o}hr}, {B{\"o}rst}, {Bojahr}, {Bolz}, {Contreras},
  {Cornils}, {Cortina}, {Denninghoff}, {Fonseca}, {Girma}, {Gonzalez},
  {G{\"o}tting}, {Heinzelmann}, {Hermann}, {Heusler}, {Hofmann}, {Horns},
  {Jung}, {Kankanyan}, {Kestel}, {Kettler}, {Kohnle}, {Konopelko}, {Kornmeyer},
  {Kranich}, {Krawczynski}, {Lampeitl}, {Lopez}, {Lorenz}, {Lucarelli},
  {Magnussen}, {Mang}, {Meyer}, {Mirzoyan}, {Moralejo}, {Ona}, {Padilla},
  {Panter}, {Plaga}, {Plyasheshnikov}, {P{\"u}hlhofer}, {Rauterberg},
  {R{\"o}hring}, {Rhode}, {Robrade}, {Rowell}, {Sahakian}, {Samorski},
  {Schilling}, {Schr{\"o}der}, {Sevilla}, {Siems}, {Stamm}, {Tluczykont},
  {V{\"o}lk}, {Wiedner}, \& {Wittek}}]{aharonian+02a}
{Aharonian}, F., {et~al.} 2002, \aap, 384, L23

\bibitem[{{Aharonian} {et~al.}(2003){Aharonian}, {Akhperjanian}, {Beilicke},
  {Bernl{\"o}hr}, {B{\"o}rst}, {Bojahr}, {Bolz}, {Coarasa}, {Contreras},
  {Cortina}, {Denninghoff}, {Fonseca}, {Girma}, {G{\"o}tting}, {Heinzelmann},
  {Hermann}, {Heusler}, {Hofmann}, {Horns}, {Jung}, {Kankanyan}, {Kestel},
  {Kohnle}, {Konopelko}, {Kornmeyer}, {Kranich}, {Lampeitl}, {Lopez}, {Lorenz},
  {Lucarelli}, {Mang}, {Meyer}, {Mirzoyan}, {Moralejo}, {Ona-Wilhelmi},
  {Panter}, {Plyasheshnikov}, {P{\"u}hlhofer}, {de los Reyes}, {Rhode},
  {Ripken}, {Robrade}, {Rowell}, {Sahakian}, {Samorski}, {Schilling}, {Siems},
  {Sobzynska}, {Stamm}, {Tluczykont}, {Vitale}, {V{\"o}lk}, {Wiedner}, \&
  {Wittek}}]{aharonian+03a}
---. 2003, \aap, 406, L9

\bibitem[{{Aharonian} {et~al.}(2005){Aharonian}, {Akhperjanian}, {Aye},
  {Bazer-Bachi}, {Beilicke}, {Benbow}, {Berge}, {Berghaus}, {Bernl{\"o}hr},
  {Boisson}, {Bolz}, {Braun}, {Breitling}, {Brown}, {Bussons Gordo},
  {Chadwick}, {Chounet}, {Cornils}, {Costamante}, {Degrange},
  {Djannati-Ata{\"i}}, {O'C.~Drury}, {Dubus}, {Emmanoulopoulos}, {Espigat},
  {Feinstein}, {Fleury}, {Fontaine}, {Fuchs}, {Funk}, {Gallant}, {Giebels},
  {Gillessen}, {Glicenstein}, {Goret}, {Hadjichristidis}, {Hauser},
  {Heinzelmann}, {Henri}, {Hermann}, {Hinton}, {Hofmann}, {Holleran}, {Horns},
  {de Jager}, {Kh{\'e}lifi}, {Komin}, {Konopelko}, {Latham}, {Le Gallou},
  {Lemi{\`e}re}, {Lemoine-Goumard}, {Leroy}, {Lohse}, {Martineau-Huynh},
  {Marcowith}, {Masterson}, {McComb}, {de Naurois}, {Nolan}, {Noutsos},
  {Orford}, {Osborne}, {Ouchrif}, {Panter}, {Pelletier}, {Pita},
  {P{\"u}hlhofer}, {Punch}, {Raubenheimer}, {Raue}, {Raux}, {Rayner},
  {Redondo}, {Reimer}, {Reimer}, {Ripken}, {Rob}, {Rolland}, {Rowell},
  {Sahakian}, {Saug{\'e}}, {Schlenker}, {Schlickeiser}, {Schuster}, {Schwanke},
  {Siewert}, {Sol}, {Steenkamp}, {Stegmann}, {Tavernet}, {Terrier},
  {Th{\'e}oret}, {Tluczykont}, {Vasileiadis}, {Venter}, {Vincent}, {V{\"o}lk},
  \& {Wagner}}]{aharonian+05a}
---. 2005, \aap, 436, L17

\bibitem[{{Aharonian} {et~al.}(2006){Aharonian}, {Akhperjanian}, {Bazer-Bachi},
  {Beilicke}, {Benbow}, {Berge}, {Bernl{\"o}hr}, {Boisson}, {Bolz}, {Borrel},
  {Braun}, {Breitling}, {Brown}, {B{\"u}hler}, {B{\"u}sching}, {Carrigan},
  {Chadwick}, {Chounet}, {Cornils}, {Costamante}, {Degrange}, {Dickinson},
  {Djannati-Ata{\"i}}, {O'C.~Drury}, {Dubus}, {Egberts}, {Emmanoulopoulos},
  {Espigat}, {Feinstein}, {Ferrero}, {Fontaine}, {Funk}, {Funk}, {Gallant},
  {Giebels}, {Glicenstein}, {Goret}, {Hadjichristidis}, {Hauser}, {Hauser},
  {Heinzelmann}, {Henri}, {Hermann}, {Hinton}, {Hofmann}, {Holleran}, {Horns},
  {Jacholkowska}, {de Jager}, {Kh{\'e}lifi}, {Komin}, {Konopelko}, {Latham},
  {Le Gallou}, {Lemi{\`e}re}, {Lemoine-Goumard}, {Lohse}, {Martin},
  {Martineau-Huynh}, {Marcowith}, {Masterson}, {McComb}, {de Naurois},
  {Nedbal}, {Nolan}, {Noutsos}, {Orford}, {Osborne}, {Ouchrif}, {Panter},
  {Pelletier}, {Pita}, {P{\"u}hlhofer}, {Punch}, {Raubenheimer}, {Raue},
  {Rayner}, {Reimer}, {Reimer}, {Ripken}, {Rob}, {Rolland}, {Rowell},
  {Sahakian}, {Saug{\'e}}, {Schlenker}, {Schlickeiser}, {Schwanke}, {Sol},
  {Spangler}, {Spanier}, {Steenkamp}, {Stegmann}, {Superina}, {Tavernet},
  {Terrier}, {Th{\'e}oret}, {Tluczykont}, {van Eldik}, {Vasileiadis}, {Venter},
  {Vincent}, {V{\"o}lk}, {Wagner}, \& {Ward}}]{aharonian+06a}
---. 2006, \aap, 455, 461

\bibitem[{{Aharonian} {et~al.}(2007{\natexlab{a}}){Aharonian}, {Akhperjanian},
  {Bazer-Bachi}, {Beilicke}, {Benbow}, {Berge}, {Bernl{\"o}hr}, {Boisson},
  {Bolz}, {Borrel}, {Braun}, {Brion}, {Brown}, {B{\"u}hler}, {B{\"u}sching},
  {Boutelier}, {Carrigan}, {Chadwick}, {Chounet}, {Coignet}, {Cornils},
  {Costamante}, {Degrange}, {Dickinson}, {Djannati-Ata{\"i}}, {O'C.~Drury},
  {Dubus}, {Egberts}, {Emmanoulopoulos}, {Espigat}, {Farnier}, {Feinstein},
  {Ferrero}, {Fiasson}, {Fontaine}, {Funk}, {Funk}, {F{\"u}{\ss}ling},
  {Gallant}, {Giebels}, {Glicenstein}, {Gl{\"u}ck}, {Goret}, {Hadjichristidis},
  {Hauser}, {Hauser}, {Heinzelmann}, {Henri}, {Hermann}, {Hinton}, {Hoffmann},
  {Hofmann}, {Holleran}, {Hoppe}, {Horns}, {Jacholkowska}, {de Jager},
  {Kendziorra}, {Kerschhaggl}, {Kh{\'e}lifi}, {Komin}, {Kosack}, {Lamanna},
  {Latham}, {Le Gallou}, {Lemi{\`e}re}, {Lemoine-Goumard}, {Lohse}, {Martin},
  {Martineau-Huynh}, {Marcowith}, {Masterson}, {Maurin}, {McComb}, {Moulin},
  {de Naurois}, {Nedbal}, {Nolan}, {Noutsos}, {Olive}, {Orford}, {Osborne},
  {Panter}, {Pelletier}, {Petrucci}, {Pita}, {P{\"u}hlhofer}, {Punch},
  {Ranchon}, {Raubenheimer}, {Raue}, {Rayner}, {Ripken}, {Rob}, {Rolland},
  {Rosier-Lees}, {Rowell}, {Sahakian}, {Santangelo}, {Saug{\'e}}, {Schlenker},
  {Schlickeiser}, {Schr{\"o}der}, {Schwanke}, {Schwarzburg}, {Schwemmer},
  {Shalchi}, {Sol}, {Spangler}, {Spanier}, {Steenkamp}, {Stegmann}, {Superina},
  {Tam}, {Tavernet}, {Terrier}, {Tluczykont}, {van Eldik}, {Vasileiadis},
  {Venter}, {Vialle}, {Vincent}, {V{\"o}lk}, {Wagner}, \&
  {Ward}}]{aharonian+07c}
---. 2007{\natexlab{a}}, \aap, 470, 475

\bibitem[{{Aharonian} {et~al.}(2007{\natexlab{b}}){Aharonian}, {Akhperjanian},
  {Barres de Almeida}, {Bazer-Bachi}, {Behera}, {Beilicke}, {Benbow},
  {Bernl{\"o}hr}, {Boisson}, {Bolz}, {Borrel}, {Braun}, {Brion}, {Brown},
  {B{\"u}hler}, {Bulik}, {B{\"u}sching}, {Boutelier}, {Carrigan}, {Chadwick},
  {Chounet}, {Clapson}, {Coignet}, {Cornils}, {Costamante}, {Dalton},
  {Degrange}, {Dickinson}, {Djannati-Ata{\"i}}, {Domainko}, {O'C.~Drury},
  {Dubois}, {Dubus}, {Dyks}, {Egberts}, {Emmanoulopoulos}, {Espigat},
  {Farnier}, {Feinstein}, {Fiasson}, {F{\"o}rster}, {Fontaine}, {Funk},
  {F{\"u}{\ss}ling}, {Gallant}, {Giebels}, {Glicenstein}, {Gl{\"u}ck}, {Goret},
  {Hadjichristidis}, {Hauser}, {Hauser}, {Heinzelmann}, {Henri}, {Hermann},
  {Hinton}, {Hoffmann}, {Hofmann}, {Holleran}, {Hoppe}, {Horns},
  {Jacholkowska}, {de Jager}, {Jung}, {Katarzy{\'n}ski}, {Kendziorra},
  {Kerschhaggl}, {Kh{\'e}lifi}, {Keogh}, {Komin}, {Kosack}, {Lamanna},
  {Latham}, {Lemi{\`e}re}, {Lemoine-Goumard}, {Lenain}, {Lohse}, {Martin},
  {Martineau-Huynh}, {Marcowith}, {Masterson}, {Maurin}, {Maurin}, {McComb},
  {Moderski}, {Moulin}, {de Naurois}, {Nedbal}, {Nolan}, {Ohm}, {Olive}, {de
  O{\~n}a Wilhelmi}, {Orford}, {Osborne}, {Ostrowski}, {Panter}, {Pedaletti},
  {Pelletier}, {Petrucci}, {Pita}, {P{\"u}hlhofer}, {Punch}, {Ranchon},
  {Raubenheimer}, {Raue}, {Rayner}, {Renaud}, {Ripken}, {Rob}, {Rolland},
  {Rosier-Lees}, {Rowell}, {Rudak}, {Ruppel}, {Sahakian}, {Santangelo},
  {Schlickeiser}, {Sch{\"o}ck}, {Schr{\"o}der}, {Schwanke}, {Schwarzburg},
  {Schwemmer}, {Shalchi}, {Sol}, {Spangler}, {Stawarz}, {Steenkamp},
  {Stegmann}, {Superina}, {Tam}, {Tavernet}, {Terrier}, {van Eldik},
  {Vasileiadis}, {Venter}, {Vialle}, {Vincent}, {Vivier}, {V{\"o}lk}, {Volpe},
  {Wagner}, {Ward}, {Zdziarski}, \& {Zech}}]{aharonian+07b}
---. 2007{\natexlab{b}}, \aap, 473, L25

\bibitem[{{Aharonian} {et~al.}(2007{\natexlab{c}}){Aharonian}, {Akhperjanian},
  {Barres de Almeida}, {Bazer-Bachi}, {Behera}, {Beilicke}, {Benbow},
  {Bernl{\"o}hr}, {Boisson}, {Bolz}, {Borrel}, {Braun}, {Brion}, {Brown},
  {B{\"u}hler}, {Bulik}, {B{\"u}sching}, {Boutelier}, {Carrigan}, {Chadwick},
  {Chounet}, {Clapson}, {Coignet}, {Cornils}, {Costamante}, {Dalton},
  {Degrange}, {Dickinson}, {Djannati-Ata{\"i}}, {Domainko}, {O'C.~Drury},
  {Dubois}, {Dubus}, {Dyks}, {Egberts}, {Emmanoulopoulos}, {Espigat},
  {Farnier}, {Feinstein}, {Fiasson}, {F{\"o}rster}, {Fontaine}, {Funk},
  {F{\"u}{\ss}ling}, {Gallant}, {Giebels}, {Glicenstein}, {Gl{\"u}ck}, {Goret},
  {Hadjichristidis}, {Hauser}, {Hauser}, {Heinzelmann}, {Henri}, {Hermann},
  {Hinton}, {Hoffmann}, {Hofmann}, {Holleran}, {Hoppe}, {Horns},
  {Jacholkowska}, {de Jager}, {Jung}, {Katarzy{\'n}ski}, {Kendziorra},
  {Kerschhaggl}, {Kh{\'e}lifi}, {Keogh}, {Komin}, {Kosack}, {Lamanna},
  {Latham}, {Lemi{\`e}re}, {Lemoine-Goumard}, {Lenain}, {Lohse}, {Martin},
  {Martineau-Huynh}, {Marcowith}, {Masterson}, {Maurin}, {Maurin}, {McComb},
  {Moderski}, {Moulin}, {de Naurois}, {Nedbal}, {Nolan}, {Ohm}, {Olive}, {de
  O{\~n}a Wilhelmi}, {Orford}, {Osborne}, {Ostrowski}, {Panter}, {Pedaletti},
  {Pelletier}, {Petrucci}, {Pita}, {P{\"u}hlhofer}, {Punch}, {Ranchon},
  {Raubenheimer}, {Raue}, {Rayner}, {Renaud}, {Ripken}, {Rob}, {Rolland},
  {Rosier-Lees}, {Rowell}, {Rudak}, {Ruppel}, {Sahakian}, {Santangelo},
  {Schlickeiser}, {Sch{\"o}ck}, {Schr{\"o}der}, {Schwanke}, {Schwarzburg},
  {Schwemmer}, {Shalchi}, {Sol}, {Spangler}, {Stawarz}, {Steenkamp},
  {Stegmann}, {Superina}, {Tam}, {Tavernet}, {Terrier}, {van Eldik},
  {Vasileiadis}, {Venter}, {Vialle}, {Vincent}, {Vivier}, {V{\"o}lk}, {Volpe},
  {Wagner}, {Ward}, {Zdziarski}, \& {Zech}}]{aharonian+07a}
---. 2007{\natexlab{c}}, \aap, 475, L9

\bibitem[{{Aharonian} {et~al.}(2008{\natexlab{a}}){Aharonian}, {Akhperjanian},
  {Barres de Almeida}, {Bazer-Bachi}, {Behera}, {Beilicke}, {Benbow},
  {Bernl{\"o}hr}, {Boisson}, {Borrel}, {Braun}, {Brion}, {Brucker},
  {B{\"u}hler}, {Bulik}, {B{\"u}sching}, {Boutelier}, {Carrigan}, {Chadwick},
  {Chaves}, {Chounet}, {Clapson}, {Coignet}, {Cornils}, {Costamante}, {Dalton},
  {Degrange}, {Dickinson}, {Djannati-Ata{\"i}}, {Domainko}, {O'C.~Drury},
  {Dubois}, {Dubus}, {Dyks}, {Egberts}, {Emmanoulopoulos}, {Espigat},
  {Farnier}, {Feinstein}, {Fiasson}, {F{\"o}rster}, {Fontaine},
  {F{\"u}{\ss}ling}, {Gabici}, {Gallant}, {Giebels}, {Glicenstein},
  {Gl{\"u}ck}, {Goret}, {Hadjichristidis}, {Hauser}, {Hauser}, {Heinzelmann},
  {Henri}, {Hermann}, {Hinton}, {Hoffmann}, {Hofmann}, {Holleran}, {Hoppe},
  {Horns}, {Jacholkowska}, {de Jager}, {Jung}, {Katarzy{\'n}ski}, {Kaufmann},
  {Kendziorra}, {Kerschhaggl}, {Khangulyan}, {Kh{\'e}lifi}, {Keogh}, {Komin},
  {Kosack}, {Lamanna}, {Latham}, {Lenain}, {Lohse}, {Martin},
  {Martineau-Huynh}, {Marcowith}, {Masterson}, {Maurin}, {McComb}, {Moderski},
  {Moulin}, {Naumann-Godo}, {de Naurois}, {Nedbal}, {Nekrassov}, {Nolan},
  {Ohm}, {Olive}, {de O{\~n}a Wilhelmi}, {Orford}, {Osborne}, {Ostrowski},
  {Panter}, {Pedaletti}, {Pelletier}, {Petrucci}, {Pita}, {P{\"u}hlhofer},
  {Punch}, {Quirrenbach}, {Raubenheimer}, {Raue}, {Rayner}, {Renaud}, {Rieger},
  {Ripken}, {Rob}, {Rosier-Lees}, {Rowell}, {Rudak}, {Ruppel}, {Sahakian},
  {Santangelo}, {Schlickeiser}, {Sch{\"o}ck}, {Schr{\"o}der}, {Schwanke},
  {Schwarzburg}, {Schwemmer}, {Shalchi}, {Sol}, {Spangler}, {Stawarz},
  {Steenkamp}, {Stegmann}, {Superina}, {Tam}, {Tavernet}, {Terrier}, {van
  Eldik}, {Vasileiadis}, {Venter}, {Vialle}, {Vincent}, {Vivier}, {V{\"o}lk},
  {Volpe}, {Wagner}, {Ward}, {Zdziarski}, \& {Zech}}]{aharonian+08a}
---. 2008{\natexlab{a}}, \aap, 481, L103

\bibitem[{{Aharonian} {et~al.}(2008{\natexlab{b}}){Aharonian}, {Akhperjanian},
  {Barres de Almeida}, {Bazer-Bachi}, {Behera}, {Beilicke}, {Benbow},
  {Bernl{\"o}hr}, {Boisson}, {Bolz}, {Borrel}, {Braun}, {Brion}, {Brown},
  {B{\"u}hler}, {Bulik}, {B{\"u}sching}, {Boutelier}, {Carrigan}, {Chadwick},
  {Chounet}, {Clapson}, {Coignet}, {Cornils}, {Costamante}, {Dalton},
  {Degrange}, {Dickinson}, {Djannati-Ata{\"i}}, {Domainko}, {O'C.~Drury},
  {Dubois}, {Dubus}, {Dyks}, {Egberts}, {Emmanoulopoulos}, {Espigat},
  {Farnier}, {Feinstein}, {Fiasson}, {F{\"o}rster}, {Fontaine}, {Funk},
  {F{\"u}{\ss}ling}, {Gallant}, {Giebels}, {Glicenstein}, {Gl{\"u}ck}, {Goret},
  {Hadjichristidis}, {Hauser}, {Hauser}, {Heinzelmann}, {Henri}, {Hermann},
  {Hinton}, {Hoffmann}, {Hofmann}, {Holleran}, {Hoppe}, {Horns},
  {Jacholkowska}, {de Jager}, {Jung}, {Katarzy{\'n}ski}, {Kendziorra},
  {Kerschhaggl}, {Kh{\'e}lifi}, {Keogh}, {Komin}, {Kosack}, {Lamanna},
  {Latham}, {Lemi{\`e}re}, {Lemoine-Goumard}, {Lenain}, {Lohse}, {Martin},
  {Martineau-Huynh}, {Marcowith}, {Masterson}, {Maurin}, {Maurin}, {McComb},
  {Moderski}, {Moulin}, {de Naurois}, {Nedbal}, {Nolan}, {Ohm}, {Olive}, {de
  O{\~n}a Wilhelmi}, {Orford}, {Osborne}, {Ostrowski}, {Panter}, {Pedaletti},
  {Pelletier}, {Petrucci}, {Pita}, {P{\"u}hlhofer}, {Punch}, {Ranchon},
  {Raubenheimer}, {Raue}, {Rayner}, {Renaud}, {Ripken}, {Rob}, {Rolland},
  {Rosier-Lees}, {Rowell}, {Rudak}, {Ruppel}, {Sahakian}, {Santangelo},
  {Schlickeiser}, {Sch{\"o}ck}, {Schr{\"o}der}, {Schwanke}, {Schwarzburg},
  {Schwemmer}, {Shalchi}, {Sol}, {Spangler}, {Stawarz}, {Steenkamp},
  {Stegmann}, {Superina}, {Tam}, {Tavernet}, {Terrier}, {van Eldik},
  {Vasileiadis}, {Venter}, {Vialle}, {Vincent}, {Vivier}, {V{\"o}lk}, {Volpe},
  {Wagner}, {Ward}, {Zdziarski}, \& {Zech}}]{aharonian+08b}
---. 2008{\natexlab{b}}, \aap, 477, 481

\bibitem[{{Aharonian} {et~al.}(2010){Aharonian}, {Akhperjanian}, {Anton},
  {Barres de Almeida}, {Bazer-Bachi}, {Becherini}, {Behera}, {Benbow},
  {Bernl{\"o}hr}, {Bochow}, {Boisson}, {Bolmont}, {Borrel}, {Brucker}, {Brun},
  {Brun}, {B{\"u}hler}, {Bulik}, {B{\"u}sching}, {Boutelier}, {Chadwick},
  {Charbonnier}, {Chaves}, {Cheesebrough}, {Chounet}, {Clapson}, {Coignet},
  {Dalton}, {Daniel}, {Davids}, {Degrange}, {Deil}, {Dickinson},
  {Djannati-Ata{\"i}}, {Domainko}, {O'C.~Drury}, {Dubois}, {Dubus}, {Dyks},
  {Dyrda}, {Egberts}, {Emmanoulopoulos}, {Espigat}, {Farnier}, {Feinstein},
  {Fiasson}, {F{\"o}rster}, {Fontaine}, {F{\"u}{\ss}ling}, {Gabici}, {Gallant},
  {G{\'e}rard}, {Gerbig}, {Giebels}, {Glicenstein}, {Gl{\"u}ck}, {Goret},
  {G{\"o}ring}, {Hauser}, {Hauser}, {Heinz}, {Heinzelmann}, {Henri}, {Hermann},
  {Hinton}, {Hoffmann}, {Hofmann}, {Holleran}, {Hoppe}, {Horns},
  {Jacholkowska}, {de Jager}, {Jahn}, {Jung}, {Katarzy{\'n}ski}, {Katz},
  {Kaufmann}, {Kendziorra}, {Kerschhaggl}, {Khangulyan}, {Kh{\'e}lifi},
  {Keogh}, {Klu{\'z}niak}, {Kneiske}, {Komin}, {Kosack}, {Lamanna}, {Lenain},
  {Lohse}, {Marandon}, {Martin}, {Martineau-Huynh}, {Marcowith}, {Masbou},
  {Maurin}, {McComb}, {Medina}, {Moderski}, {Moulin}, {Naumann-Godo}, {de
  Naurois}, {Nedbal}, {Nekrassov}, {Nicholas}, {Niemiec}, {Nolan}, {Ohm},
  {Olive}, {de O{\~n}a Wilhelmi}, {Orford}, {Ostrowski}, {Panter}, {Paz
  Arribas}, {Pedaletti}, {Pelletier}, {Petrucci}, {Pita}, {P{\"u}hlhofer},
  {Punch}, {Quirrenbach}, {Raubenheimer}, {Raue}, {Rayner}, {Renaud}, {Rieger},
  {Ripken}, {Rob}, {Rosier-Lees}, {Rowell}, {Rudak}, {Rulten}, {Ruppel},
  {Sahakian}, {Santangelo}, {Schlickeiser}, {Sch{\"o}ck}, {Schr{\"o}der},
  {Schwanke}, {Schwarzburg}, {Schwemmer}, {Shalchi}, {Sikora}, {Skilton},
  {Sol}, {Spangler}, {Stawarz}, {Steenkamp}, {Stegmann}, {Stinzing},
  {Superina}, {Szostek}, {Tam}, {Tavernet}, {Terrier}, {Tibolla}, {Tluczykont},
  {van Eldik}, {Vasileiadis}, {Venter}, {Venter}, {Vialle}, {Vincent},
  {Vivier}, {V{\"o}lk}, {Volpe}, {Wagner}, {Ward}, {Zdziarski}, \&
  {Zech}}]{aharonian+10a}
---. 2010, \aap, 521, 69

\bibitem[{{Albert} {et~al.}(2006){Albert}, {Aliu}, {Anderhub}, {Antoranz},
  {Armada}, {Asensio}, {Baixeras}, {Barrio}, {Bartko}, {Bastieri}, {Becker},
  {Bednarek}, {Berger}, {Bigongiari}, {Biland}, {Bisesi}, {Bock}, {Bordas},
  {Bosch-Ramon}, {Bretz}, {Britvitch}, {Camara}, {Carmona}, {Chilingarian},
  {Ciprini}, {Coarasa}, {Commichau}, {Contreras}, {Cortina}, {Curtef},
  {Danielyan}, {Dazzi}, {De Angelis}, {de los Reyes}, {De Lotto},
  {Domingo-Santamar{\'{\i}}a}, {Dorner}, {Doro}, {Errando}, {Fagiolini},
  {Ferenc}, {Fern{\'a}ndez}, {Firpo}, {Flix}, {Fonseca}, {Font}, {Fuchs},
  {Galante}, {Garczarczyk}, {Gaug}, {Giller}, {Goebel}, {Hakobyan},
  {Hayashida}, {Hengstebeck}, {H{\"o}hne}, {Hose}, {Hsu}, {Jacon}, {Kalekin},
  {Kosyra}, {Kranich}, {Laatiaoui}, {Laille}, {Lenisa}, {Liebing}, {Lindfors},
  {Lombardi}, {Longo}, {L{\'o}pez}, {L{\'o}pez}, {Lorenz}, {Majumdar},
  {Maneva}, {Mannheim}, {Mansutti}, {Mariotti}, {Mart{\'{\i}}nez}, {Mazin},
  {Merck}, {Meucci}, {Meyer}, {Miranda}, {Mirzoyan}, {Mizobuchi}, {Moralejo},
  {Nilsson}, {Ninkovic}, {O{\~n}a-Wilhelmi}, {Ordu{\~n}a}, {Otte}, {Oya},
  {Paneque}, {Paoletti}, {Paredes}, {Pasanen}, {Pascoli}, {Pauss}, {Pegna},
  {Persic}, {Peruzzo}, {Piccioli}, {Poller}, {Prandini}, {Raymers}, {Rhode},
  {Rib{\'o}}, {Rico}, {Riegel}, {Rissi}, {Robert}, {R{\"u}gamer}, {Saggion},
  {S{\'a}nchez}, {Sartori}, {Scalzotto}, {Scapin}, {Schmitt}, {Schweizer},
  {Shayduk}, {Shinozaki}, {Shore}, {Sidro}, {Sillanp{\"a}{\"a}}, {Sobczynska},
  {Stamerra}, {Stark}, {Takalo}, {Temnikov}, {Tescaro}, {Teshima}, {Tonello},
  {Torres}, {Torres}, {Turini}, {Vankov}, {Vitale}, {Wagner}, {Wibig},
  {Wittek}, {Zanin}, \& {Zapatero}}]{albert+06a}
{Albert}, J., {et~al.} 2006, \apjl, 648, L105

\bibitem[{{Albert} {et~al.}(2007{\natexlab{a}}){Albert}, {Aliu}, {Anderhub},
  {Antoranz}, {Armada}, {Baixeras}, {Barrio}, {Bartko}, {Bastieri}, {Becker},
  {Bednarek}, {Berger}, {Bigongiari}, {Biland}, {Bock}, {Bordas},
  {Bosch-Ramon}, {Bretz}, {Britvitch}, {Camara}, {Carmona}, {Chilingarian},
  {Coarasa}, {Commichau}, {Contreras}, {Cortina}, {Costado}, {Curtef},
  {Danielyan}, {Dazzi}, {De Angelis}, {Delgado}, {de los Reyes}, {De Lotto},
  {Domingo-Santamar{\'{\i}}a}, {Dorner}, {Doro}, {Errando}, {Fagiolini},
  {Ferenc}, {Fern{\'a}ndez}, {Firpo}, {Flix}, {Fonseca}, {Font}, {Fuchs},
  {Galante}, {Garc{\'{\i}}a-L{\'o}pez}, {Garczarczyk}, {Gaug}, {Giller},
  {Goebel}, {Hakobyan}, {Hayashida}, {Hengstebeck}, {Herrero}, {H{\"o}hne},
  {Hose}, {Hsu}, {Jacon}, {Jogler}, {Kosyra}, {Kranich}, {Kritzer}, {Laille},
  {Lindfors}, {Lombardi}, {Longo}, {L{\'o}pez}, {L{\'o}pez}, {Lorenz},
  {Majumdar}, {Maneva}, {Mannheim}, {Mansutti}, {Mariotti}, {Mart{\'{\i}}nez},
  {Mazin}, {Merck}, {Meucci}, {Meyer}, {Miranda}, {Mirzoyan}, {Mizobuchi},
  {Moralejo}, {Nilsson}, {Ninkovic}, {O{\~n}a-Wilhelmi}, {Otte}, {Oya},
  {Paneque}, {Panniello}, {Paoletti}, {Paredes}, {Pasanen}, {Pascoli}, {Pauss},
  {Pegna}, {Persic}, {Peruzzo}, {Piccioli}, {Poller}, {Prandini}, {Puchades},
  {Raymers}, {Rhode}, {Rib{\'o}}, {Rico}, {Rissi}, {Robert}, {R{\"u}gamer},
  {Saggion}, {S{\'a}nchez}, {Sartori}, {Scalzotto}, {Scapin}, {Schmitt},
  {Schweizer}, {Shayduk}, {Shinozaki}, {Shore}, {Sidro}, {Sillanp{\"a}{\"a}},
  {Sobczynska}, {Stamerra}, {Stark}, {Takalo}, {Temnikov}, {Tescaro},
  {Teshima}, {Tonello}, {Torres}, {Turini}, {Vankov}, {Vitale}, {Wagner},
  {Wibig}, {Wittek}, {Zandanel}, {Zanin}, \& {Zapatero}}]{albert+07b}
---. 2007{\natexlab{a}}, \apjl, 666, L17

\bibitem[{{Albert} {et~al.}(2007{\natexlab{b}}){Albert}, {Aliu}, {Anderhub},
  {Antoranz}, {Armada}, {Baixeras}, {Barrio}, {Bartko}, {Bastieri}, {Becker},
  {Bednarek}, {Berger}, {Bigongiari}, {Biland}, {Bock}, {Bordas},
  {Bosch-Ramon}, {Bretz}, {Britvitch}, {Camara}, {Carmona}, {Chilingarian},
  {Coarasa}, {Commichau}, {Contreras}, {Cortina}, {Costado}, {Curtef},
  {Danielyan}, {Dazzi}, {De Angelis}, {Delgado}, {de los Reyes}, {De Lotto},
  {Domingo-Santamar{\'{\i}}a}, {Dorner}, {Doro}, {Errando}, {Fagiolini},
  {Ferenc}, {Fern{\'a}ndez}, {Firpo}, {Flix}, {Fonseca}, {Font}, {Fuchs},
  {Galante}, {Garc{\'{\i}}a-L{\'o}pez}, {Garczarczyk}, {Gaug}, {Giller},
  {Goebel}, {Hakobyan}, {Hayashida}, {Hengstebeck}, {Herrero}, {H{\"o}hne},
  {Hose}, {Hsu}, {Jacon}, {Jogler}, {Kosyra}, {Kranich}, {Kritzer}, {Laille},
  {Lindfors}, {Lombardi}, {Longo}, {L{\'o}pez}, {L{\'o}pez}, {Lorenz},
  {Majumdar}, {Maneva}, {Mannheim}, {Mansutti}, {Mariotti}, {Mart{\'{\i}}nez},
  {Mazin}, {Merck}, {Meucci}, {Meyer}, {Miranda}, {Mirzoyan}, {Mizobuchi},
  {Moralejo}, {Nieto}, {Nilsson}, {Ninkovic}, {O{\~n}a-Wilhelmi}, {Otte},
  {Oya}, {Paneque}, {Panniello}, {Paoletti}, {Paredes}, {Pasanen}, {Pascoli},
  {Pauss}, {Pegna}, {Perlman}, {Persic}, {Peruzzo}, {Piccioli}, {Prandini},
  {Puchades}, {Raymers}, {Rhode}, {Rib{\'o}}, {Rico}, {Rissi}, {Robert},
  {R{\"u}gamer}, {Saggion}, {Saito}, {S{\'a}nchez}, {Sartori}, {Scalzotto},
  {Scapin}, {Schmitt}, {Schweizer}, {Shayduk}, {Shinozaki}, {Shore}, {Sidro},
  {Sillanp{\"a}{\"a}}, {Sobczynska}, {Stamerra}, {Stark}, {Takalo},
  {Tavecchio}, {Temnikov}, {Tescaro}, {Teshima}, {Torres}, {Turini}, {Vankov},
  {Vitale}, {Wagner}, {Wibig}, {Wittek}, {Zandanel}, {Zanin}, \&
  {Zapatero}}]{albert+07a}
---. 2007{\natexlab{b}}, \apjl, 667, L21

\bibitem[{{Albert} {et~al.}(2007{\natexlab{c}}){Albert}, {Aliu}, {Anderhub},
  {Antoranz}, {Armada}, {Baixeras}, {Barrio}, {Bartko}, {Bastieri}, {Becker},
  {Bednarek}, {Berger}, {Bigongiari}, {Biland}, {Bock}, {Bordas},
  {Bosch-Ramon}, {Bretz}, {Britvitch}, {Camara}, {Carmona}, {Chilingarian},
  {Ciprini}, {Coarasa}, {Commichau}, {Contreras}, {Cortina}, {Costado},
  {Curtef}, {Danielyan}, {Dazzi}, {De Angelis}, {Delgado}, {de los Reyes}, {De
  Lotto}, {Domingo-Santamar{\'{\i}}a}, {Dorner}, {Doro}, {Errando},
  {Fagiolini}, {De Angelis}, {Ferenc}, {Fern{\'a}ndez}, {Firpo}, {Flix},
  {Fonseca}, {Font}, {Fuchs}, {Galante}, {Garc{\'{\i}}a-L{\'o}pez},
  {Garczarczyk}, {Gaug}, {Giller}, {Goebel}, {Hakobyan}, {Hayashida},
  {Hengstebeck}, {Herrero}, {H{\"o}hne}, {Hose}, {Hsu}, {Jacon}, {Jogler},
  {Kalekin}, {Kosyra}, {Kranich}, {Kritzer}, {Laille}, {Liebing}, {Lindfors},
  {Lombardi}, {Longo}, {L{\'o}pez}, {L{\'o}pez}, {Lorenz}, {Majumdar},
  {Maneva}, {Mannheim}, {Mansutti}, {Mariotti}, {Mart{\'{\i}}nez}, {Mazin},
  {Merck}, {Meucci}, {Meyer}, {Miranda}, {Mirzoyan}, {Mizobuchi}, {Moralejo},
  {Nilsson}, {Ninkovic}, {On{\~n}a-Wilhelmi}, {Otte}, {Oya}, {Paneque},
  {Panniello}, {Paoletti}, {Paredes}, {Pasanen}, {Pascoli}, {Pauss}, {Pegna},
  {Persic}, {Peruzzo}, {Piccioli}, {Poller}, {Prandini}, {Puchades}, {Raymers},
  {Rhode}, {Rib{\'o}}, {Rico}, {Rissi}, {Robert}, {R{\"u}gamer}, {Saggion},
  {S{\'a}nchez}, {Sartori}, {Scalzotto}, {Scapin}, {Schmitt}, {Schweizer},
  {Shayduk}, {Shinozaki}, {Shore}, {Sidro}, {Sillanp{\"a}{\"a}}, {Sobczynska},
  {Stamerra}, {Stark}, {Takalo}, {Temnikov}, {Tescaro}, {Teshima}, {Tonello},
  {Torres}, {Turini}, {Vankov}, {Vitale}, {Wagner}, {Wibig}, {Wittek},
  {Zandanel}, {Zanin}, \& {Zapatero}}]{albert+07c}
---. 2007{\natexlab{c}}, \apj, 662, 892

\bibitem[{{Aleksi{\'c}} {et~al.}(2010){Aleksi{\'c}}, {Antonelli}, {Antoranz},
  {Backes}, {Barrio}, {Bastieri}, {Becerra Gonz{\'a}lez}, {Bednarek},
  {Berdyugin}, {Berger}, {Bernardini}, {Biland}, {Blanch}, {Bock}, {Boller},
  {Bonnoli}, {Bordas}, {Borla Tridon}, {Bosch-Ramon}, {Bose}, {Braun}, {Bretz},
  {Camara}, {Ca{\~n}ellas}, {Carmona}, {Carosi}, {Colin}, {Colombo},
  {Contreras}, {Cortina}, {Cossio}, {Covino}, {Dazzi}, {De Angelis}, {De Cea
  del Pozo}, {De Lotto}, {De Maria}, {De Sabata}, {Delgado Mendez}, {Diago
  Ortega}, {Doert}, {Dom{\'{\i}}nguez}, {Dominis Prester}, {Dorner}, {Doro},
  {Elsaesser}, {Errando}, {Ferenc}, {Fonseca}, {Font}, {Garc{\'{\i}}a
  L{\'o}pez}, {Garczarczyk}, {Giavitto}, {Godinovi{\'c}}, {Hadasch}, {Herrero},
  {Hildebrand}, {H{\"o}hne-M{\"o}nch}, {Hose}, {Hrupec}, {Jogler}, {Klepser},
  {Kr{\"a}henb{\"u}hl}, {Kranich}, {Krause}, {La Barbera}, {Leonardo},
  {Lindfors}, {Lombardi}, {Longo}, {L{\'o}pez}, {Lorenz}, {Majumdar},
  {Makariev}, {Maneva}, {Mankuzhiyil}, {Mannheim}, {Maraschi}, {Mariotti},
  {Mart{\'{\i}}nez}, {Mazin}, {Meucci}, {Miranda}, {Mirzoyan}, {Miyamoto},
  {Mold{\'o}n}, {Moralejo}, {Nieto}, {Nilsson}, {Orito}, {Oya}, {Paoletti},
  {Paredes}, {Partini}, {Pasanen}, {Pauss}, {Pegna}, {Perez-Torres}, {Persic},
  {Peruzzo}, {Pochon}, {Prada}, {Prada Moroni}, {Prandini}, {Puchades},
  {Puljak}, {Reichardt}, {Reinthal}, {Rhode}, {Rib{\'o}}, {Rico},
  {R{\"u}gamer}, {Saggion}, {Saito}, {Saito}, {Salvati}, {S{\'a}nchez-Conde},
  {Satalecka}, {Scalzotto}, {Scapin}, {Schultz}, {Schweizer}, {Shayduk},
  {Shore}, {Sierpowska-Bartosik}, {Sillanp{\"a}{\"a}}, {Sitarek}, {Sobczynska},
  {Spanier}, {Spiro}, {Stamerra}, {Steinke}, {Storz}, {Strah}, {Struebig},
  {Suric}, {Takalo}, {Tavecchio}, {Temnikov}, {Terzi{\'c}}, {Tescaro},
  {Teshima}, {Torres}, {Vankov}, {Wagner}, {Weitzel}, {Zabalza}, {Zandanel},
  {Zanin}, {Neronov}, {Pfrommer}, {Pinzke}, {Semikoz}, \& {MAGIC
  Collaboration}}]{alek_etal:10}
{Aleksi{\'c}}, J., {et~al.} 2010, \apjl, 723, L207

\bibitem[{{Aliu} {et~al.}(2009){Aliu}, {Anderhub}, {Antonelli}, {Antoranz},
  {Backes}, {Baixeras}, {Balestra}, {Barrio}, {Bartko}, {Bastieri}, {Becerra
  Gonz{\'a}lez}, {Becker}, {Bednarek}, {Berger}, {Bernardini}, {Biland},
  {Bock}, {Bonnoli}, {Bordas}, {Borla Tridon}, {Bosch-Ramon}, {Bretz},
  {Britvitch}, {Camara}, {Carmona}, {Chilingarian}, {Commichau}, {Contreras},
  {Cortina}, {Costado}, {Covino}, {Curtef}, {Dazzi}, {DeAngelis}, {DeCea del
  Pozo}, {de los Reyes}, {DeLotto}, {DeMaria}, {DeSabata}, {Delgado Mendez},
  {Dominguez}, {Dorner}, {Doro}, {Elsaesser}, {Errando}, {Ferenc},
  {Fern{\'a}ndez}, {Firpo}, {Fonseca}, {Font}, {Galante}, {Garc{\'{\i}}a
  L{\'o}pez}, {Garczarczyk}, {Gaug}, {Goebel}, {Hadasch}, {Hayashida},
  {Herrero}, {H{\"o}hne-M{\"o}nch}, {Hose}, {Hsu}, {Huber}, {Jogler},
  {Kranich}, {La Barbera}, {Laille}, {Leonardo}, {Lindfors}, {Lombardi},
  {Longo}, {L{\'o}pez}, {Lorenz}, {Majumdar}, {Maneva}, {Mankuzhiyil},
  {Mannheim}, {Maraschi}, {Mariotti}, {Mart{\'{\i}}nez}, {Mazin}, {Meucci},
  {Meyer}, {Miranda}, {Mirzoyan}, {Mold{\'o}n}, {Moles}, {Moralejo}, {Nieto},
  {Nilsson}, {Ninkovic}, {Otte}, {Oya}, {Paoletti}, {Paredes}, {Pasanen},
  {Pascoli}, {Pauss}, {Pegna}, {Perez-Torres}, {Persic}, {Peruzzo}, {Prada},
  {Prandini}, {Puchades}, {Raymers}, {Rhode}, {Rib{\'o}}, {Rico}, {Rissi},
  {Robert}, {R{\"u}gamer}, {Saggion}, {Saito}, {Salvati}, {Sanchez-Conde},
  {Sartori}, {Satalecka}, {Scalzotto}, {Scapin}, {Schweizer}, {Shayduk},
  {Shinozaki}, {Shore}, {Sidro}, {Sierpowska-Bartosik}, {Sillanp{\"a}{\"a}},
  {Sitarek}, {Sobczynska}, {Spanier}, {Stamerra}, {Stark}, {Takalo},
  {Tavecchio}, {Temnikov}, {Tescaro}, {Teshima}, {Tluczykont}, {Torres},
  {Turini}, {Vankov}, {Venturini}, {Vitale}, {Wagner}, {Wittek}, {Zabalza},
  {Zandanel}, {Zanin}, \& {Zapatero}}]{aliu+09}
{Aliu}, E., {et~al.} 2009, \apjl, 692, L29

\bibitem[{{Ando} \& {Kusenko}(2010)}]{Ando+10}
{Ando}, S., \& {Kusenko}, A. 2010, \apjl, 722, L39

\bibitem[{{Barrow} {et~al.}(1997){Barrow}, {Ferreira}, \& {Silk}}]{Barrow+97}
{Barrow}, J.~D., {Ferreira}, P.~G., \& {Silk}, J. 1997, Physical Review
  Letters, 78, 3610

\bibitem[{{Boyd} \& {Sanderson}(2003)}]{Boyd-Sand:03}
{Boyd}, T.~J.~M., \& {Sanderson}, J.~J. 2003, {The Physics of Plasmas}
  (Cambridge: Cambridge University Press)

\bibitem[{{Bret}(2009)}]{Bret:09}
{Bret}, A. 2009, \apj, 699, 990

\bibitem[{{Bret} {et~al.}(2004){Bret}, {Firpo}, \&
  {Deutsch}}]{Bret-Firp-Deut:04}
{Bret}, A., {Firpo}, M., \& {Deutsch}, C. 2004, \pre, 70, 046401

\bibitem[{{Bret} {et~al.}(2005{\natexlab{a}}){Bret}, {Firpo}, \&
  {Deutsch}}]{Bret-Firp-Deut:05}
---. 2005{\natexlab{a}}, Physical Review Letters, 94, 115002

\bibitem[{{Bret} {et~al.}(2005{\natexlab{b}}){Bret}, {Firpo}, \&
  {Deutsch}}]{Bret-Firp-Deut:05b}
---. 2005{\natexlab{b}}, \pre, 72, 016403

\bibitem[{{Bret} {et~al.}(2010{\natexlab{a}}){Bret}, {Gremillet}, \&
  {B{\'e}nisti}}]{Bret-Grem-Beni:10}
{Bret}, A., {Gremillet}, L., \& {B{\'e}nisti}, D. 2010{\natexlab{a}}, \pre, 81,
  036402

\bibitem[{{Bret} {et~al.}(2010{\natexlab{b}}){Bret}, {Gremillet}, \&
  {Dieckmann}}]{Bret-Grem-Diec:10}
{Bret}, A., {Gremillet}, L., \& {Dieckmann}, M.~E. 2010{\natexlab{b}}, Physics
  of Plasmas, 17, 120501

\bibitem[{{Cavadini} {et~al.}(2011){Cavadini}, {Salvaterra}, \&
  {Haardt}}]{Cavadini+2011}
{Cavadini}, M., {Salvaterra}, R., \& {Haardt}, F. 2011, arXiv:1105.4613

\bibitem[{{Chandra} {et~al.}(2010){Chandra}, {Yadav}, {Rannot}, {Singh},
  {Tickoo}, {Sharma}, {Venugopal}, {Bhat}, {Bhatt}, {Bhattacharyya},
  {Chanchalani}, {Dhar}, {Godambe}, {Goyal}, {Kothari}, {Kotwal}, {Koul},
  {Koul}, \& {Sahaynathan}}]{Chandra+10}
{Chandra}, P., {et~al.} 2010, Journal of Physics G Nuclear Physics, 37, 125201

\bibitem[{{Chang} {et~al.}(2012){Chang}, {Broderick}, \& {Pfrommer}}]{CBP}
{Chang}, P., {Broderick}, A.~E., \& {Pfrommer}, C. 2012, \apj~in print,
  arXiv:1106.5504

\bibitem[{{Chang} {et~al.}(2008){Chang}, {Spitkovsky}, \& {Arons}}]{Chang+08}
{Chang}, P., {Spitkovsky}, A., \& {Arons}, J. 2008, \apj, 674, 378

\bibitem[{{Costamante} {et~al.}(2007){Costamante}, {Aharonian}, \&
  {Khangulyan}}]{Costamante+2007}
{Costamante}, L., {Aharonian}, F., \& {Khangulyan}, D. 2007, in American
  Institute of Physics Conference Series, Vol. 921, The First GLAST Symposium,
  ed. {S.~Ritz, P.~Michelson, \& C.~A.~Meegan}, 157--159

\bibitem[{{CTA Consortium}(2010)}]{CTA:10}
{CTA Consortium}, T. 2010, arXiv:1008.3703

\bibitem[{{Dai} \& {Lu}(2002)}]{Dai-Lu:02}
{Dai}, Z.~G., \& {Lu}, T. 2002, \apj, 580, 1013

\bibitem[{{Daniel} {et~al.}(2005){Daniel}, {Badran}, {Bond}, {Boyle},
  {Bradbury}, {Buckley}, {Carter-Lewis}, {Catanese}, {Celik}, {Cogan}, {Cui},
  {D'Vali}, {de la Calle Perez}, {Duke}, {Falcone}, {Fegan}, {Fegan}, {Finley},
  {Fortson}, {Gaidos}, {Gammell}, {Gibbs}, {Gillanders}, {Grube}, {Hall},
  {Hall}, {Hanna}, {Hillas}, {Holder}, {Horan}, {Humensky}, {Jarvis}, {Jordan},
  {Kenny}, {Kertzman}, {Kieda}, {Kildea}, {Knapp}, {Kosack}, {Krawczynski},
  {Krennrich}, {Lang}, {Le Bohec}, {Linton}, {Lloyd-Evans}, {Milovanovic},
  {Moriarty}, {M{\"u}ller}, {Nagai}, {Nolan}, {Ong}, {Pallassini}, {Petry},
  {Power-Mooney}, {Quinn}, {Quinn}, {Ragan}, {Rebillot}, {Reynolds}, {Rose},
  {Schroedter}, {Sembroski}, {Swordy}, {Syson}, {Vassiliev}, {Wakely},
  {Walker}, {Weekes}, \& {Zweerink}}]{Dani_etal:05}
{Daniel}, M.~K., {et~al.} 2005, \apj, 621, 181

\bibitem[{{Davidson} {et~al.}(1972){Davidson}, {Hammer}, {Haber}, \&
  {Wagner}}]{Davidson+72}
{Davidson}, R.~C., {Hammer}, D.~A., {Haber}, I., \& {Wagner}, C.~E. 1972,
  Physics of Fluids, 15, 317

\bibitem[{{de Angelis} {et~al.}(2008){de Angelis}, {Persic}, \&
  {Roncadelli}}]{de_Angelis+08}
{de Angelis}, A., {Persic}, M., \& {Roncadelli}, M. 2008, Modern Physics
  Letters A, 23, 315

\bibitem[{{Dermer} {et~al.}(2011){Dermer}, {Cavadini}, {Razzaque}, {Finke},
  {Chiang}, \& {Lott}}]{Derm_etal:10}
{Dermer}, C.~D., {Cavadini}, M., {Razzaque}, S., {Finke}, J.~D., {Chiang}, J.,
  \& {Lott}, B. 2011, \apjl, 733, L21

\bibitem[{{Dolag} {et~al.}(2011){Dolag}, {Kachelriess}, {Ostapchenko}, \&
  {Tom{\`a}s}}]{Dola_etal:11}
{Dolag}, K., {Kachelriess}, M., {Ostapchenko}, S., \& {Tom{\`a}s}, R. 2011,
  \apjl, 727, L4

\bibitem[{{Franceschini} {et~al.}(2008){Franceschini}, {Rodighiero}, \&
  {Vaccari}}]{Fran-Rodi-Vacc:08}
{Franceschini}, A., {Rodighiero}, G., \& {Vaccari}, M. 2008, \aap, 487, 837

\bibitem[{{Frederiksen} {et~al.}(2004){Frederiksen}, {Hededal}, {Haugb{\o}lle},
  \& {Nordlund}}]{Frederiksen+04}
{Frederiksen}, J.~T., {Hededal}, C.~B., {Haugb{\o}lle}, T., \& {Nordlund},
  {\AA}. 2004, \apjl, 608, L13

\bibitem[{{Fried}(1959)}]{Frie:59}
{Fried}, B.~D. 1959, Physics of Fluids, 2, 337

\bibitem[{{Ghisellini}(1999)}]{Ghisellini+1999}
{Ghisellini}, G. 1999, Astrophysical Letters Communications, 39, 17

\bibitem[{{Ghisellini}(2011)}]{Ghis:11}
---. 2011, arXiv: 1104.0006

\bibitem[{{Ghisellini} {et~al.}(2009){Ghisellini}, {Maraschi}, \&
  {Tavecchio}}]{Ghisellini+2009}
{Ghisellini}, G., {Maraschi}, L., \& {Tavecchio}, F. 2009, \mnras, 396, L105

\bibitem[{{Ghisellini} \& {Tavecchio}(2008)}]{Ghisellini+2008}
{Ghisellini}, G., \& {Tavecchio}, F. 2008, \mnras, 387, 1669

\bibitem[{{Gould} \& {Schr{\'e}der}(1967)}]{Goul-Schr:67}
{Gould}, R.~J., \& {Schr{\'e}der}, G.~P. 1967, Physical Review, 155, 1408

\bibitem[{{Guetta} \& {Granot}(2003)}]{Guet-Gran:03}
{Guetta}, D., \& {Granot}, J. 2003, \apj, 585, 885

\bibitem[{{HESS Collaboration} {et~al.}(2010){HESS Collaboration},
  {Abramowski}, {Acero}, {Aharonian}, {Akhperjanian}, {Anton}, {Barres de
  Almeida}, {Bazer-Bachi}, {Becherini}, {Benbow}, {Bernl{\"o}hr}, {Bochow},
  {Boisson}, {Bolmont}, {Borrel}, {Brucker}, {Brun}, {Brun}, {B{\"u}hler},
  {Bulik}, {B{\"u}sching}, {Boutelier}, {Chadwick}, {Charbonnier}, {Chaves},
  {Cheesebrough}, {Chounet}, {Clapson}, {Coignet}, {Conrad}, {Costamante},
  {Dalton}, {Daniel}, {Davids}, {Degrange}, {Deil}, {Dickinson},
  {Djannati-Ata{\"i}}, {Domainko}, {O'C.~Drury}, {Dubois}, {Dubus}, {Dyks},
  {Dyrda}, {Egberts}, {Eger}, {Espigat}, {Fallon}, {Farnier}, {Fegan},
  {Feinstein}, {Fernandes}, {Fiasson}, {F{\"o}rster}, {Fontaine},
  {F{\"u}{\ss}ling}, {Gabici}, {Gallant}, {G{\'e}rard}, {Gerbig}, {Giebels},
  {Glicenstein}, {Gl{\"u}ck}, {Goret}, {G{\"o}ring}, {Hampf}, {Hauser},
  {Heinz}, {Heinzelmann}, {Henri}, {Hermann}, {Hinton}, {Hoffmann}, {Hofmann},
  {Hofverberg}, {Holleran}, {Hoppe}, {Horns}, {Jacholkowska}, {de Jager},
  {Jahn}, {Jung}, {Katarzy{\'n}ski}, {Katz}, {Kaufmann}, {Kerschhaggl},
  {Khangulyan}, {Kh{\'e}lifi}, {Keogh}, {Klochkov}, {Klu{\'z}niak}, {Kneiske},
  {Komin}, {Kosack}, {Kossakowski}, {Lamanna}, {Lenain}, {Lohse}, {Lu},
  {Marandon}, {Marcowith}, {Masbou}, {Maurin}, {McComb}, {Medina},
  {M{\'e}hault}, {Moderski}, {Moulin}, {Naumann-Godo}, {de Naurois}, {Nedbal},
  {Nekrassov}, {Nguyen}, {Nicholas}, {Niemiec}, {Nolan}, {Ohm}, {Olive}, {de
  O{\~n}a Wilhelmi}, {Opitz}, {Orford}, {Ostrowski}, {Panter}, {Paz Arribas},
  {Pedaletti}, {Pelletier}, {Petrucci}, {Pita}, {P{\"u}hlhofer}, {Punch},
  {Quirrenbach}, {Raubenheimer}, {Raue}, {Rayner}, {Reimer}, {Renaud}, {de los
  Reyes}, {Rieger}, {Ripken}, {Rob}, {Rosier-Lees}, {Rowell}, {Rudak},
  {Rulten}, {Ruppel}, {Ryde}, {Sahakian}, {Santangelo}, {Schlickeiser},
  {Sch{\"o}ck}, {Sch{\"o}nwald}, {Schwanke}, {Schwarzburg}, {Schwemmer},
  {Shalchi}, {Sushch}, {Sikora}, {Skilton}, {Sol}, {Stawarz}, {Steenkamp},
  {Stegmann}, {Stinzing}, {Superina}, {Szostek}, {Tam}, {Tavernet}, {Terrier},
  {Tibolla}, {Tluczykont}, {Valerius}, {van Eldik}, {Vasileiadis}, {Venter},
  {Venter}, {Vialle}, {Viana}, {Vincent}, {Vivier}, {V{\"o}lk}, {Volpe},
  {Vorobiov}, {Wagner}, {Ward}, {Zdziarski}, {Zech}, \& {Zechlin}}]{hess+10a}
{HESS Collaboration} {et~al.} 2010, \aap, 520, 83

\bibitem[{{Hopkins} {et~al.}(2007){Hopkins}, {Richards}, \&
  {Hernquist}}]{Hopkins+07}
{Hopkins}, P.~F., {Richards}, G.~T., \& {Hernquist}, L. 2007, \apj, 654, 731

\bibitem[{{Huang} {et~al.}(2009){Huang}, {Konopelko}, \& {for the VERITAS
  collaboration}}]{Huang+09}
{Huang}, D., {Konopelko}, A., \& {for the VERITAS collaboration}. 2009, arXiv:
  0912.3772

\bibitem[{{Inoue} \& {Totani}(2009)}]{Inou-Tota:09}
{Inoue}, Y., \& {Totani}, T. 2009, \apj, 702, 523

\bibitem[{{Kandus} {et~al.}(2011){Kandus}, {Kunze}, \& {Tsagas}}]{Kandus+11}
{Kandus}, A., {Kunze}, K.~E., \& {Tsagas}, C.~G. 2011, \physrep, 505, 1

\bibitem[{{Kneiske} {et~al.}(2004){Kneiske}, {Bretz}, {Mannheim}, \&
  {Hartmann}}]{Knei_etal:04}
{Kneiske}, T.~M., {Bretz}, T., {Mannheim}, K., \& {Hartmann}, D.~H. 2004, \aap,
  413, 807

\bibitem[{{Kneiske} \& {Mannheim}(2008)}]{Knei-Mann:08}
{Kneiske}, T.~M., \& {Mannheim}, K. 2008, \aap, 479, 41

\bibitem[{{Komatsu} {et~al.}(2011){Komatsu}, {Smith}, {Dunkley}, {Bennett},
  {Gold}, {Hinshaw}, {Jarosik}, {Larson}, {Nolta}, {Page}, {Spergel},
  {Halpern}, {Hill}, {Kogut}, {Limon}, {Meyer}, {Odegard}, {Tucker}, {Weiland},
  {Wollack}, \& {Wright}}]{WMAP7_2011}
{Komatsu}, E., {et~al.} 2011, \apjs, 192, 18

\bibitem[{{Kotera} \& {Lemoine}(2008)}]{Kotera+08}
{Kotera}, K., \& {Lemoine}, M. 2008, \prd, 77, 123003

\bibitem[{{Lemoine} \& {Pelletier}(2010)}]{Lemo-Pell:10}
{Lemoine}, M., \& {Pelletier}, G. 2010, \mnras, 402, 321

\bibitem[{{Lesch} \& {Schlickeiser}(1987)}]{Lesch+87}
{Lesch}, H., \& {Schlickeiser}, R. 1987, \aap, 179, 93

\bibitem[{{MAGIC Collaboration} {et~al.}(2008{\natexlab{a}}){MAGIC
  Collaboration}, {Albert}, {Aliu}, {Anderhub}, {Antonelli}, {Antoranz},
  {Backes}, {Baixeras}, {Barrio}, {Bartko}, {Bastieri}, {Becker}, {Bednarek},
  {Berger}, {Bernardini}, {Bigongiari}, {Biland}, {Bock}, {Bonnoli}, {Bordas},
  {Bosch-Ramon}, {Bretz}, {Britvitch}, {Camara}, {Carmona}, {Chilingarian},
  {Commichau}, {Contreras}, {Cortina}, {Costado}, {Covino}, {Curtef}, {Dazzi},
  {De Angelis}, {Cea del Pozo}, {de los Reyes}, {De Lotto}, {De Maria}, {De
  Sabata}, {Mendez}, {Dominguez}, {Dorner}, {Doro}, {Errando}, {Fagiolini},
  {Ferenc}, {Fern{\'a}ndez}, {Firpo}, {Fonseca}, {Font}, {Galante},
  {Garc{\'{\i}}a L{\'o}pez}, {Garczarczyk}, {Gaug}, {Goebel}, {Hayashida},
  {Herrero}, {H{\"o}hne}, {Hose}, {Hsu}, {Huber}, {Jogler}, {Kneiske},
  {Kranich}, {La Barbera}, {Laille}, {Leonardo}, {Lindfors}, {Lombardi},
  {Longo}, {L{\'o}pez}, {Lorenz}, {Majumdar}, {Maneva}, {Mankuzhiyil},
  {Mannheim}, {Maraschi}, {Mariotti}, {Mart{\'{\i}}nez}, {Mazin}, {Meucci},
  {Meyer}, {Miranda}, {Mirzoyan}, {Mizobuchi}, {Moles}, {Moralejo}, {Nieto},
  {Nilsson}, {Ninkovic}, {Otte}, {Oya}, {Panniello}, {Paoletti}, {Paredes},
  {Pasanen}, {Pascoli}, {Pauss}, {Pegna}, {Perez-Torres}, {Persic}, {Peruzzo},
  {Piccioli}, {Prada}, {Prandini}, {Puchades}, {Raymers}, {Rhode}, {Rib{\'o}},
  {Rico}, {Rissi}, {Robert}, {R{\"u}gamer}, {Saggion}, {Saito}, {Salvati},
  {Sanchez-Conde}, {Sartori}, {Satalecka}, {Scalzotto}, {Scapin}, {Schmitt},
  {Schweizer}, {Shayduk}, {Shinozaki}, {Shore}, {Sidro}, {Sierpowska-Bartosik},
  {Sillanp{\"a}{\"a}}, {Sobczynska}, {Spanier}, {Stamerra}, {Stark}, {Takalo},
  {Tavecchio}, {Temnikov}, {Tescaro}, {Teshima}, {Tluczykont}, {Torres},
  {Turini}, {Vankov}, {Venturini}, {Vitale}, {Wagner}, {Wittek}, {Zabalza},
  {Zandanel}, {Zanin}, \& {Zapatero}}]{magic+08}
{MAGIC Collaboration} {et~al.} 2008{\natexlab{a}}, Science, 320, 1752

\bibitem[{{MAGIC Collaboration} {et~al.}(2008{\natexlab{b}}){MAGIC
  Collaboration}, {Albert}, {Aliu}, {Anderhub}, {Antonelli}, {Antoranz},
  {Backes}, {Baixeras}, {Barrio}, {Bartko}, {Bastieri}, {Becker}, {Bednarek},
  {Berger}, {Bernardini}, {Bigongiari}, {Biland}, {Bock}, {Bonnoli}, {Bordas},
  {Bosch-Ramon}, {Bretz}, {Britvitch}, {Camara}, {Carmona}, {Chilingarian},
  {Commichau}, {Contreras}, {Cortina}, {Costado}, {Covino}, {Curtef}, {Dazzi},
  {De Angelis}, {Cea del Pozo}, {de los Reyes}, {De Lotto}, {De Maria}, {De
  Sabata}, {Mendez}, {Dominguez}, {Dorner}, {Doro}, {Errando}, {Fagiolini},
  {Ferenc}, {Fern{\'a}ndez}, {Firpo}, {Fonseca}, {Font}, {Galante},
  {Garc{\'{\i}}a L{\'o}pez}, {Garczarczyk}, {Gaug}, {Goebel}, {Hayashida},
  {Herrero}, {H{\"o}hne}, {Hose}, {Hsu}, {Huber}, {Jogler}, {Kneiske},
  {Kranich}, {La Barbera}, {Laille}, {Leonardo}, {Lindfors}, {Lombardi},
  {Longo}, {L{\'o}pez}, {Lorenz}, {Majumdar}, {Maneva}, {Mankuzhiyil},
  {Mannheim}, {Maraschi}, {Mariotti}, {Mart{\'{\i}}nez}, {Mazin}, {Meucci},
  {Meyer}, {Miranda}, {Mirzoyan}, {Mizobuchi}, {Moles}, {Moralejo}, {Nieto},
  {Nilsson}, {Ninkovic}, {Otte}, {Oya}, {Panniello}, {Paoletti}, {Paredes},
  {Pasanen}, {Pascoli}, {Pauss}, {Pegna}, {Perez-Torres}, {Persic}, {Peruzzo},
  {Piccioli}, {Prada}, {Prandini}, {Puchades}, {Raymers}, {Rhode}, {Rib{\'o}},
  {Rico}, {Rissi}, {Robert}, {R{\"u}gamer}, {Saggion}, {Saito}, {Salvati},
  {Sanchez-Conde}, {Sartori}, {Satalecka}, {Scalzotto}, {Scapin}, {Schmitt},
  {Schweizer}, {Shayduk}, {Shinozaki}, {Shore}, {Sidro}, {Sierpowska-Bartosik},
  {Sillanp{\"a}{\"a}}, {Sobczynska}, {Spanier}, {Stamerra}, {Stark}, {Takalo},
  {Tavecchio}, {Temnikov}, {Tescaro}, {Teshima}, {Tluczykont}, {Torres},
  {Turini}, {Vankov}, {Venturini}, {Vitale}, {Wagner}, {Wittek}, {Zabalza},
  {Zandanel}, {Zanin}, \& {Zapatero}}]{MAGIC+2008}
---. 2008{\natexlab{b}}, Science, 320, 1752

\bibitem[{{Mariotti} \& {MAGIC Collaboration}(2010)}]{Mariotti+2010}
{Mariotti}, M., \& {MAGIC Collaboration}. 2010, The Astronomer's Telegram,
  2916, 1

\bibitem[{{Massaro} {et~al.}(2009){Massaro}, {Giommi}, {Leto}, {Marchegiani},
  {Maselli}, {Perri}, {Piranomonte}, \& {Sclavi}}]{Massaro+2009}
{Massaro}, E., {Giommi}, P., {Leto}, C., {Marchegiani}, P., {Maselli}, A.,
  {Perri}, M., {Piranomonte}, S., \& {Sclavi}, S. 2009, \aap, 495, 691

\bibitem[{{Medvedev} \& {Loeb}(1999)}]{Medv-Loeb:99}
{Medvedev}, M.~V., \& {Loeb}, A. 1999, \apj, 526, 697

\bibitem[{{Melrose}(1980)}]{Melrose80}
{Melrose}, D.~B. 1980, {Plasma astrophysics: Nonthermal processes in diffuse
  magnetized plasmas. Volume 2 - Astrophysical applications}

\bibitem[{{Mose Mariotti}(2010)}]{Mariotti_FSRQ2010}
{Mose Mariotti}, M. 2010, The Astronomer's Telegram, 2684, 1

\bibitem[{{Nakar} {et~al.}(2011){Nakar}, {Bret}, \& {Milosavljevic}}]{Nakar+11}
{Nakar}, E., {Bret}, A., \& {Milosavljevic}, M. 2011, ArXiv 1104.5249

\bibitem[{{Narumoto} \& {Totani}(2006)}]{Naru-Tota:06}
{Narumoto}, T., \& {Totani}, T. 2006, \apj, 643, 81

\bibitem[{{Neronov} \& {Semikoz}(2009)}]{Nero-Semi:09}
{Neronov}, A., \& {Semikoz}, D.~V. 2009, \prd, 80, 123012

\bibitem[{{Neronov} {et~al.}(2011){Neronov}, {Semikoz}, {Tinyakov}, \&
  {Tkachev}}]{Nero-Semi-Tiny-Tkac:11}
{Neronov}, A., {Semikoz}, D.~V., {Tinyakov}, P.~G., \& {Tkachev}, I.~I. 2011,
  \aap, 526, 90

\bibitem[{{Neronov} \& {Vovk}(2010)}]{Nero-Vovk:10}
{Neronov}, A., \& {Vovk}, I. 2010, Science, 328, 73

\bibitem[{{Oh}(2001)}]{Oh2001}
{Oh}, S.~P. 2001, \apj, 553, 25

\bibitem[{{Padovani} \& {Giommi}(1995)}]{Padovani+1995}
{Padovani}, P., \& {Giommi}, P. 1995, \apj, 444, 567

\bibitem[{{Pfrommer} {et~al.}(2012){Pfrommer}, {Chang}, \& {Broderick}}]{PCB}
{Pfrommer}, C., {Chang}, P., \& {Broderick}, A.~E. 2012, \apj~in print,
  arXiv:1106.5505

\bibitem[{{Plaga}(1995)}]{Plag:95}
{Plaga}, R. 1995, \nat, 374, 430

\bibitem[{{Puchwein} {et~al.}(2011){Puchwein}, {Pfrommer}, {Springel},
  {Broderick}, \& {Chang}}]{Puchwein+2011}
{Puchwein}, E., {Pfrommer}, C., {Springel}, V., {Broderick}, A.~E., \& {Chang},
  P. 2011, arXiv:1107.3837

\bibitem[{{Raue} {et~al.}(2010){Raue}, {Lenain}, {Aharonian}, {Becherini},
  {Boisson}, {Clapson}, {Costamante}, {G{\'e}rard}, {Medina}, {de Naurois},
  {Punch}, {Rieger}, {So}, {Stawarz}, {Zech}, \&
  {H.~E.~S.~S.~Collaboration}}]{raue+10}
{Raue}, M., {et~al.} 2010, in Astronomical Society of the Pacific Conference
  Series, Vol. 427, Astronomical Society of the Pacific Conference Series, ed.
  {L.~Maraschi, G.~Ghisellini, R.~Della Ceca, \& F.~Tavecchio}, 302

\bibitem[{{Salamon} \& {Stecker}(1998)}]{Sala-Stec:98}
{Salamon}, M.~H., \& {Stecker}, F.~W. 1998, \apj, 493, 547

\bibitem[{{Silva} {et~al.}(2003){Silva}, {Fonseca}, {Tonge}, {Dawson}, {Mori},
  \& {Medvedev}}]{Silva+03}
{Silva}, L.~O., {Fonseca}, R.~A., {Tonge}, J.~W., {Dawson}, J.~M., {Mori},
  W.~B., \& {Medvedev}, M.~V. 2003, \apjl, 596, L121

\bibitem[{{Spitkovsky}(2008)}]{Spitkovsky+08}
{Spitkovsky}, A. 2008, \apjl, 682, L5

\bibitem[{{Sreekumar} {et~al.}(1998){Sreekumar}, {Bertsch}, {Dingus},
  {Esposito}, {Fichtel}, {Hartman}, {Hunter}, {Kanbach}, {Kniffen}, {Lin},
  {Mayer-Hasselwander}, {Michelson}, {von Montigny}, {Muecke}, {Mukherjee},
  {Nolan}, {Pohl}, {Reimer}, {Schneid}, {Stacy}, {Stecker}, {Thompson}, \&
  {Willis}}]{EGRET_EGRB1998}
{Sreekumar}, P., {et~al.} 1998, \apj, 494, 523

\bibitem[{{Stecker} \& {Venters}(2011)}]{Stec-Vent:11}
{Stecker}, F., \& {Venters}, T.~M. 2011, \apj, 736, 40

\bibitem[{{Stecker} {et~al.}(1992){Stecker}, {de Jager}, \&
  {Salamon}}]{Stec-deJa-Sala:92}
{Stecker}, F.~W., {de Jager}, O.~C., \& {Salamon}, M.~H. 1992, \apjl, 390, L49

\bibitem[{{Strong} {et~al.}(2004){Strong}, {Moskalenko}, \&
  {Reimer}}]{Stro-Mosk-Reim:04}
{Strong}, A.~W., {Moskalenko}, I.~V., \& {Reimer}, O. 2004, \apj, 613, 956

\bibitem[{{Sturrock}(1994)}]{Sturrock94}
{Sturrock}, P.~A. 1994, {Plasma Physics, An Introduction to the Theory of
  Astrophysical, Geophysical and Laboratory Plasmas} (Cambridge: Cambridge
  University Press)

\bibitem[{{Takahashi} {et~al.}(2012){Takahashi}, {Mori}, {Ichiki}, \&
  {Inoue}}]{Taka_etal:11}
{Takahashi}, K., {Mori}, M., {Ichiki}, K., \& {Inoue}, S. 2012, \apjl, 744, L7

\bibitem[{{Takahashi} {et~al.}(2008){Takahashi}, {Murase}, {Ichiki}, {Inoue},
  \& {Nagataki}}]{Taka_etal:08}
{Takahashi}, K., {Murase}, K., {Ichiki}, K., {Inoue}, S., \& {Nagataki}, S.
  2008, \apjl, 687, L5

\bibitem[{{Tavecchio} \& {Ghisellini}(2008)}]{Tavecchio+2008}
{Tavecchio}, F., \& {Ghisellini}, G. 2008, \mnras, 386, 945

\bibitem[{{Tavecchio} {et~al.}(2011){Tavecchio}, {Ghisellini}, {Bonnoli}, \&
  {Foschini}}]{Tave_etal:10b}
{Tavecchio}, F., {Ghisellini}, G., {Bonnoli}, G., \& {Foschini}, L. 2011,
  \mnras, 414, 3566

\bibitem[{{Tavecchio} {et~al.}(2010){Tavecchio}, {Ghisellini}, {Foschini},
  {Bonnoli}, {Ghirlanda}, \& {Coppi}}]{Tave_etal:10a}
{Tavecchio}, F., {Ghisellini}, G., {Foschini}, L., {Bonnoli}, G., {Ghirlanda},
  G., \& {Coppi}, P. 2010, \mnras, 406, L70

\bibitem[{{Taylor} {et~al.}(2011){Taylor}, {Vovk}, \&
  {Neronov}}]{Tayl-Vovk-Nero:11}
{Taylor}, A.~M., {Vovk}, I., \& {Neronov}, A. 2011, \aap, 529, A144

\bibitem[{{Tramacere} {et~al.}(2009){Tramacere}, {Giommi}, {Perri},
  {Verrecchia}, \& {Tosti}}]{Tram_etal:09}
{Tramacere}, A., {Giommi}, P., {Perri}, M., {Verrecchia}, F., \& {Tosti}, G.
  2009, \aap, 501, 879

\bibitem[{{Urry} \& {Padovani}(1995)}]{Urry+1995}
{Urry}, C.~M., \& {Padovani}, P. 1995, \pasp, 107, 803

\bibitem[{{Vall{\'e}e}(2011)}]{Vallee11}
{Vall{\'e}e}, J.~P. 2011, \nar, 55, 91

\bibitem[{{Venters}(2010)}]{Vent:10}
{Venters}, T.~M. 2010, \apj, 710, 1530

\bibitem[{{Waxman} \& {Miralda-Escude}(1996)}]{Waxman+96}
{Waxman}, E., \& {Miralda-Escude}, J. 1996, \apjl, 472, L89+

\bibitem[{{Weibel}(1959)}]{Weib:59}
{Weibel}, E.~S. 1959, Physical Review Letters, 2, 83

\bibitem[{{Widrow}(2002)}]{Widrow2002}
{Widrow}, L.~M. 2002, Reviews of Modern Physics, 74, 775

\bibitem[{{Yoon} \& {Davidson}(1987)}]{Yoon-Davi:87}
{Yoon}, P.~H., \& {Davidson}, R.~C. 1987, \pra, 35, 2718

\end{thebibliography}

\end{document}